\documentclass[a4paper,11pt]{article}
\usepackage[DIV12]{typearea}
\usepackage{longtable}
\usepackage{booktabs,colortbl}
\usepackage{multirow}
\usepackage{amsmath}
\usepackage{amssymb}
\usepackage{bbm}
\usepackage[utf8]{inputenc}
\usepackage{pdflscape}
\usepackage{xspace}
\usepackage[caption=false]{subfig}
\usepackage{nicefrac}
\usepackage{graphicx}
\usepackage{cite}  
\usepackage[small]{caption2}
\usepackage{mathtools}
\usepackage{nccfoots}
\usepackage{tikz}
\usetikzlibrary{calc,positioning}
\usepackage[normalem]{ulem}

\usepackage[all,cmtip]{xy}
\newlength{\xywd}
\newcommand{\xyrightarrow}[2][]{%
  \sbox{0}{$\scriptstyle#1$}%
  \xywd=\wd0
  \sbox{0}{$\scriptstyle#2$}%
  \ifdim\wd0>\xywd \xywd=\wd0 \fi
  \xymatrix@C\dimexpr\xywd+1em\relax{{}\ar[r]^{#2}_{#1}&{}}%
}

\DeclareMathOperator{\re}{Re}
\DeclareMathOperator{\im}{Im}

\newcommand{\rep}[1]{\ensuremath\boldsymbol{#1}}
\newcommand{\crep}[1]{\ensuremath\bar{\boldsymbol{#1}}}
\newcommand{\Z}[1]{\ensuremath{\mathbbm{Z}_{#1}}} 

\newcommand{\SU}[1]{\ensuremath{\mathrm{SU}(#1)}}
\newcommand{\SL}[1]{\ensuremath{\mathrm{SL}(#1)}}
\newcommand{\U}[1]{\ensuremath{\mathrm{U}(#1)}}
\newcommand{\E}[1]{\ensuremath{\mathrm{E}_{#1}}}

\newcommand{\I}{\mathrm{i}}
\newcommand{\Id}{\mathbbm{1}}

\newcommand{\CP}{\ensuremath{\mathcal{CP}}\xspace}
\newcommand{\x}{\ensuremath{\times}}
\newcommand{\vev}[1]{\ensuremath{\langle{#1}\rangle}}

\newcommand{\w}{\ensuremath{\omega}}

\usepackage{xcolor}
\definecolor{darkgreen}{HTML}{109930}
\definecolor{pink}{rgb}{0.858, 0.188, 0.478}
\definecolor{bluelink}{HTML}{0000EE} 

\advance \headheight by 3.0truept       
\usepackage[pdftex,hidelinks]{hyperref}
\hypersetup{
    pdftitle = {The first string-derived eclectic flavor model with realistic phenomenology},
    pdfauthor = {Baur, Nilles, Ramos-Sanchez, Trautner, Vaudrevange}
}

\begin{document}

\begin{titlepage}

\begin{flushright}
\normalsize{TUM-HEP 1408/22}
\end{flushright}

\vspace*{1.0cm}
\begin{center}%
{\Large\textbf{\boldmath The first string-derived eclectic flavor model \\ with realistic phenomenology\unboldmath}} \\[0.3cm]

\vspace{1cm}
\textbf{Alexander~Baur}$^{a,b}$,
\textbf{Hans~Peter~Nilles}$^{c}$,
\textbf{Sa\'ul Ramos--S\'anchez}$^{b}$, 
\textbf{Andreas Trautner}$^{d}$, \\ and 
\textbf{Patrick~K.S.~Vaudrevange}$^{a}$
\Footnote{*}{%
\href{mailto:alexander.baur@tum.de;nilles@th.physik.uni-bonn.de;ramos@fisica.unam.mx;trautner@mpi-hd.mpg.de;patrick.vaudrevange@tum.de}{\textcolor{bluelink}{\underline{\tt Electronic addresses} }}
}
\\[5mm]
\textit{$^a$\small Physik Department, Technische Universit\"at M\"unchen,\\ James-Franck-Stra\ss e 1, 85748 Garching, Germany}
\\[2mm]
\textit{$^b$\small Instituto de F\'isica, Universidad Nacional Aut\'onoma de M\'exico,\\ POB 20-364, Cd.Mx. 01000, M\'exico}
\\[2mm]
\textit{$^c$\small Bethe Center for Theoretical Physics and Physikalisches Institut der Universit\"at Bonn,\\ Nussallee 12, 53115 Bonn, Germany}
\\[2mm]
\textit{$^d$\small Max-Planck-Institut f\"ur Kernphysik, \\ Saupfercheckweg 1, 69117 Heidelberg, Germany}
\end{center}

\vspace{1cm}
\vspace*{1.0cm}

\begin{abstract}
Eclectic flavor groups arising from string compactifications combine the power of modular and 
traditional flavor symmetries to address the flavor puzzle. This top-down scheme determines the 
representations and modular weights of all matter fields, imposing strict constraints on 
the structure of the effective potential, which result in controlled corrections. We study
the lepton and quark flavor phenomenology of an explicit, potentially realistic example model based on a  
$\mathbbm T^6/\Z3\x\Z3$ orbifold compactification of the heterotic string that gives rise to 
an $\Omega(2)$ eclectic flavor symmetry. We find that the interplay of flavon alignment and the 
localization of the modulus in the vicinity of a symmetry-enhanced point leads to 
naturally protected fermion mass hierarchies, favoring normal-ordered neutrino 
masses arising from a see-saw mechanism. We show that our model can reproduce all observables 
in the lepton sector with a small number of parameters and deliver predictions for so far undetermined 
neutrino observables. Furthermore, we extend the fit to quarks and find that K\"ahler corrections are
instrumental in obtaining a successful simultaneous fit to the quark and lepton sectors.
\end{abstract}

\end{titlepage}

\newpage

\section{Introduction}

Top-down (TD) model building from string theory leads to the
concept of the eclectic flavor group~\cite{Nilles:2020nnc,Nilles:2020kgo,Nilles:2020tdp,Nilles:2020gvu} 
that includes traditional and modular flavor symmetries in the framework of
``Local Flavor Unification''~\cite{Baur:2019iai,Baur:2019kwi}. Any discussion 
of the flavor problem should consider both, traditional and modular flavor
symmetries, as they give important restrictions on the K\"ahler potential
and superpotential of the theory. Spontaneous breaking of the
eclectic flavor group exhibits a subtle interplay of the
vacuum expectation values (VEVs) of flavon and moduli
fields~\cite{Baur:2021bly} that allow for a hierarchical pattern of masses
and mixing angles of quarks and leptons. While the appearance
of the eclectic flavor group is automatic in the
TD approach, it could also be discussed within the bottom-up (BU)
approach, where potential modular symmetries are contained
in the outer automorphisms of the traditional flavor group~\cite{Nilles:2020nnc,Baur:2019iai,Baur:2019kwi}.
In general, only part of the eclectic flavor group is linearly realized and the
traditional flavor symmetry is enhanced at certain points or
sub-loci in moduli space. This provides the basis of
``Local Flavor Unification'' at these regions of enhanced
symmetry. Ultimately, this does lead to a flavor scheme that
incorporates both the quark and lepton sectors. 

Since their introduction in BU constructions~\cite{Feruglio:2017spp}, 
most of the attempts for a description of flavor with modular flavor symmetries
have concentrated on the lepton sector alone, see e.g.~\cite{Kobayashi:2018vbk,Penedo:2018nmg,Criado:2018thu,Kobayashi:2018scp,Novichkov:2018yse,Ding:2019xna,Liu:2019khw,deMedeirosVarzielas:2019cyj,Feruglio:2019ybq,Ding:2019zxk,Novichkov:2020eep,Liu:2020msy,Yao:2020zml,Liu:2021gwa,Behera:2020sfe}
and references therein. Even though apparently more difficult to accomodate, there have been some fits of the flavor 
parameters that include the quark sector, see 
e.g.~\cite{deAnda:2018ecu,Okada:2018yrn,Kobayashi:2019rzp,Lu:2019vgm,King:2020qaj,Okada:2020ukr,Ding:2020zxw,Abbas:2020qzc,Zhao:2021jxg,Chen:2021zty,King:2021fhl,Kuranaga:2021ujd,Ding:2021eva,Nomura:2022boj}. 
Yet no clearly favored scheme has emerged.
There are many choices of flavor groups,
representations of these groups as well as parameters in the
action that provide reasonable fits, but one still did not find a
baseline theory or a fundamental principle through the
BU considerations. Furthermore, the predictivity of these BU models may 
be challenged by the arbitrariness of their K\"ahler potential~\cite{Chen:2019ewa}.
The TD approach is much more restrictive
and it remains to be seen whether a realistic fit to the data
can be achieved at all. The present paper is meant to be a
first attempt for a global description of flavor in the
quark and lepton sector from a TD perspective. It will also
serve as a benchmark scheme that allows a comparison to
previous BU constructions as it will indicate which properties
of the construction and choice of parameters will be most
relevant. We shall see, for example, that nontrivial
parameters in the K\"ahler potential (usually ignored 
in the BU approach) might play an important role.

To initiate a TD construction of flavor we select a most
promising scheme of a string compactification with an
elliptic fibration based on the $\mathbbm T^2/\Z3$ orbifold~\cite{Baur:2019iai,Nilles:2020kgo,Nilles:2020tdp}.
It leads to the traditional flavor group $\Delta(54)$,
the discrete modular flavor group 
$\Gamma'_3\cong T'=\mathrm{SL}(2,3)\cong[24,3]$ with eclectic
flavor group $\Omega(2)\cong[1944,3448]$. Matter fields appear
in twisted sectors with nontrivial representations of
$\Delta(54)$ and $T'$. Full details of this general
flavor scheme can be found in table 2 of our previous
paper~\cite{Baur:2021bly}. The choice of the possible representations is
quite restricted, as in other TD 
scenarios~\cite{Kobayashi:2018rad,Ohki:2020bpo,Kikuchi:2020frp,Almumin:2021fbk}. 
It is therefore difficult to compare
this approach to BU constructions where, even for the
same group $T'$, typically different representations
have been chosen~\cite{Liu:2019khw,Lu:2019vgm,Kuranaga:2021ujd}.

The next step in our program is the choice of a
(semi-)realistic string construction with Standard Model
gauge group $\SU3\x\SU2\x\U1$, three
families of quarks and leptons and suitable Higgs-doublets.
Here we concentrate on the constructions of ref.~\cite{Carballo-Perez:2016ooy,Olguin-Trejo:2018wpw}
based on  $\mathbbm T^6/\Z3\x\Z3$ orbifolds where the gauge and flavor
structure has been explicitly worked out. Several
classes of models with eclectic flavor group $\Omega(2)$
have been identified, as shown in table 3 of ref.~\cite{Baur:2021bly}. We
choose here the simplest example (class A) with properties
displayed in table~\ref{tab:MSSM-model447}. Twisted fields all have the same
modular weight $n=\nicefrac{-2}3$, transform as $\rep3_2$ representations
of $\Delta(54)$ and $\rep1\oplus\rep2'$ representations of the
$T'$ modular group.\footnote{This is the simplest class of models as we have
only representations of class $\rep3_2$, and none of $\rep3_1$
and very restrictive values for modular weights both in
twisted and untwisted sector.} 
The pattern of the spontaneous breaking of the eclectic flavor group has
been discussed in our earlier paper~\cite{Baur:2021bly} (see Tables
1, 2 and 3 there). The simplicity of the scheme leads to
severe restrictions on superpotential and K\"ahler potential
as we shall discuss later in sections~\ref{sec:mass-matrices} and~\ref{sec:Kaehler}. 
Still, it has to be stressed that the K\"ahler potential is not
diagonal (as usually assumed in the BU approach, with some 
exceptions~\cite{Chen:2019ewa,Lu:2019vgm}) and this
will be relevant for the global fit to the data.

The model allows for a successful fit of flavor both in the
quark and lepton sector. It predicts a see-saw mechanism
in the lepton sector and a ``normal hierarchy'' for
neutrino masses. Hierarchies for masses and mixing angles
appear from a subtle interplay of aligned flavon VEVs and
the location of the modular parameter in the vicinity of
fixed points, as a result of ``Local Flavor Unification''.

The paper is structured as follows. In section~\ref{sec:DiscussionModelDefinition} 
we present the explicit string model, matter representations (table~\ref{tab:MSSM-model447}),
superpotential (section~\ref{sec:mass-matrices}) and K\"ahler potential (\ref{sec:Kaehler}).
Section~\ref{sec:StepwiseBreaking} contains the step-wise symmetry breaking and
the resulting hierarchical structure in a qualitative form. 
Section~\ref{sec:NumericalAnalysisLepton} 
will be devoted to the numerical analysis of the lepton
sector, which will be completed to include also quarks in
section~\ref{sec:NumericalAnalysisBoth}. In section~\ref{sec:conclusions} 
we shall summarize our results and give an outlook to future developments.
Our appendices include details on the structure of the K\"ahler
corrections, our numerical analysis and the full massless matter spectrum 
of our model.
\vspace*{1mm}

\section{A string theory model with eclectic flavor symmetries}
\label{sec:DiscussionModelDefinition}

\subsection{Model definition}
\label{sec:model}
\begin{table}[t!]
{\centering
\resizebox{\textwidth}{!}{ 
\begin{tabular}{|c||c|c|c|c|c|c||c|c||c|c|c|c|c|c|c|c|}
\hline
             & \multicolumn{6}{c||}{quarks and leptons} & \multicolumn{2}{c||}{Higgs fields} & \multicolumn{8}{c|}{flavons} \\
\hline\hline
label        & $q$           & $\bar u$       & $\bar d$      & $\ell$         & $\bar e$  & $\bar\nu$ & $H_\mathrm{u}$         & $H_\mathrm{d}$          
             & $\varphi_\mathrm{e}$ & $\varphi_\mathrm{u}$ & $\varphi_\nu$ & $\phi^0$ & $\phi^0_\mathrm{M}$& $\phi^0_\mathrm{e}$& $\phi^0_\mathrm{u}$ & $\phi^0_\mathrm{d}$\\
\hline
$\SU3_c$     & $\rep3$       & $\crep3$       & $\crep3$      & $\rep1$        & $\rep1$   & $\rep1$   & $\rep1$       & $\rep1$        
             & $\rep1$       & $\rep1$        & $\rep1$       & $\rep1$        & $\rep1$   & $\rep1$   & $\rep1$       &  $\rep1$\\
$\SU2_L$     & $\rep2$       & $\rep1$        & $\rep1$       & $\rep2$        & $\rep1$   & $\rep1$   & $\rep2$       & $\rep2$        
             & $\rep1$       & $\rep1$        & $\rep1$       & $\rep1$        & $\rep1$   & $\rep1$   & $\rep1$       &  $\rep1$\\
$\U1_Y$      & $\nicefrac16$ & $-\nicefrac23$ & $\nicefrac13$ & $-\nicefrac12$ & $1$       & $0$       & $\nicefrac12$ & $-\nicefrac12$ 
             & $0$           & $0$            & $0$           & $0$            & $0$       & $0$       & $0$           & $0$\\
$\Delta(54)$ & $\rep3_2$     & $\rep3_2$      & $\rep3_2$     & $\rep3_2$      & $\rep3_2$ & $\rep3_2$ & $\rep1$       & $\rep1$        
             & $\rep3_2$     & $\rep3_2$      & $\rep3_2$     & $\rep1$        & $\rep1$   & $\rep1$   & $\rep1$       &  $\rep1$\\
$T'$         & $\rep2'\oplus\rep1$ & $\rep2'\oplus\rep1$ & $\rep2'\oplus\rep1$ & $\rep2'\oplus\rep1$ & $\rep2'\oplus\rep1$ & $\rep2'\oplus\rep1$ & $\rep1$       & $\rep1$        
             & $\rep2'\oplus\rep1$ & $\rep2'\oplus\rep1$ & $\rep2'\oplus\rep1$ & $\rep1$ & $\rep1$   & $\rep1$   & $\rep1$       &  $\rep1$\\
$\Z9^R$      & $1$           & $1$            & $1$           & $1$            & $1$       & $1$       & $0$           & $0$ 
             & $1$           & $1$            & $1$           & $0$            & $0$       & $0$       & $0$           & $0$\\
$n$          & $-\nicefrac23$ & $-\nicefrac23$ & $-\nicefrac23$ & $-\nicefrac23$ &  $-\nicefrac23$ &  $-\nicefrac23$ & $0$ & $0$ 
             & $-\nicefrac23$ & $-\nicefrac23$ & $-\nicefrac23$ & $0$            & $0$       & $0$       & $0$           & $0$\\
\hline
$\Z3^{}$     & $1$           & $1$            & $\omega$      & $\omega$       & $1$       & $1$       & $1$           & $1$
             & $1$           & $\omega$       & $\omega^2$    & $1$            & $1$       & $\omega^2$& $\omega^2$    & $\omega^2$\\
$\Z3^{}$     & $\omega^2$    & $\omega^2$     & $1$           & $1$            & $\omega^2$& $\omega^2$& $1$           & $1$
             & $\omega^2$    & $1$            & $\omega$      & $1$            & $1$       & $\omega^2$& $\omega^2$    & $\omega^2$\\
$\Z3^{}$     & $1$           & $1$            & $\omega$      & $1$            & $1$       & $1$       & $1$           & $1$
             & $1$           & $\omega^2$     & $1$           & $1$            & $1$       & $1$       & $\omega$      & $\omega^2$\\
\hline
\end{tabular}
}
\caption{
MSSM matter and flavon states of a \Z3\x\Z3 heterotic orbifold realization of a model endowed with 
$\Omega(2)$ eclectic flavor symmetry. We display quantum numbers with respect to the SM gauge group, 
the traditional flavor symmetry $\Delta(54)$, the finite modular symmetry $T'$, the modular weights 
$n$ and the $\Z9^R$ discrete $R$-symmetry arising from the full 10D orbifold compactification. We use 
the results from refs.~\cite{Nilles:2020kgo,Nilles:2020gvu} to identify these quantum numbers.
We provide the additional unbroken $\Z3\x\Z3\x\Z3$ symmetries (with $\omega:=e^{\nicefrac{2\pi\I}3}$), 
arising from the compact dimensions orthogonal to the $\mathbbm{T}^2/\Z3$ sector where $\Omega(2)$ is realized.
As shown in appendix~\ref{app:spectrum}, the fields carry additional gauge \U1 charges that 
distinguish e.g.~$\phi^0$ and $\phi^0_M$. The subindices $\mathrm{e},\nu,\mathrm{u},\mathrm{d}$ 
label the flavons associated with the respective leptons and quarks. The electron and down-quark 
sectors share the same flavon triplet $\varphi_\mathrm{e}$, as discussed in section~\ref{sec:mass-matrices}.
Besides these relevant matter states, the model contains the vectorlike exotic fields 
shown in table~\ref{tab:exotics-model447}.
\label{tab:MSSM-model447}
}
}
\end{table}

Let us consider a fully consistent model based on the \E8\x\E8 heterotic string 
containing an eclectic flavor symmetry $\Omega(2) \cong [1944,3448]$, consisting 
of the traditional flavor group $\Delta(54)$, the finite modular group $T'$ and 
a $\Z{9}^R$ $R$-symmetry. As usual, there is an additional $\Z2^\CP$ \CP-like
modular symmetry that acts as a simultaneous outer automorphism on all of these groups 
and enlarges the eclectic flavor symmetry of this setting to order $3888$.
The \CP-like transformation is generally spontaneously broken by the VEV of the modulus 
as well as by the VEVs of flavon fields thereby giving rise to \CP violation at low energies.
It has been known that $\mathbbm T^6/\Z3\x\Z3$ $(1,1)$ orbifold 
compactifications\footnote{See ref.~\cite{Fischer:2012qj} for orbifold nomenclature.}
of the heterotic string with some vanishing Wilson lines can yield an MSSM-like
massless spectrum equipped with a $\Delta(54)$ traditional flavor 
symmetry~\cite{Kobayashi:2006wq,Carballo-Perez:2016ooy,Olguin-Trejo:2018wpw}. This symmetry 
arises from a two-dimensional $\mathbbm T^2/\Z3$ orbifold sector, whose modular symmetries 
complete the eclectic scenario~\cite{Baur:2019iai,Baur:2019kwi,Nilles:2020nnc,Nilles:2020kgo,Nilles:2020tdp,Nilles:2020gvu}.
It leads to a picture where the $\Omega(2)$ eclectic symmetry of this sector is 
extended by three extra \Z3 symmetries arising from the other compact dimensions, 
which can be regarded as ``shaping symmetries''.

We consider a particular string orbifold defined by the background gauge-lattice shifts
\begin{subequations}
\begin{eqnarray}
  V_1 &=& \left(-\tfrac{1}{2}, \tfrac{1}{6}, \tfrac{1}{6}, \tfrac{1}{6}, \tfrac{1}{6}, \tfrac{1}{6}, \tfrac{1}{6}, \tfrac{1}{6}\right),  \left(-\tfrac{1}{6}, -\tfrac{1}{6}, \tfrac{1}{6}, \tfrac{1}{6}, \tfrac{1}{6}, \tfrac{1}{6}, \tfrac{1}{2}, \tfrac{7}{6}\right)\;,\\
  V_2 &=& \left(-\tfrac{2}{3}, -\tfrac{2}{3}, -\tfrac{1}{3},     0,     0,     0,     1, \tfrac{4}{3}\right),  \left(-\tfrac{5}{6}, \tfrac{5}{6}, \tfrac{1}{6}, \tfrac{1}{6}, \tfrac{1}{2}, \tfrac{7}{6}, -\tfrac{5}{6}, \tfrac{5}{6}\right)\;,
\end{eqnarray}
and Wilson lines
\begin{eqnarray}
  A_1 = A_2 & = & \left(   -1, \tfrac{1}{3}, -\tfrac{1}{3},    -1,     0,     0, \tfrac{4}{3}, -\tfrac{2}{3}\right),  \left(\tfrac{3}{2}, -\tfrac{1}{2}, -\tfrac{1}{6}, \tfrac{1}{2}, \tfrac{5}{6}, \tfrac{5}{6}, -\tfrac{5}{6}, -\tfrac{1}{6}\right)\;,\\
  A_3 = A_4 & = & \left(-\tfrac{1}{3}, -\tfrac{2}{3},     1, \tfrac{4}{3}, \tfrac{1}{3}, \tfrac{4}{3}, \tfrac{2}{3},    -1\right),  \left(\tfrac{1}{2}, \tfrac{1}{2}, \tfrac{1}{2}, \tfrac{1}{2}, \tfrac{1}{2}, \tfrac{3}{2}, -\tfrac{1}{2}, \tfrac{1}{2}\right)\;.
\end{eqnarray}
\end{subequations}
The Wilson lines associated with the last two compact dimensions are chosen to be trivial, 
i.e.\ $A_5=A_6=0$. This is the condition for this $\mathbbm T^2/\Z3$ orbifold sector to 
yield the eclectic flavor symmetry $\Omega(2)$. One can further show that the three extra 
$\Z3$ discrete symmetries that are left unbroken from the orbifold action on the first
four compact dimensions, are orthogonal to the $\Omega(2)$ eclectic group. 
From the gauge degrees of freedom, the unbroken 4D gauge group of this model is 
$\SU3_c\x\SU2_L\x\U1_Y\x[\SU4\x\U1^9]$. By using e.g.\ the \texttt{orbifolder}~\cite{Nilles:2011aj}, 
one finds that the $\mathcal N=1$ massless matter spectrum includes three generations of 
quark and lepton superfields as well as a pair of Higgs fields and various flavons, all 
listed in table~\ref{tab:MSSM-model447}. Additionally, this model includes several 
vectorlike exotics summarized separately in table~\ref{tab:exotics-model447}, which decouple 
from the low-energy dynamics when some singlets $s_i$ develop VEVs close to the string scale. 
Details of the entire massless spectrum are given in appendix~\ref{app:spectrum}. We provide 
the SM gauge quantum numbers, as well as the discrete flavor charges for all phenomenologically 
relevant matter states in table~\ref{tab:MSSM-model447}, which we discuss in the following.

\begin{table}[t!]
{\centering
\begin{tabular}{|r|c|l||r|c|l|}
\hline
   \# & irrep  & labels & \# & irrep  & labels \\
\hline\hline
  101 & $\left(\rep{1},\rep{1}\right)_0$                       & $s_i$       &&&\\
   51 & $\left(\rep{1},\rep{1}\right)_{-\nicefrac{1}{3}}$      & $V_i$       & 51 & $\left(\rep{1},\rep{1}\right)_{\nicefrac{1}{3}}$       & $\bar V_i$  \\
   14 & $\left(\rep{1},\rep{1}\right)_{-\nicefrac{2}{3}}$      & $X_i$       & 14 & $\left(\rep{1},\rep{1}\right)_{\nicefrac{2}{3}}$       & $\bar X_i$  \\
   10 & $\left(\rep{1},\rep{2}\right)_{-\nicefrac{1}{2}}$      & $ L_i$      & 10 & $\left(\rep{1},\rep{2}\right)_{\nicefrac{1}{2}}$       & $ \bar L_i$ \\
    9 & $\left(\crep{3},\rep{1}\right)_{\nicefrac{1}{3}}$      & $\bar D_i$  &  9 & $\left(\rep{3},\rep{1}\right)_{-\nicefrac{1}{3}}$      & $D_i$       \\    
    8 & $\left(\rep{1},\rep{2}\right)_{-\nicefrac{1}{6}}$      & $W_i$       &  8 & $\left(\rep{1},\rep{2}\right)_{\nicefrac{1}{6}}$       & $\bar W_i$  \\
    2 & $\left(\crep{3},\rep{1}\right)_{-\nicefrac{2}{3}}$     & $\bar U_i$  &  2 & $\left(\rep{3},\rep{1}\right)_{\nicefrac{2}{3}}$       & $U_i$       \\   
    4 & $\left(\crep{3},\rep{1}\right)_0$                      & $Z_i$       &  4 & $\left(\rep{3},\rep{1}\right)_0$                       & $\bar Z_i$  \\
    1 & $\left(\crep{3},\rep{1}\right)_{-\nicefrac{1}{3}}$     & $Y$         &  1 & $\left(\rep{3},\rep{1}\right)_{\nicefrac{1}{3}}$       & $\bar Y$    \\
\hline
\end{tabular}
\caption{Vectorlike exotic matter states of a \Z3\x\Z3 heterotic orbifold realization of a model endowed with $\Omega(2)$ 
eclectic flavor symmetry. In parenthesis, we display the gauge quantum number under $\SU3_c\x\SU2_L$ and the subindices 
denote the hypercharges. \label{tab:exotics-model447}}
}
\end{table}

\subsection{Flavor symmetry representations}
\label{sec:flavorReps}

This model belongs to the category A of the models classified in table 3 of ref.~\cite{Baur:2021bly}.
The assignment of symmetry representations under the $\Omega(2)$ eclectic flavor symmetry is fairly simple
because it is entirely determined by the modular weight $n$ of each field under the $\SL{2,\Z{}}_T$ 
group of modular transformations of the K\"ahler modulus $T$~\cite{Baur:2021bly}.\footnote{
As pointed out in \cite{Baur:2021bly}, the fact that the flavor symmetry representations are entirely fixed 
by knowing the modular weight might be conjectured to be a general feature of TD constructions.
Other examples for this are~\cite{Baur:2020jwc,Baur:2021mtl,Kikuchi:2021ogn,Almumin:2021fbk,Ishiguro:2021ccl,Kikuchi:2022txy}, 
while virtually all BU constructions violate this rule.} 
We follow the notation of~\cite{Nilles:2020kgo} and denote generic fields by $\Phi_{n}$ to indicate 
their transformation behavior under $\Omega(2)$.
Quarks, leptons, and flavons $\varphi_{\mathrm{u}, \mathrm{e},\nu}$ correspond to 
$\Phi_{\nicefrac{-2}3}$ fields with modular weights $n=\nicefrac{-2}3$, while the Higgs 
fields and flavons $\phi^0$ form $\Phi_0$ fields with trivial modular weights. While $\Phi_0$ 
fields are trivial singlets under all flavor symmetries, $\Phi_{\nicefrac{-2}3}$ are flavor 
triplets transforming simultaneously as $\rep{3}_2$ of the traditional flavor group $\Delta(54)$, 
as well as $\rep{2'}\oplus\rep{1}$ of the finite modular group $T'$~\cite{Baur:2019kwi,Baur:2019iai}.
In addition, $\Phi_{\nicefrac{-2}{3}}$ fields have $\Z{9}^R$-charge $1$~\cite{Nilles:2020gvu}. 

Next to the expectation value of the modulus $\langle T\rangle$ also the VEVs of the flavon triplets 
$\varphi_\mathrm{i}$ contribute to the breaking of the flavor symmetries of the model leading 
to the patterns described in our previous work~\cite{Baur:2021bly}.

The generators of the three-dimensional representation $\rep{3}_2$ of the traditional $\Delta(54)$ 
flavor symmetry are given by the matrices
\begin{equation}
\label{eq:32TraditionalGenerators}
\rho_{\rep{3}_2}(\mathrm{A}) ~:=~ \left(\begin{array}{ccc} 0 & 1 & 0 \\ 0 & 0 & 1 \\ 1 & 0 & 0 \end{array}\right)\;, \quad 
\rho_{\rep{3}_2}(\mathrm{B}) ~:=~ \left(\begin{array}{ccc} 1 & 0 & 0 \\ 0 & \omega & 0 \\ 0 & 0 & \omega^2 \end{array}\right) \quad\mathrm{and,}\quad 
\rho_{\rep{3}_2}(\mathrm{C}) ~:=~ -\left(\begin{array}{ccc} 1 & 0 & 0 \\ 0 & 0 & 1 \\ 0 & 1 & 0 \end{array}\right),
\end{equation}
where $\omega:=\exp(\nicefrac{2\pi\I}{3})$, such that for $g \in \Delta(54)$,
\begin{equation}
\Phi_{\nicefrac{-2}{3}} ~\stackrel{g}{\longrightarrow}~ \rho_{\rep{3}_2}(g)\,\Phi_{\nicefrac{-2}{3}} \;.
\end{equation}
Furthermore, the superpotential $\mathcal{W}$ transforms under $\mathrm{C}$ as 
$\mathcal{W} \stackrel{\mathrm{C}}{\rightarrow} -\mathcal{W}$, such that the $\Z{2}$ subgroup of 
$\Delta(54)$ generated by $\mathrm{C}$ corresponds to an $R$-symmetry. This also
implies that the superpotential transforms as a $\Delta(54)$ nontrivial singlet $\rep1'$,
see also~\cite[Table 2]{Baur:2021bly}.

For modular transformations,
\begin{equation}
\gamma ~=~ \begin{pmatrix}a&b\\ c&d\end{pmatrix} ~\in~ \SL{2,\Z{}}_T\;,
\end{equation}
the transformations of the relevant matter fields and the superpotential are given by
\begin{equation}\label{eq:PhiModularTrafo}
\Phi_{\nicefrac{-2}3} ~\stackrel{\gamma}{\longrightarrow}~ (c\,T+d)^{\nicefrac{-2}3}\, \rho(\gamma)\,\Phi_{\nicefrac{-2}3} \qquad\mathrm{and}\qquad 
          \mathcal{W} ~\stackrel{\gamma}{\longrightarrow}~ (c\,T+d)^{-1}\, \mathcal{W}\;,
\end{equation}
with explicit representation matrices for the generators $\mathrm{S}$ and $\mathrm{T}$ 
of the modular group
\begin{equation}
\label{eq:32ModularGenerators}
\rho(\mathrm{S}) ~:=~ \frac{\I}{\sqrt{3}}\left(\begin{array}{ccc} 1 & 1 & 1 \\ 1 & \omega^2 &\omega \\ 1 & \omega & \omega^2 \end{array}\right) \qquad\mathrm{and}\qquad
\rho(\mathrm{T}) ~:=~ \left(\begin{array}{ccc} \omega^2 & 0 & 0 \\ 0 & 1 &0 \\ 0 & 0 & 1 \end{array}\right)\;.
\end{equation}
The $\Z{9}^R$ $R$-symmetry generated by the sublattice rotation $\hat{\mathrm{R}}$ 
(see~\cite{Nilles:2020gvu} for details) acts as
\begin{equation}\label{eq:Z9R}
\Phi_{\nicefrac{-2}{3}} ~\stackrel{\hat{\mathrm{R}}}{\longrightarrow}~ \exp(\nicefrac{2\pi\I}{9})\,\Phi_{\nicefrac{-2}{3}} 
\qquad\mathrm{and}\qquad
\mathcal{W} ~\stackrel{\hat{\mathrm{R}}}{\longrightarrow}~ \omega\,\mathcal{W}\;.
\end{equation}
Finally, the $\Z3\x\Z3\x\Z3$ charges shown in table~\ref{tab:MSSM-model447} can be 
understood by the localization of the fields in the compact dimensions orthogonal to the 
$\mathbbm T^2/\Z3$ orbifold sector, supporting the geometric intuition of the eclectic 
picture. For completeness, let us recall that the generator of the additional $\Z2^\CP$ \CP-like symmetry
of our TD eclectic scenario acts on the modulus as $T\stackrel{\CP}{\longrightarrow}-\bar T$ while mapping 
$\Phi_{\nicefrac{-2}3}\stackrel{\CP}{\longrightarrow}\bar\Phi_{\nicefrac{-2}3}$~\cite{Baur:2019kwi,Baur:2019iai},
where bars denote complex conjugation (in agreement with results in the BU approach~\cite{Novichkov:2019sqv}).\footnote{
In general, transformations of the \CP-type are accompanied by a non-trivial representation 
matrix and an automorphy factor, see e.g.\ \cite[eq.~(3)]{Baur:2021bly}.}

\subsection[T' modular forms]{\boldmath $T'$ modular forms\unboldmath}
\label{sec:modularForms}

In order to determine the structure of the effective action of the model, let us recall
the properties of the modular forms that are relevant to build the couplings among the 
matter fields of table~\ref{tab:MSSM-model447}. 
For the leading terms in the superpotential we only need the modular forms of 
level 3 and weight 1, which form a doublet representations of $\Gamma'_3 \cong T'$ 
and can be expressed as~\cite{Liu:2019khw,Nilles:2020kgo}
\begin{equation}\label{eq:ModularFormY1}
\hat{Y}^{(1)}(T) ~=~ 
\begin{pmatrix} \hat{Y}_1(T) \\ \hat{Y}_2(T) \end{pmatrix} ~=~
\begin{pmatrix}
-3\sqrt{2}\,\frac{\eta^3(3\,T)}{\eta(T)}\\
3\frac{\eta^3(3\,T)}{\eta(T)} + \frac{\eta^3(T/3)}{\eta(T)}
\end{pmatrix}\;,
\end{equation}
where $\eta(T)$ is the Dedekind $\eta$ function.
Under a modular transformation $\gamma \in \SL{2,\Z{}}_T$, this transforms as
\begin{equation}
\hat{Y}^{(1)}(T) \stackrel{\gamma}{\longrightarrow} (c\,T+d)\,\rho_{\rep{2}''}(\gamma)\,\hat{Y}^{(1)}(T)\;,
\end{equation}
where $\rho_{\rep{2}''}(\gamma)$ denotes the $\rep{2}''$ representation of $T'$, which can be generated by
\begin{equation}
\rho_{\rep{2}''}(\mathrm{S}) ~=~ -\dfrac{\I}{\sqrt{3}}\begin{pmatrix}1&\sqrt{2}\\\sqrt{2}&1\end{pmatrix}
\qquad\mathrm{and}\qquad
\rho_{\rep{2}''}(\mathrm{T}) ~=~ \begin{pmatrix}\omega&0\\0&1\end{pmatrix}\;.
\end{equation}
Using $q:=\exp\left(2\pi\I\,T\right)$, we will make use of the ``$q$-expansion'' of $\hat{Y}^{(1)}(T)$ given by
\begin{subequations}\label{eq:Qexpansion}
	\begin{align}
	\hat{Y}_1(T) ~&=~ -3\,\sqrt{2}\,q^{\nicefrac{1}{3}}\,(1+q+2q^2+2q^4+q^5+2q^6+q^8+2q^9+\dots)\;, \\
	\hat{Y}_2(T) ~&=~ 1 + 6q+6q^3+6q^4+12q^7+6q^9+\dots\;.
	\end{align}
\end{subequations}
From these expansions, the behavior of the modular forms for large $\im T$ can be read off: 
$\hat Y_2(T)\to1$ while $\hat{Y}_1(T)\to0$. Hence, for large $\im T$, the modular form of 
weight $1$ is hierarchically structured.

Let us mention here the appearance of an approximate accidental symmetry because of
the special behavior of these modular forms under the transformations 
$T\rightarrow T+\nicefrac{3}{4}$ and $T\rightarrow T + \nicefrac{3}{2}$. Using
\begin{subequations}
\begin{align}
T \rightarrow T+\nicefrac{3}{4}&:\quad q ~\rightarrow~ \exp\left(2 \pi \I (T + \nicefrac{3}{4})\right) ~=~ -\I\,q\;, \\
T \rightarrow T+\nicefrac{3}{2}&:\quad q ~\rightarrow~ \exp\left(2 \pi \I (T + \nicefrac{3}{2})\right) ~=~ -q\;,
\end{align}
\end{subequations}
and the $q$-expansions of eqs.~\eqref{eq:Qexpansion}, we find the approximate transformations
\begin{subequations}\label{eq:ModformApproximateSymmetry}
	\begin{align}
	T \rightarrow T+\nicefrac{3}{4}:&\quad 
	\begin{pmatrix}\hat{Y}_1(T)\\\hat{Y}_2(T)\end{pmatrix} ~\rightarrow~ \begin{pmatrix}-\I\,\hat{Y}_1(T)\\\hat{Y}_2(T)\end{pmatrix} + \mathcal{O}(q)\;,\\
	T \rightarrow T+\nicefrac{3}{2}:&\quad 
	\begin{pmatrix}\hat{Y}_1(T)\\\hat{Y}_2(T)\end{pmatrix} ~\rightarrow~ \begin{pmatrix}-\hat{Y}_1(T)\\\hat{Y}_2(T)\end{pmatrix} + \mathcal{O}(q)\;.
	\end{align}
\end{subequations}
These relations will be useful to interpret some of our phenomenological observations in
section~\ref{sec:NumericalAnalysisLepton}.

We note that, under the generator of the $\Z2^\CP$ \CP-like symmetry, both components of the modular 
form get complex conjugated, i.e.\
\begin{equation}
T \stackrel{\CP}{\longrightarrow} -\bar{T}:\qquad 
\hat{Y}^{(1)}(T) ~\stackrel{\CP}{\longrightarrow}~ \hat{Y}^{(1)}(-\bar T)~=~\left(\hat{Y}^{(1)}(T)\right)^*\,.
\end{equation}

\subsection{Superpotential and mass matrices}
\label{sec:mass-matrices}

Respecting gauge invariance\footnote{Recall that there are additional \U1 gauge symmetries 
with charges not listed in table~\ref{tab:MSSM-model447} but given in appendix~\ref{app:spectrum}.} 
as well as the correct transformation behavior under the eclectic flavor symmetries of the 
model (see table~\ref{tab:MSSM-model447}),\footnote{We stress that superpotential operators 
invariant under these symmetries also respect all string-theory selection 
rules~\cite{Hamidi:1986vh,Dixon:1986qv,Font:1988tp,Lauer:1989ax,Lauer:1990tm,Kobayashi:2004ya,Buchmuller:2006ik,Kobayashi:2011cw,Nilles:2013lda,CaboBizet:2013hms}.} 
the effective superpotential to leading order in operator mass dimension is given by
\begin{equation} 
\label{eq:superpot}
  \mathcal W ~=~ \hat Y^{(1)}(T)\left\{
                 \phi^0 \left[ \phi^0_\mathrm{u} H_\mathrm{u} \bar{u} q \varphi_\mathrm{u} 
                             + \phi^0_\mathrm{d} H_\mathrm{d} \bar{d} q \varphi_\mathrm{e}
                             + \phi^0_\mathrm{e} H_\mathrm{d} \bar{e} \ell \varphi_\mathrm{e}
                             + H_\mathrm{u} \bar\nu \ell \varphi_\nu \right]
                + \phi^0_\mathrm{M} \bar\nu\bar\nu \varphi_\mathrm{e}\right\}\;,
\end{equation}
where henceforth we use Planck units. Here, $\hat Y^{(1)}(T)$ are the modular forms discussed 
in section~\ref{sec:modularForms} and, for brevity, we do not include the symmetry invariant overall couplings 
of each term. Note that by plain effective-field-theory (EFT) power counting, the neutrino Majorana 
mass term induced by the flavon VEV is hierarchically larger than the Dirac masses for 
all other quarks and leptons. A see-saw mechanism is thus a prediction of the model.
In addition, we remark that down-quark and charged-lepton Yukawa couplings, as well as the Majorana mass term, all 
are accompanied by the same flavon triplet $\varphi_\mathrm{e}$, suggesting that our model exhibits a particular kind 
of bottom-tau unification.

Owing to the highly constraining symmetries, all superpotential terms in eq.~\eqref{eq:superpot} 
have the generic structure
\begin{equation}
\Phi_0 \dots \Phi_0 \,\hat{Y}^{(1)}(T)\, \Phi^{1}_{\nicefrac{-2}{3}}\,\Phi^{2}_{\nicefrac{-2}{3}}\,\Phi^{3}_{\nicefrac{-2}{3}}\;,
\end{equation}
where the triplets $\Phi_{\nicefrac{-2}{3}}^{1}$ and $\Phi_{\nicefrac{-2}{3}}^{2}$ denote 
SM matter fields, $\Phi_{\nicefrac{-2}{3}}^{3}$ is a flavon triplet, and the series of $\Phi_0$'s 
includes a varying number of flavon singlets and the MSSM Higgs fields.
Considering that the superpotential must transform as a nontrivial singlet $\rep1'$ of $\Delta(54)$,
see~\cite[Table 2]{Baur:2021bly},
the explicit form of each mass term can be written as~\cite{Nilles:2020kgo,Nilles:2020gvu}
\begin{equation}
\left(\Phi^{1}_{\nicefrac{-2}{3}}\right)^\mathrm{T} ~ M(T, c, \Phi_{\nicefrac{-2}{3}}^{3}) ~ \Phi^{2}_{\nicefrac{-2}{3}}\;,
\end{equation}
where 
\begin{equation}
M(T,c,\Phi_{\nicefrac{-2}{3}}^{3}) ~:=~ c~ \begin{pmatrix}
\hat{Y}_2(T) \,X                 &- \dfrac{\hat{Y}_1(T)}{\sqrt{2}}\, Z&- \dfrac{\hat{Y}_1(T)}{\sqrt{2}} \, Y\\[6pt]
- \dfrac{\hat{Y}_1(T)}{\sqrt{2}} \,Z &\hat{Y}_2(T) \,Y &- \dfrac{\hat{Y}_1(T)}{\sqrt{2}} \, X\\[6pt]
- \dfrac{\hat{Y}_1(T)}{\sqrt{2}}  \,Y&- \dfrac{\hat{Y}_1(T)}{\sqrt{2}}  \,X&\hat{Y}_2(T) \,Z
\end{pmatrix}\;.
\label{eq:MassMatrix}
\end{equation}
Here, we have expressed the three components of the flavon triplet as $\Phi_{\nicefrac{-2}{3}}^{3}=(X,Y,Z)^\mathrm{T}$
and introduced $c$ to denote the overall coefficient of the terms.

As an  example, let us illustrate here how the charged lepton mass matrix $M_\mathrm{e}$ obeys the general texture described 
by eq.~\eqref{eq:MassMatrix}. For the charged lepton sector we find the following term in the superpotential of eq.~\eqref{eq:superpot}
\begin{equation} 
\label{eq:We}
\mathcal{W}_\mathrm{e} ~=~ c_\mathrm{e}\,\phi^0\,\phi^0_\mathrm{e}\,H_\mathrm{d}\,
   \left(\hat{Y}^{(1)}(T)\,\bar{e}\,\ell\,\varphi_\mathrm{e}\right)_{\rep{1}'}\;.
\end{equation} 
Here, we have explicitly included the symmetry-invariant overall coefficient $c_\mathrm{e}$, which we take as
a free parameter because its direct determination by string computations is still beyond our reach. 
After inserting the VEV $v_\mathrm{d}$ of the $H_\mathrm{d}$ Higgs field as well as all flavon VEVs, the mass matrix 
is given by 
\begin{equation}
M_\mathrm{e} ~=~ M(T,\Lambda_\mathrm{e}, \vev{\tilde\varphi_\mathrm{e}})\;,
\qquad \text{with}\qquad 
\Lambda_\mathrm{e} ~=~ c_\mathrm{e}\,v_\mathrm{d}\,\vev{\phi^0}\,\vev{\phi^0_\mathrm{e}}\,\Lambda_{\varphi_{\mathrm{e}}}\;
\label{eq:ElectronMassMatrix}
\end{equation}
denoting the overall global scale which, effectively, is the only dimensionful parameter of the mass matrix.  
Here we have introduced the dimensionless flavon triplet $\tilde\varphi_\mathrm{e}$ 
and its VEV, defined by
\begin{equation}
\varphi_\mathrm{e} ~=:~ \Lambda_{\varphi_{\mathrm{e}}} ~ \tilde\varphi_\mathrm{e}
\qquad\text{with}\qquad 
\tilde\varphi_\mathrm{e} ~:=~ \left(\tilde\varphi_{\mathrm{e},1},\,\tilde\varphi_{\mathrm{e},2},\,1\right)^\mathrm{T}.
\label{eq:ScalelessVarphiE}
\end{equation}
Without loss of generality, we can assume that the components of the (dimensionless) flavon
triplet VEV have the hierarchical structure\footnote{Such an ordering can always be achieved 
for exactly one flavon VEV by using the symmetry transformations of the $\mathrm{S}_3$ subgroup of 
$\Delta(54)$.}
\begin{equation}
0 ~\leq~ |\vev{\tilde\varphi_{\mathrm{e},1}}| ~\leq~ |\vev{\tilde\varphi_{\mathrm{e},2}}| ~\leq~ 1\;.
\label{eq:VarphiEChoice}
\end{equation}

Likewise, the neutrino masses are determined by the superpotential terms
\begin{equation}
\label{eq:Wnu}
\mathcal{W}_\nu ~=~ 
c_\mathrm{D}\,\phi^0\,H_\mathrm{u}\,\left(\hat{Y}^{(1)}(T)\,\bar{\nu}\,\ell\,\varphi_\nu\right)_{\rep{1}'} 
~+~ c_\mathrm{M}\,\phi^0_{\mathrm{M}}\,\left(\hat{Y}^{(1)}(T)\,\bar{\nu}\,\bar{\nu}\,\varphi_\mathrm{e}\right)_{\rep{1}'}\;,
\end{equation}
where we have explicitly included the symmetry-invariant coefficients $c_\mathrm{D}$ and $c_\mathrm{M}$,
and indicated that we have to take the $\Delta(54)$ nontrivial singlet contraction $\rep1'$ of each term. 
$\mathcal W_\nu$ predicts a type-I see-saw mechanism for neutrino masses. Hence, the light neutrino mass 
matrix is given by
\begin{equation}
M_\nu ~=~ -\dfrac{1}{2}\,M_\mathrm{D}\,M_\mathrm{M}^{-1}\,M_\mathrm{D}^{\mathrm{T}}\;,
\label{eq:NeutrinoMassMatrix}
\end{equation}
where
\begin{equation}
M_\mathrm{D}~=~ M(T,\Lambda_\mathrm{D},\vev{\tilde\varphi_\nu}) \qquad\text{and}\qquad
M_\mathrm{M}~=~ M(T,\Lambda_\mathrm{M},\vev{\tilde\varphi_\mathrm{e}})
\label{eq:DiracAndMajornaMassMatrices}
\end{equation}
are the Dirac and Majorana neutrino mass matrices which again follow the general form~\eqref{eq:MassMatrix}. 
Analogously to eq.~\eqref{eq:ScalelessVarphiE}, we have defined the dimensionless flavon triplet $\tilde\varphi_\nu$ through
\begin{equation}
\varphi_\nu ~=:~ \Lambda_{\varphi_{\nu}} ~ \tilde\varphi_\nu
\qquad\text{with}\qquad 
\tilde\varphi_\nu ~:=~ \left(\tilde\varphi_{\nu,1},\,\tilde\varphi_{\nu,2},\,1\right)^\mathrm{T}\;.
\label{eq:ScalelessVarphiNu}
\end{equation}
From the structure of the superpotential contribution~\eqref{eq:Wnu} and the see--saw neutrino masses
\eqref{eq:NeutrinoMassMatrix}, we see that the overall scale of the light neutrino mass matrix is given by
\begin{equation}
\Lambda_\nu ~=~ \dfrac{\Lambda_{\mathrm{D}}^2}{\Lambda_\mathrm{M}} ~=~ 
\dfrac{\left(c_\mathrm{D}\,v_\mathrm{u}\,\vev{\phi^0}\,\Lambda_{\varphi_{\nu}}\right)^2}{c_\mathrm{M}\,\vev{\phi^0_\mathrm{M}}\,\Lambda_{\varphi_{\mathrm{e}}}}\;,
\end{equation}
where $v_\mathrm{u}$ stands for the VEV of the up-type Higgs $H_\mathrm{u}$.

In complete analogy with the charged-lepton sector, from the Yukawa couplings for the up and down-quark sectors,
we find that the corresponding mass matrices follow the structure of eq.~\eqref{eq:MassMatrix} depending
as follows on the different parameters
\begin{subequations}
\label{eqs:QuarkMassMatrices}
\begin{align}
   M_\mathrm{u} ~&=~ M(T,\Lambda_\mathrm{u}, \vev{\tilde\varphi_\mathrm{u}}) && \text{with} && \Lambda_\mathrm{u} ~=~ c_\mathrm{u}\,v_\mathrm{u}\,\vev{\phi^0}\,\vev{\phi^0_\mathrm{u}}\,\Lambda_{\varphi_\mathrm{u}}\;, \\
   M_\mathrm{d} ~&=~ M(T,\Lambda_\mathrm{d}, \vev{\tilde\varphi_\mathrm{e}}) && \text{with} && \Lambda_\mathrm{d} ~=~ c_\mathrm{d}\,v_\mathrm{d}\,\vev{\phi^0}\,\vev{\phi^0_\mathrm{d}}\,\Lambda_{\varphi_\mathrm{e}}\;.
\end{align}\label{eq:QuarkMassMatrix}
\end{subequations}
Analogously to the previous cases, $c_\mathrm{u}$ and $c_\mathrm{d}$ denote the unconstrained symmetry-invariant
coefficients of the up and down-quark Yukawa couplings, respectively. Furthermore, 
\begin{equation}
\varphi_\mathrm{u} ~=:~ \Lambda_{\varphi_{\mathrm{u}}} ~ \tilde\varphi_\mathrm{u}
\qquad\text{with}\qquad 
\tilde\varphi_\mathrm{u} ~:=~ \left(\tilde\varphi_{\mathrm{u},1},\,\tilde\varphi_{\mathrm{u},2},\,1\right)^\mathrm{T}\;.
\label{eq:DimensionlessVarphiU}
\end{equation}

In summary, the superpotential contributions to the lepton masses include the following parameters: 
the global mass scales $\Lambda_\mathrm{e}$ for charged leptons and $\Lambda_\nu$ for neutrinos, 
the VEV $\vev T$ of the complex K\"ahler modulus,
and the free components, $\vev{\tilde\varphi_{\mathrm{e},1}}$, $\vev{\tilde\varphi_{\mathrm{e},2}}$, $\vev{\tilde\varphi_{\nu,1}}$ 
and $\vev{\tilde\varphi_{\nu,2}}$, of the flavon VEVs. As we shall see, a subtle interplay among the 
modulus and flavon VEVs can explain the observed lepton-mass hierarchies (cf.\ section~\ref{sec:hierarchies})
and even yield a fit of lepton flavor data with interesting predictions (cf.\ section~\ref{sec:NumericalAnalysisLepton}).
We will see that it suffices to consider real flavon VEVs to arrive at those results, 
which implies that the modulus VEV $\vev T$ is the only source of \CP violation in the lepton sector.
Finally, since we aim at a global fit of flavor in both lepton and quark sectors, note that 
up-quark Yukawa couplings introduce additional parameters: the global up and down-quark
mass scales $\Lambda_\mathrm{u}$ and $\Lambda_\mathrm{d}$ as well as
the flavon components $\vev{\tilde\varphi_{\mathrm{u},1}}$ 
and $\vev{\tilde\varphi_{\mathrm{u},2}}$. Down-quark Yukawas in the superpotential of our model,
eq.~\eqref{eq:superpot}, share the charged-lepton flavon $\tilde\varphi_{\mathrm{e}}$, avoiding 
extra parameters but also imposing thereby severe constraints. In fact, these restrictions challenge
the compatibility of our model with observations. Fortunately, as we shall see in 
section~\ref{sec:NumericalAnalysisBoth}, this issue can be addressed by including
K\"ahler corrections, which we now discuss.

\subsection{K\"ahler corrections to the mass matrices}
\label{sec:Kaehler}

In contrast to the most common assumption of BU model building, the K\"ahler potential is, 
in general, nontrivial.\footnote{The phenomenological consequences of noncanonical contributions 
to the K\"ahler potential have been considered in BU models of traditional flavor symmetries (see
\cite{Antusch:2007ib,Antusch:2007vw} for a special case and~\cite{Chen:2012ha,Chen:2013aya} for the general case)
as well as modular flavor symmetries~\cite{Chen:2019ewa,Lu:2019vgm}.} 
In string-derived TD models, we have to include 
the phenomenological consequences of this fact.
At leading order in the EFT expansion of the matter fields and flavons, the K\"ahler potential 
of the model introduced in section~\ref{sec:model} is given by~\cite{Nilles:2020kgo}
\begin{equation}
\begin{aligned}
  K ~\supset
    &~ -\log(-\I T+\I \bar T) \\
    &~ + \sum_{\Psi}\left[(-\I T+ \I\bar T)^{\nicefrac{-2}3}
        + (-\I T+ \I\bar T)^{\nicefrac13}|\hat Y^{(1)}(T)|^2\right]|\Psi|^2\\
    &~ + \sum_{\varphi}\left[(-\I T+ \I\bar T)^{\nicefrac{-2}3}
    + (-\I T+ \I\bar T)^{\nicefrac13}|\hat Y^{(1)}(T)|^2\right]|\varphi|^2\;.\\
\end{aligned}
\label{eq:KahlerGeneral}
\end{equation}
Here we again suppress all symmetry-invariant coupling parameters, and the respective summations 
run over all MSSM matter fields, $\Psi\in\{q,\bar u, \bar d, \ell,\bar e,\bar\nu\}$, and the various 
flavon triplets of the model, $\varphi\in\{\varphi_\mathrm{e},\varphi_\mathrm{u},\varphi_\nu,\ldots\}$, 
see table~\ref{tab:MSSM-model447}. Interestingly, the canonical form of the K\"ahler potential
at this level is preserved in models endowed with eclectic symmetries because matter fields are charged 
under a traditional flavor symmetry~\cite{Nilles:2020kgo}, $\Delta(54)$ in our case,
avoiding the loss of predictivity that challenges models exclusively based on modular 
symmetries~\cite{Chen:2019ewa}. Consequently, corrections to this canonical K\"ahler potential only
appear if the traditional flavor symmetry is spontaneously broken by flavons. Couplings between flavons 
and matter fields induce additional terms in the K\"ahler potential of the form
\begin{equation}
K~\supset~ \sum_{\Psi,\varphi} \left[(-\I T+\I \bar T)^{\nicefrac{-4}3}\sum_a |\Psi\varphi|^2_{\rep1,a}
                          +(-\I T+\I\bar T)^{\nicefrac{-1}3}\sum_a |\hat Y^{(1)}(T)\Psi\varphi|^2_{\rep1,a}\right]\;,
\label{eq:KahlerCorrections}
\end{equation}
where the subindex ${\rep1,a}$ refers to the $a$th invariant singlet contraction with respect to
the whole eclectic flavor symmetry. Since the terms in eq.~\eqref{eq:KahlerCorrections} are 
proportional to the ratio of flavon VEVs to the fundamental scale, they represent small corrections 
to the leading-order K\"ahler potential~\eqref{eq:KahlerGeneral}. For simplicity,\footnote{In
principle, one might also consider contributions from modular forms with higher modular weights. 
These forms are powers of $\hat Y^{(1)}(T)$ and, hence, we expect that the term considered in 
eq.~\eqref{eq:KahlerCorrections} captures the structure of the corrections.} 
we restrict ourselves here to the modular forms $\hat Y^{(1)}(T)$ that naturally appear also in 
the superpotential $\mathcal W$.

Since the (Planck suppressed) next-to-leading order terms, given in eq.~\eqref{eq:KahlerCorrections}, 
can yield noncanonical contributions if the flavons develop VEVs, let us briefly discuss 
the consequences of such contributions. 
As pointed out in~\cite{Chen:2019ewa}, noncanonical terms can 
be relevant for the mass matrices of a model.
Hence, studying the K\"ahler potential is important to correctly determine the 
phenomenology  of a model. In order to canonically normalize the fields,
the K\"ahler metric associated with $\Psi$
\begin{equation}
K_{ij} ~=~ \dfrac{\partial^2 K}{\partial \Psi_i ~ \partial \Psi_{j}^*}
\end{equation}
needs to be diagonalized, such that
\begin{equation}
K_{ij} ~=~ \left(U_K^\dagger \,D^2\,U_K\right)_{ij} \;,
\end{equation}
where $U_K$ is unitary and $D$ is diagonal and positive. Then, the canonically normalized 
fields $\hat{\Psi}$ read
\begin{equation}
\hat\Psi ~=~ D\,U_K\,\Psi\;.
\end{equation}
Assuming a superpotential mass term
\begin{equation}
\left(\Psi^{(1)}\right)^\mathrm{T} M \, \Psi^{(2)}\;,
\end{equation}
we need to consider the correct normalization of each field, i.e.
\begin{equation}
\hat\Psi^{(1)} ~=~ D^{(1)}\,U_K^{(1)}\,\Psi^{(1)} 
\qquad \text{and} \qquad
\hat\Psi^{(2)} ~=~ D^{(2)}\,U_K^{(2)}\,\Psi^{(2)}\;.
\end{equation}
When applying these transformations to the mass term one obtains
\begin{equation}
\left(\hat{\Psi}^{(1)}\right)^\mathrm{T} \, \hat{M} \, \hat{\Psi}^{(2)} \;,
\end{equation}
with a mass matrix for the canonically normalized (i.e.\ ``physical'') fields that reads
\begin{equation}
\hat{M} ~=~ \left(D^{(1)}\right)^{-1} \, \left(U_K^{(1)}\right)^* \, M \, \left(U_K^{(2)}\right)^\dagger \, \left(D^{(2)}\right)^{-1} \;.
\end{equation}
Note that since $D^{(1)}$ and $D^{(2)}$ are not unitary, the normalization of the 
right-handed fields {\it does} affect the mixing matrices and should, therefore, {\it not} 
be neglected. That is, $\hat{M}\,\hat{M}^\dagger$ depends on the normalization of both fields,
$\Psi^{(1)}$ and $\Psi^{(2)}$.

In our specific case, the mass matrices~\eqref{eq:ElectronMassMatrix}, \eqref{eq:NeutrinoMassMatrix}, 
and~\eqref{eqs:QuarkMassMatrices} obtained solely from the superpotential will pick up corrections
from the noncanonical K\"ahler potential eq.~\eqref{eq:KahlerCorrections}. 
Since both, the superpotential and K\"ahler potential, are expansions in powers of fields, we may 
also analyze the corrections in a perturbative manner. Let us consider the part $K_\Psi\subset K$ 
associated with a field $\Psi$.
Explicitly introducing the symmetry-invariant coefficients $\kappa^{(0)}$
and $\kappa^{(Y)}$ in eq.~\eqref{eq:KahlerGeneral}, 
the leading order, i.e.\ bilinear, contributions are given by
\begin{equation}
K_\Psi ~\supset~ \left[(-\I T+ \I\bar T)^{\nicefrac{-2}3} \,\kappa^{(0)} + (-\I T+ \I\bar T)^{\nicefrac13} \, \kappa^{(Y)}\,|\hat Y^{(1)}(T)|^2\right]|\Psi|^2\;.
\label{eq:KahlerFirstOrder}
\end{equation}
These terms have already been studied in~\cite{Nilles:2020kgo}. It was found that the traditional
flavor symmetry restricts them in such a strong manner that the K\"ahler metric becomes 
proportional to the identity matrix, i.e.
\begin{equation}
\label{eq:KmetricIdterm}
K_{ij}^{(\mathrm{id})} ~=~ \chi ~ \delta_{ij}\;,
\end{equation} 
where $\delta_{ij}$ denotes the Kronecker delta and
\begin{equation}
\chi ~:=~ \left[(-\I T+ \I\bar T)^{\nicefrac{-2}3} \,\kappa^{(0)} + (-\I T+ \I\bar T)^{\nicefrac13} \, \kappa^{(Y)}\,|\hat Y^{(1)}(T)|^2\right]\;.
\end{equation}
Therefore, the K\"ahler potential is indeed (apart from normalization) canonical at leading order. 
That is, there are no corrections to \textit{the structure} of the mass matrices at this order 
(as a result of the traditional flavor symmetry).

The next-to-leading order K\"ahler contributions do yield corrections to the structure of the mass matrices.
From eq.~\eqref{eq:KahlerCorrections}, restoring coefficients, the relevant terms of the K\"ahler potential are
\begin{equation}
K_\Psi \supset \sum_{\varphi} \left[(-\I T+\I \bar{T})^{\nicefrac{-4}3}\sum_a \zeta_a^{(\varphi)} \, |\Psi\varphi|^2_{\rep1,a}
+(-\I T+\I \bar{T})^{\nicefrac{-1}3}\sum_a \zeta_a^{(Y\varphi)} \,|\hat Y^{(1)}(T)\Psi\varphi|^2_{\rep1,a}\right],
\label{eq:KahlerSecondOrder}
\end{equation}
where the first sum runs over all flavon triplets $\varphi$ of the theory that develop VEVs, and 
the second sum over $a$ runs over all invariant singlet contractions of the tensor products.
The coefficients $\zeta_a^{(\varphi)}$ and $\zeta_a^{(Y\varphi)}$ cannot 
be fixed by symmetry. It may, however, be argued that they should be $\mathcal{O}(1)$.
The explicit tensor products are given in appendix~\ref{app:Kahler}.
Some of them yield canonical contributions to the K\"ahler metric, proportional to the identity matrix. 
These can be absorbed in the overall normalization and, hence, would only modify $\chi$.
However, other terms, generically denoted as $K_{ij}^{(\mathrm{non-id})}$, yield noncanonical contributions 
to the K\"ahler metric, which will be essential for phenomenology, as we will see below. These noncanonical 
terms depend on the flavon VEVs and are given in eq.~\eqref{eq:KahlerMetric2}.

Hence, the K\"ahler metric of a generic matter field is given by a canonical contribution
$K_{ij}^{(\mathrm{id})}$ and various noncanonical terms,
\begin{equation}
K_{ij} ~=~ K_{ij}^{(\mathrm{id})} + \sum_\varphi K_{ij}^{(\mathrm{non-id})}\;.
\end{equation}

Using the matrices $A_{ij}$ and $B_{ij}$ which are functions only of the flavon triplets $\varphi$ and the modulus 
$T$, and whose explicit forms are given in eqs.~\eqref{eq:Aij} and \eqref{eq:Bij}, the K\"ahler metric 
can be parametrized as\footnote{The relation~\eqref{eq:KahlerMetricParameterized} is approximate
because, as discussed in appendix~\ref{app:Kahler}, $\chi$ receives small contributions from the 
K\"ahler corrections that we neglect.}
\begin{equation}
K_{ij} ~\approx~ \chi ~\left(\delta_{ij} ~+~ \sum_\varphi \lambda_\varphi\, \left(A_{ij}(\varphi) + \kappa_\varphi\, B_{ij}(\varphi)\right) \right)\;.
\label{eq:KahlerMetricParameterized}
\end{equation}
We note that the overall factor $\chi$ can (and will) be eliminated by a simple rescaling of $\Psi$.
Here, $\lambda_\varphi$ is the ratio
\begin{equation}
\lambda_\varphi ~=~ (-\I T+\I\bar T)^{\nicefrac{-2}3}\, 
\dfrac{|\hat{Y}^{(1)}(T)|^2 \,\zeta_1^{(Y\varphi)} + (-\I T+\I\bar T)^{-1} \,\zeta_1^{(\varphi)}}{\kappa^{(Y)}\,|\hat Y^{(1)}(T)|^2 + (-\I T+ \I\bar T)^{-1} \,\kappa^{(0)}}\;,
\end{equation}
which parametrizes the relative size of the correction with respect to the leading-order term~\eqref{eq:KahlerFirstOrder}.
In addition,
\begin{equation}
\kappa_\varphi ~=~ \dfrac{\zeta_2^{(Y\varphi)}}{|\hat{Y}^{(1)}(T)|^2 \,\zeta_1^{(Y\varphi)} + (-\I T+\I\bar T)^{-1} \,\zeta_1^{(\varphi)}}\;
\end{equation}
parametrizes the ratio of the two linearly independent corrections associated with $A_{ij}(\varphi)$ 
and $B_{ij}(\varphi)$. In the limit $T\rightarrow \I\infty$, up to $\mathcal{O}(1)$ factors, 
$\lambda_\varphi$ scales as $\lambda_\varphi \approx \left(-\I T+\I\bar T\right)^{\nicefrac{-2}3}$ 
while $\kappa_\varphi$ is $\mathcal{O}(1)$ just as $|\hat{Y}^{(1)}(T)|$.
This limit will be important in our phenomenological considerations below.

Importantly, note that all occurring flavon triplet representations $\varphi$ enter the 
K\"ahler metric in exactly the same way, cf.\ eq.~\eqref{eq:KahlerMetricParameterized}. 
Hence, in order to capture the effect of all flavons on the K\"ahler metric in the most efficient way without
parameter degeneracies, we define two effective flavons
\begin{equation}
\varphi_{\mathrm{eff}}^{(A)} ~=:~ \Lambda_{\varphi_{\mathrm{eff}}^{(A)}} ~ \tilde\varphi_{\mathrm{eff}}^{(A)}
\qquad\text{with}\qquad 
\tilde\varphi_{\mathrm{eff}}^{(A)} ~:=~ \left(\tilde\varphi_{\mathrm{eff},1}^{(A)},\,\tilde\varphi_{\mathrm{eff},2}^{(A)},\,1\right)^\mathrm{T}\;,
\label{eq:DimensionlessPhiEffA}
\end{equation}
and 
\begin{equation}
\varphi_{\mathrm{eff}}^{(B)} ~=:~ \Lambda_{\varphi_{\mathrm{eff}}^{(B)}} ~ \tilde\varphi_{\mathrm{eff}}^{(B)}
\qquad\text{with}\qquad 
\tilde\varphi_{\mathrm{eff}}^{(B)} ~:=~ \left(\tilde\varphi_{\mathrm{eff},1}^{(B)},\,\tilde\varphi_{\mathrm{eff},2}^{(B)},\,1\right)^\mathrm{T}\;.
\label{eq:DimensionlessPhiEffB}
\end{equation}
These are sufficient to represent all $\varphi$'s in the sense that, by definition,
\begin{subequations}
\label{eqs:effectiveParams}
	\begin{align}
	\sum_\varphi \lambda_\varphi \, A_{ij}(\varphi)                ~&=:~ \lambda_{\varphi_{\mathrm{eff}}} \, A_{ij}(\tilde\varphi_{\mathrm{eff}}^{(A)})\;,\\
	\sum_\varphi \lambda_\varphi \, \kappa_\varphi \, B_{ij}(\varphi) ~&=:~ \lambda_{\varphi_{\mathrm{eff}}} \, \kappa_{\varphi_{\mathrm{eff}}} \, B_{ij}(\tilde\varphi_{\mathrm{eff}}^{(B)}) \;.
	\end{align}
\end{subequations}
The expansion parameter $\lambda_{\varphi_{\mathrm{eff}}}$ will now be roughly 
$(-\I T+\I\bar T)^{\nicefrac{-2}3} \sum_\varphi \Lambda_{\varphi}^2$ in the 
$T\rightarrow\I\infty$ region, where we used 
\begin{equation}
\varphi ~=:~ \Lambda_{\varphi} ~ \tilde\varphi
\qquad\text{with}\qquad 
\tilde\varphi~:=~ \left(\tilde\varphi_1,\,\tilde\varphi_2,\,1\right)^\mathrm{T}\;,
\label{eq:DimensionlessPhi}
\end{equation}
while $\kappa_{\varphi_{\mathrm{eff}}}$ should still be $\mathcal{O}(1)$.

\section{Eclectic breaking and charged-lepton mass hierarchies}
\label{sec:StepwiseBreaking}

Let us now turn to the spontaneous breaking of the eclectic flavor symmetry in detail and 
its consequences for the model introduced in section~\ref{sec:DiscussionModelDefinition}.
We study the breaking in two stages. First, the modulus $T$ is stabilized 
at or near to a fixed point in moduli space where the traditional flavor symmetry is enhanced;
and second, one or more flavon fields develop VEVs. 

\paragraph{\boldmath Breaking by \vev{T}. \unboldmath}
As we have studied before~\cite{Baur:2021bly}, depending on the value of $\vev{T}$, the 
$\Delta(54)$ traditional flavor symmetry is enhanced to the following two linearly realized 
unified flavor groups:
\begin{equation}
  \Omega(2) ~\xrightarrow{\vev T = \I}~\Xi(2,2) \cong [324,111]\qquad\text{or}\qquad
  \Omega(2) ~\xrightarrow{\vev T = 1,\I\infty,\omega}~H(3,2,1) \cong [486,125]\;.
\end{equation}
In these cases, also a $\Z{2}^{\CP}$ \CP-like symmetry is left unbroken. Including 
this symmetry, the enhanced traditional symmetry at the fixed points in moduli space are either
$H(3,2,1)\rtimes\Z2^\CP\cong[972,469]$ at $\vev T = 1,\I\infty,\omega$ or
$\Xi(2,2)\rtimes \Z2^\CP\cong [648,548]$ at $\vev T = \I$.

\paragraph{\boldmath Breaking by flavon VEVs. \unboldmath}

In our model, all (matter and) flavon fields transform as triplets $\rep3_2$ of the traditional flavor 
symmetry $\Delta(54)$ and have modular weight $\nicefrac{-2}3$, see table~\ref{tab:MSSM-model447}. 
This scenario significantly reduces the number of possible breaking patterns.
At the moduli point $\vev T=\I$, the possible breakings read~\cite{Baur:2021bly}
\begin{equation}
\Z2~\xleftarrow{\vev{\Phi_{\nicefrac{-2}3}}}~\Xi(2,2)~\xrightarrow{\vev{\Phi_{\nicefrac{-2}3}}}~\Z3^{(i)}\;,\qquad i=1,2\;,
\end{equation}
where the two different $\Z3^{(i)}$ correspond to inequivalent \Z3 subgroups of $\Xi(2,2)$, 
associated with different directions of flavon VEVs. On the other hand, at $\vev T = 1,\I\infty,\omega$, 
all possible breaking patterns are described by
\begin{subequations}
\label{eqs:H321breakdown}
\begin{eqnarray}
\Z6~\xleftarrow{\vev{\Phi_{\nicefrac{-2}3}}}&H(3,2,1)&\xrightarrow{\vev{\Phi_{\nicefrac{-2}3}}}~\Z3^{(i)}\;,\qquad i=1,4\,
\qquad\text{or}\\
\label{eq:OurH321breakdown}
&H(3,2,1)&\xrightarrow{\vev{\Phi_{\nicefrac{-2}3}}}~\Z3^{(2)}\x\Z3^{(3)}~\xrightarrow{\vev{\Phi_{\nicefrac{-2}3}}}~\Z3^{(3)}\;.
\end{eqnarray}
\end{subequations}
Whether or not the $\Z2^\CP$ \CP-like symmetry is broken, depends not only on the structure of the flavon 
VEVs discussed  here, but also on their global phases, cf.~\cite{Baur:2021bly}. 
Nevertheless, considering the flavon VEVs to be real ensures that the $\Z2^\CP$ \CP-like symmetry 
is preserved for $\vev T = \I\infty$.

\subsection{A pattern of eclectic breaking}
\label{sec:EclecticBreakdown}

In this work we choose the modulus to be fixed in the vicinity of $\vev T =\I\infty$, i.e.\
we assume that moduli stabilization leads to $H(3,2,1)$ as unified flavor group. Hence, only 
the breaking patterns described in eqs.~\eqref{eqs:H321breakdown} are relevant in our case. 
Furthermore, we focus on the breaking pattern described by eq.~\eqref{eq:OurH321breakdown}.
In order to better understand this breaking, let us consider the $H(3,2,1)$ generators 
and the flavon VEVs that lead to this breaking pattern. The generators of the unified flavor group
at $\vev T =\I\infty$ are $\{\mathrm{A,B,C,T,\hat{R},\CP}\}$; the modular generator $\mathrm{S}$
is excluded because it does not leave the modulus invariant. For generic flavon fields
$\varphi$ of type $\Phi_{\nicefrac{-2}3}$, such as those listed for our model in 
table~\ref{tab:MSSM-model447}, the representations of the generators are given by the 
traditional group matrices~\eqref{eq:32TraditionalGenerators}, $\rho(\mathrm{T})$ in
eq.~\eqref{eq:32ModularGenerators} (including the automorphy factor equals one), and 
$\rho(\mathrm{\hat R})=\exp(\nicefrac{2\pi\I}9)\Id_3$ from eq.~\eqref{eq:Z9R}.

\begin{figure}
	\begin{tikzpicture}[node distance=0.2cm and 2.27cm, rounded corners, >=stealth]
	
	\node[minimum height=22pt, draw, rectangle,fill=lightgray] (Omega2) {$\Omega(2)$ };
	\node[minimum height=22pt, draw, rectangle, right=of Omega2,fill=lightgray] (H321) { $H(3,2,1)$ };
	\node[minimum height=22pt, draw, rectangle, right=of H321,fill=lightgray] (Z3xZ3) { $\;\!\Z{3}^{(2)} \!\times \Z{3}^{(3)}$ };
	\node[minimum height=22pt, draw, rectangle, right = of Z3xZ3,fill=lightgray] (Z3) { $\;\!\Z3^{(3)}$ };
	\node[minimum height=22pt, draw, rectangle, above right=of Z3,fill=lightgray] (empty1) { $~\emptyset~$ };
	\node[minimum height=22pt, draw, rectangle, below right=of Z3,fill=lightgray] (empty2) { $~\emptyset~$ };
	
	\draw[->, >=stealth, shorten >=2pt, shorten <=2pt] (Omega2.east) -- node [fill=white,rectangle,midway,align=center] {\resizebox*{30pt}{!}{$\vev{T}=\I\infty$}} (H321.west);
	\draw[->, >=stealth, shorten >=2pt, shorten <=2pt] (H321.east) -- node [fill=white,rectangle,midway,align=center] {\resizebox*{35pt}{!}{$\vev{\tilde\varphi}=\begin{pmatrix}0\\0\\1\end{pmatrix}$}} (Z3xZ3.west);
	\draw[->, >=stealth, shorten >=2pt, shorten <=2pt] (Z3xZ3.east) -- node [fill=white,rectangle,midway,align=center] {\resizebox*{38pt}{!}{$\vev{\tilde\varphi}=\begin{pmatrix}0\\\lambda_2\\1\end{pmatrix}$}} (Z3.west);
	\draw[->, >=stealth, shorten >=2pt, shorten <=2pt] (Z3.east) -- node [fill=white,rectangle,midway,align=center] {\resizebox*{40pt}{!}{$\vev{\tilde\varphi}=\begin{pmatrix}\lambda_1\\\lambda_2\\1\end{pmatrix}$}}(empty1.west);
	\draw[->, >=stealth, shorten >=2pt, shorten <=2pt] (Z3.east) -- node [fill=white,rectangle,midway,align=center] {\resizebox*{55pt}{!}{$\epsilon=\mathrm{e}^{2\pi\I\vev{T}} \neq 0$}} (empty2.west);
	\end{tikzpicture}
	\caption{Breaking pattern of the eclectic flavor symmetry $\Omega(2)$ of a $\mathbbm T^2/\Z3$ orbifold model
                 triggered by the VEVs of the modulus $T$ and (dimensionless) flavons $\tilde\varphi$. All flavons transform 
                 in the $\rep3_2$ representation of $\Delta(54)$, see table~\ref{tab:MSSM-model447}.
		\label{fig:BreakdownPattern}}
\end{figure}
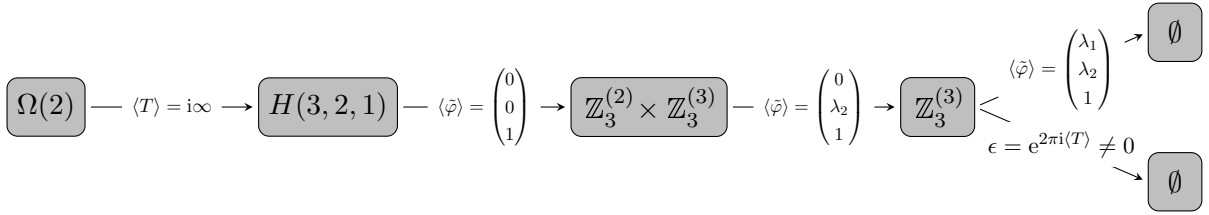

As before, it is convenient to use the dimensionless flavon $\tilde\varphi$ instead 
of $\varphi$, which are related by eq.~\eqref{eq:DimensionlessPhi},
since an overall factor would not affect the breaking pattern of the eclectic flavor symmetry.
The first step 
in the breaking chain~\eqref{eq:OurH321breakdown} is achieved by setting the dimensionless 
flavon VEV $\vev{\tilde\varphi} = (0,0,1)^\mathrm{T}$. This VEV is left invariant only by
the generators
\begin{equation}
\label{eq:UnbrokenGensStep1}
\rho(\mathrm{ABA^2})~=~ \begin{pmatrix}\w&0&0\\0&\w^2&0\\0&0&1\end{pmatrix}\qquad\text{and}\qquad
\rho(\mathrm{T})~=~ \begin{pmatrix}\w^2&0&0\\0&1&0\\0&0&1\end{pmatrix}\;,
\end{equation}
i.e.\ one traditional and one modular generator. Both of them are of order three
and generate the group $\Z3^{(2)}\times\Z3^{(3)}$. In a second step, one can choose
a misalignment of the flavon VEV $\vev{\tilde\varphi} = (0,\lambda_2,1)^\mathrm{T}$ with $\lambda_2\neq0$,
which breaks the traditional $\Z3^{(2)}$ symmetry generated by $\rho(\mathrm{ABA^2})$,
leaving only the modular $\Z3^{(3)}$ symmetry unbroken. Finally, $\Z3^{(3)}$ can be broken too
by perturbing either the modulus VEV or the flavon VEV. In moduli space, one must simply
get slightly away from the moduli enhanced point $\vev T =\I\infty$, such that 
$\epsilon:=\vev q = \exp(2\pi\I\vev T)$ is small but does not vanish. Note that this perturbation
breaks the $\Z2^\CP$ \CP-like symmetry too. In flavon space, $\Z3^{(3)}$ is broken by 
considering the VEV $\vev{\tilde\varphi} = (\lambda_1,\lambda_2,1)^\mathrm{T}$, 
which is no longer left invariant by $\rho(\mathrm{T})$. This breaking process is 
illustrated in figure~\ref{fig:BreakdownPattern}.

Using this information, we realize that some useful hierarchies can arise in the model by choosing
appropriately the parameters $\epsilon,\lambda_1$ and $\lambda_2$. From our previous discussion, we 
notice that the vanishing of any of these parameters corresponds to a symmetry enhancement at certain 
points in moduli and flavon space, where the symmetries displayed in figure~\ref{fig:BreakdownPattern} 
are left intact. If the VEV parameters are small, i.e.\ $|\epsilon|,|\lambda_1|,|\lambda_2|\ll1$, one 
can find that the subgroups $\Z{3}^{(2)}$ and $\Z{3}^{(3)}$ of $H(3,2,1)$ are approximately realized. 
If, in addition, those parameters have very different values, then the three groups may correspond 
to hierarchically different symmetries of the model, providing thereby a plausible explanation of the 
nontrivial textures of masses and mixing of particle physics. We shall focus in the following on the 
possibility of arriving at a hierarchical mass structure in both the quark and lepton sector of the SM.  
For phenomenological reasons, we shall assume that the flavon VEVs follow this symmetry breaking pattern 
and satisfy
\begin{equation}
0 ~<~ |\lambda_1| ~<~ |\lambda_2| ~<~ 1\;.
\end{equation}
Depending on the sector, we will consider the relevant flavon $\varphi$ from table~\ref{tab:MSSM-model447}.
For example, in the lepton sector, the flavon fields $\varphi$ that we can use are the 
$\Delta(54)$ triplets $\varphi_{\mathrm{e}}$ and $\varphi_\nu$.

\subsection{Hierarchical masses from approximate symmetries}
\label{sec:hierarchies}

Let us now study the hierarchical structure of fermion masses that arise in the vicinity of 
the symmetry-enhanced points. Following the discussion of~\cite{Marzocca:2014tga,Novichkov:2021evw}, 
we make use of the following relation valid for any $n\x n$ complex matrix $M$:
\begin{equation}
\sum_{i_1<\dots<i_p} m_{i_1}^2\,\cdots\,m_{i_p}^2 ~=~ \sum \left|\mathrm{det}\,M_{p\times p}\right|^2\;,
\end{equation}
where $m_i$ are the singular values of $M$, $p=1,\dots,n$ is fixed, and the sum on the right-hand 
side goes over all possible $p\times p$ submatrices $M_{p\x p}$ of $M$. This relation can be 
used to extract the physical masses $m_i$, $i\in\{\mathrm{I},\mathrm{II},\mathrm{III}\}$, as singular 
values of the $3\x3$ mass matrices of our model. Moreover, we shall assume the observed hierarchical 
pattern $m_\mathrm{I} \ll m_\mathrm{II} \ll m_\mathrm{III}$, which implies
\begin{subequations}
	\begin{align}
	m_\mathrm{III}^2 ~&\approx~ \sum_{i,j} |M_{ij}|^2 = \mathrm{Tr}\,M^\dagger\,M\;, \hspace*{-20pt}& \\
	m_\mathrm{II}^2\,m_\mathrm{III}^2 ~&\approx~ \sum |\mathrm{det}\,M_{2\times 2}|^2 &\Rightarrow~~ m_\mathrm{II}^2 ~\approx~ \dfrac{\sum |\mathrm{det}\,M_{2\times 2}|^2}{ \mathrm{Tr}\,M^\dagger\,M}\;, \\
	m_\mathrm{I}^2\,m_\mathrm{II}^2\,m_\mathrm{III}^2 ~&=~ |\mathrm{det}\,M|^2 & \Rightarrow~~ m_\mathrm{I}^2 ~\approx~ \dfrac{|\mathrm{det}\,M|^2}{\sum |\mathrm{det}\,M_{2\times 2}|^2}\;.
	\end{align}\label{eq:PetcovMassRelations}
\end{subequations}

\subsubsection{Charged-lepton and quark mass hierarchies}
\label{sec:AnalyticalLeptons}

The explicit forms of the charged-lepton and quark mass matrices that arise 
from the superpotential~\eqref{eq:superpot} are given in eqs.~\eqref{eq:ElectronMassMatrix} 
and \eqref{eq:QuarkMassMatrix}, respectively. We see that the resulting mass textures
are equal for charged leptons, up-type quarks, and down-type quarks, but 
the specific masses in each sector depend on the values of the VEV parameters 
of the respective flavons. 
Hence, the results derived in this section apply to all three sectors.
 
For a generic sector, in terms of the small VEV parameters $\lambda_1$, $\lambda_2$, 
and $\epsilon$, the structure of the mass matrices reads
\begin{equation}
\label{eq:MassMatrixAtInftyFlavonScenario}
M(\vev T,\Lambda,\vev\varphi)~=~\Lambda\,
\begin{pmatrix}
\lambda_1                    & 3\,\epsilon^{1/3}            & 3\,\lambda_2\,\epsilon^{1/3}\\
3\,\epsilon^{1/3}            & \lambda_2                    & 3\,\lambda_1\,\epsilon^{1/3}\\
3\,\lambda_2\,\epsilon^{1/3} & 3\,\lambda_1\,\epsilon^{1/3} & 1
\end{pmatrix}
\,+\,\mathcal{O}(\epsilon)\;.
\end{equation} 
Here we have used the $q$-expansions~\eqref{eq:Qexpansion} for the modular 
forms $\hat Y_1$ and $\hat Y_2$, valid in our case because $|\epsilon|=|\vev{q}|\ll1$ 
in the vicinity of $\vev T=\I\infty$. Using eqs.~\eqref{eq:PetcovMassRelations} and taking
$|\epsilon|,|\lambda_1|,|\lambda_2|\ll1$, we identify the physical masses 
\begin{subequations}
\label{eq:MassApproxApplied}
    \begin{alignat}{2}
	m_\mathrm{III}^2 ~&\approx~ \mathrm{Tr}\,M^\dagger M ~&&\approx~ \Lambda^2\;, \\
	m_\mathrm{II}^2  ~&\approx~ \dfrac{\sum\left|\mathrm{det}\,M_{2\times2}\right|^2}{\mathrm{Tr}\,M^\dagger M} ~&&
			\approx~ \Lambda^2 \left( | \lambda_1|^2 + |\lambda_2|^2 + 18\,|\epsilon^{2/3}| \right) \;, \\
	m_\mathrm{I}^2   ~&\approx~ \dfrac{|\mathrm{det}\,M|^2}{\sum\left|\mathrm{det}\,M_{2\times2}\right|^2} ~&&\approx~ \Lambda^2 \dfrac{|\lambda_1\lambda_2-9\epsilon^{2/3}|^2}{|\lambda_1|^2 + |\lambda_2|^2 + 18\,|\epsilon^{2/3}|} \;.
   \end{alignat}
\end{subequations}
Depending on the relations among $\lambda_1$, $\lambda_2$, and 
$\epsilon$, our model leads to three possible mass hierarchies:
\begin{equation}
\label{eq:HierarchyPatterns}
 (m_\mathrm{I},\,m_\mathrm{II},\,m_\mathrm{III}) ~\approx~ \Lambda~\left\{ \begin{array}{ll}
 \left(\frac{3}{\sqrt{2}}\,|\epsilon^{1/3}|,\,3\sqrt{2}\,|\epsilon^{1/3}|,\,1\right) & \quad\text{for}~~~|\lambda_1|^2 < |\lambda_2|^2 \ll |\epsilon^{2/3}|\;, \\[8pt]
 \left(|\lambda_1|,\,|\lambda_2|,\,1\right)           & \quad\text{for}~~~|\epsilon^{2/3}| \ll |\lambda_1\lambda_2| \ll |\lambda_2|^2\;, \\[8pt]
 \left(9\,\left|\frac{\epsilon^{2/3}}{\lambda_2}\right|,\,|\lambda_2|,\,1\right) & \quad\text{for}~~~|\lambda_1\lambda_2| \ll |\epsilon^{2/3}| \ll |\lambda_2|^2\;.
 \end{array}\right.
\end{equation}
Recall that we assume $|\lambda_1| < |\lambda_2| < 1$ and aim at the observed
mass hierarchies $m_\mathrm{I}\ll m_\mathrm{II}\ll m_\mathrm{III}$. Clearly, 
the first mass configuration in eq.~\eqref{eq:HierarchyPatterns} does not 
satisfy the condition of hierarchical masses. The other two scenarios are compatible 
with our assumptions.

In the valid cases, we find the mass ratios
\begin{subequations}
\label{eqs:massRatios}
  \begin{align}
  \label{eq:massRatios1}
    \dfrac{m_\mathrm{I}}{m_\mathrm{II}} &\approx~\left|\dfrac{\lambda_1}{\lambda_2}\right| && \text{and} &
    \dfrac{m_\mathrm{II}}{m_\mathrm{III}} &\approx~ |\lambda_2| && \text{for}& |\epsilon^{2/3}| \ll |\lambda_1\lambda_2| \ll |\lambda_2|^2\;,\\
  \label{eq:MassRatiosMixedScenario}
    \dfrac{m_\mathrm{I}}{m_\mathrm{II}} &\approx~ 9\,\left|\dfrac{\epsilon^{2/3}}{\lambda_2^2}\right| && \text{and} &
    \dfrac{m_\mathrm{II}}{m_\mathrm{III}} &\approx~ |\lambda_2| && \text{for}& |\lambda_1\lambda_2| \ll |\epsilon^{2/3}| \ll |\lambda_2|^2\,.
  \end{align}
\end{subequations}
Interestingly enough, in both cases the ratio of the heavier masses depends only on $|\lambda_2|$
that, as we saw in section~\ref{sec:EclecticBreakdown}, measures the amount by which the 
$\Z3^{(2)}$ approximate symmetry is broken. On the other hand, the hierarchy $m_\mathrm{I}/m_\mathrm{II}$ 
is governed by the breaking of the modular $\Z3^{(3)}$ approximate symmetry,\footnote{This
situation is similar to the BU scenarios~\cite{Okada:2020ukr,Feruglio:2021dte,Novichkov:2021evw}.}
which is broken either by the flavon parameter $\lambda_1$ or by the modulus parameter $\epsilon$. 
Since in string constructions both moduli and flavons acquire VEVs roughly around the same scales, 
we consider the hierarchy pattern described by eq.~\eqref{eq:MassRatiosMixedScenario} 
to be more appropriate to our scenario.

Let us concentrate now on the lepton sector.
Applying eq.~\eqref{eq:MassRatiosMixedScenario} to charged leptons
(with $m_\mathrm{I}\to m_\mathrm{e}$, $m_\mathrm{II}\to m_\mu$ and $m_\mathrm{III}\to m_\tau$) 
and comparing with their measured mass values (see section~\ref{sec:NumericalAnalysisLepton} 
for the experimental values
of lepton observables), we can fit the flavon VEV as
\begin{equation}\label{eq:Varphi2Prediction}
 \vev{\tilde{\varphi}_{\mathrm{e},2}} ~=~ |\lambda_{\mathrm e,2}| ~\approx~ \dfrac{m_\mu}{m_\tau} ~=~ 0.0586\;.
\end{equation}
Analogously, the modulus VEV is constrained to be approximately
\begin{equation}\label{eq:ImtauPrediction}
    |\epsilon^{\nicefrac{1}{3}}| ~\approx~ \sqrt{\dfrac{|\lambda_{\mathrm e,2}|^2}{9}\dfrac{m_\mathrm{e}}{m_\mu}} ~\approx~ 0.00134
  \qquad \Rightarrow \qquad \im\, \vev T ~\approx~ 3.16
\end{equation}
in order to yield the correct hierarchy for the two light charged 
leptons. We shall see in section~\ref{sec:NumericalAnalysisLepton}
that this approximate analytical result is compatible with a more complete numerical analysis.

As already mentioned, the uncovered pattern for charged leptons applies equally in our model
also to the up and down-quark sector separately. This symmetric structure has its 
root in the spectrum of our model, see table~\ref{tab:MSSM-model447}, which leads to
the superpotential~\eqref{eq:superpot}. We notice that the only difference among
the Yukawas is that the flavons are different fields but have identical quantum
numbers. Even more, the appearance of $\varphi_\mathrm{e}$ in the down-quark and 
charged-lepton Yukawas reveals identical mass relations in both sectors.
These symmetries are interesting but challenge the phenomenological viability of our model.
As we shall shortly see, corrections to the K\"ahler potential arising from
flavon VEVs alleviate this issue.

\subsubsection{Neutrino mass hierarchies}

Light neutrino masses occur in our model via a seesaw mechanism.
The corresponding light neutrino mass matrix $M_\nu$ 
has been defined in eq.~\eqref{eq:NeutrinoMassMatrix}.
In order to write down the explicit mass matrix, we need a closed 
form expression for the inverse of the Majorana mass matrix 
eq.~\eqref{eq:DiracAndMajornaMassMatrices}. This is up to an overall factor given by
\begin{equation}
 M^{-1}_\mathrm{M}
 ~\sim~
 \begin{pmatrix}
 \lambda_{\mathrm{e},2} & -3 \,\epsilon^{\nicefrac{1}{3}} & -3 \,\lambda_{\mathrm{e},2}^2\,\epsilon^{\nicefrac{1}{3}} \\
 -3 \,\epsilon^{\nicefrac{1}{3}} & \lambda_{\mathrm{e},1} & -3 \,\lambda_{\mathrm{e},1}^2\,\epsilon^{\nicefrac{1}{3}} \\
 -3 \,\lambda_{\mathrm{e},2}^2\,\epsilon^{\nicefrac{1}{3}} & -3\,\lambda_{\mathrm{e},1}^2\,\epsilon^{\nicefrac{1}{3}} & \lambda_{\mathrm{e},1}\,\lambda_{\mathrm{e},2}
 -9\,\epsilon^{\nicefrac{2}{3}}
 \end{pmatrix}\,+\,\mathcal{O}(\lambda_{\mathrm{e},1}\,\epsilon^{\nicefrac{2}{3}})\;.
\end{equation}
Since two flavons appear in the light neutrino mass matrix, we have to distinguish
between the components $\lambda_{\mathrm{e},1}$, $\lambda_{\mathrm{e},2}$ from 
$\vev{\tilde{\varphi}_{\mathrm{e}}}$, and $\lambda_{\nu,1}$, $\lambda_{\nu,2}$ 
from $\vev{\tilde{\varphi}_\nu}$ in the following.
The structure of the light neutrino mass matrix is then given by
\begin{equation}
 M_\nu
 ~\sim~
 \begin{pmatrix}
 \Delta_1 & \Sigma_3 \,\epsilon^{\nicefrac{1}{3}} & \Sigma_2 \,\epsilon^{\nicefrac{1}{3}} \\
 \Sigma_3 \,\epsilon^{\nicefrac{1}{3}} & \Delta_2 & \Sigma_1 \,\epsilon^{\nicefrac{1}{3}} \\
 \Sigma_2 \,\epsilon^{\nicefrac{1}{3}} & \Sigma_1 \,\epsilon^{\nicefrac{1}{3}} & \Delta_3
 \end{pmatrix}\,+\,\mathcal{O}(\epsilon^{\nicefrac{2}{3}})\;,
\label{eq:NeutrinoMassMatrixAtInfty}
\end{equation}
where
\begin{eqnarray}
 \Delta_1 = \lambda_{\nu,1}^2 \,\lambda_{\mathrm{e},2}\;,\quad
 \Delta_2 = \lambda_{\mathrm{e},1} \lambda_{\nu,2}^2\;, \quad
 \Delta_3 = \lambda_{\mathrm{e},1} \lambda_{\mathrm{e},2}\;,
\end{eqnarray}
and
\begin{subequations} 
 	\begin{align}
 	\Sigma_1 ~&=~ 3\,\lambda_{\mathrm{e},1} \,\Big(\lambda_{\nu,1} \lambda_{\nu,2} + \lambda_{\nu,1} \lambda_{\mathrm{e},2} - \lambda_{\mathrm{e},1} \lambda_{\nu,2} \Big)\;, \\
 	\Sigma_2 ~&=~ 3\,\lambda_{\mathrm{e},2} \,\Big(\lambda_{\nu,1} \lambda_{\nu,2} - \lambda_{\nu,1} \lambda_{\mathrm{e},2} + \lambda_{\mathrm{e},1} \lambda_{\nu,2} \Big)\;, \\
 	\Sigma_3 ~&=~ 3 ~\Big(-\lambda_{\nu,1} \lambda_{\nu,2} + \lambda_{\nu,1} \lambda_{\mathrm{e},2} + \lambda_{\mathrm{e},1} \lambda_{\nu,2} \Big)\;.
 	\end{align}
\end{subequations}
 
By using eq.~\eqref{eq:PetcovMassRelations}, one might find approximate (rather long) 
expressions for the neutrino masses, which depend on the various hierarchy configurations
of the small parameters $\lambda_i$ and $\epsilon$. A full classification of the large 
number of these hierarchies is not very enlightening.
Instead, let us focus here on the more appealing scenario given by the VEV relations
\begin{equation}
\label{eq:NeutrinoParameterOrdering}
 |\lambda_{\mathrm{e},1}\lambda_{\mathrm{e},2}| ~\approx~ |\lambda_{\nu,1}|^2 ~\ll~ |\lambda_{\mathrm{e},1}| 
    ~\ll~ |\lambda_{\nu,1}| ~\approx~ |\epsilon^{\nicefrac{1}{3}}| ~\ll~ |\lambda_{\mathrm{e},2}| ~\ll~ |\lambda_{\nu,2}| \approx 1\;.
\end{equation}
For this specific case, the neutrino masses, up to their overall mass scale, 
approximately read
\begin{equation}
 \left(m_1,\,m_2,\,m_3\right) ~\sim~ \left(9\dfrac{|\epsilon^{\nicefrac{2}{3}}\,\lambda_{\nu,1}^2|}{|\lambda_{\mathrm{e},1}|} , \,|\lambda_{\mathrm{e},1}\,\lambda_{\mathrm{e},2}| , \,|\lambda_{\mathrm{e},1}|\right)\;,
\end{equation}
where we still satisfy that $m_1\ll m_2\ll m_3$. The mass ratios turn out to be
\begin{equation}\label{eq:NeutrinoMassRationsMixedScenario}
 \dfrac{m_1}{m_2} ~\approx~ 9\,\left|\dfrac{\epsilon^{2/3}}{\lambda_{\mathrm{e},1}}\right|\qquad\text{and}\qquad
 \dfrac{m_2}{m_3} ~\approx~ |\lambda_{\mathrm{e},2}|\;.
\end{equation}
Hence, just as in the charged-lepton sector, the hierarchies in the neutrino masses 
are governed by the amount by which the $\Z3^{(2)}\x\Z3^{(3)}$ approximate symmetry is broken.
Indeed, the relation between $m_2$ and $m_3$ coincides approximately
with the hierarchy of the heavier charged leptons, eq.~\eqref{eq:Varphi2Prediction}.
Furthermore, a direct consequence of the VEV configuration~\eqref{eq:NeutrinoParameterOrdering} 
is that the difference between the lightest neutrino $m_1$ and $m_2$ is smaller than the 
difference between the heaviest neutrino $m_3$ and $m_2$, i.e.
\begin{equation}
\label{eq:NOneutrinos}
 \dfrac{m_2-m_1}{m_3-m_2} ~\approx~ |\lambda_{\mathrm{e},2}| ~<~ 1\;,
\end{equation}
which corresponds to a normal-ordered neutrino spectrum.
As the subsequent numerical analysis will show, the specific VEV relations 
of eq.~\eqref{eq:NeutrinoParameterOrdering} are in fact  compatible with the best-fit 
scenario that allows us to reproduce all observations in the lepton sector.

\section{Numerical analysis of the lepton sector}
\label{sec:NumericalAnalysisLepton}

\begin{table}[t]
	\centering
	\begin{tabular}{ll}
		\toprule
		observables                                  & best fit values  \\
		\midrule
		$m_\mathrm{e}/m_\mu$                         & $0.00474\pm0.00004$ \\[4pt]
		$m_\mu/m_\tau$                               & $0.0586^{+0.0004}_{-0.0005}$ \\
		\arrayrulecolor{lightgray}\midrule\arrayrulecolor{black}
		$\Delta m_{21}^2 / 10^{-5} ~[\mathrm{eV}^2]$ & $7.42^{+0.21}_{-0.20}$\\[4pt]
		$\Delta m_{31}^2 / 10^{-3} ~[\mathrm{eV}^2]$ & $2.510^{+0.027}_{-0.027}$\\[4pt]
		$\sin^2\theta_{12}$                          & $0.304^{+0.012}_{-0.012}$ \\[4pt]
		$\sin^2\theta_{13}$                          & $0.02246^{+0.00062}_{-0.00062}$\\[4pt]
		$\sin^2\theta_{23}$                          & $0.450^{+0.019}_{-0.016}$\\[4pt]
		$\delta_{\mathrm{\CP}}^{\ell}/\pi$           & $1.28^{+0.20}_{-0.14}$\\[2pt]
		\bottomrule
	\end{tabular}
	\caption{Observed masses and mixing angles of the lepton sector. We show the best 
		fit and $1\sigma$ errors for the charged-lepton mass ratios at the GUT scale, assuming 
		$\tan\beta=10$, $M_\mathrm{SUSY}=10$\,TeV, and $\bar\eta_b=0.09375$; 
		taken from~\cite{Antusch:2013jca}. We also present the 
		best-fit values and $1\sigma$ errors for the neutrino oscillation parameters 
		given by the global analysis NuFIT v5.1~\cite{Esteban:2020cvm} with Super-Kamiokande
		data for normal ordering. \label{tab:ExpDataLeptons}}
\end{table}

Let us now fit the parameters of our model such that it reproduces observations in the lepton sector.
We aim at the experimental observables summarized in table~\ref{tab:ExpDataLeptons}. 
In the top block, we show the current values of the mass ratios and $1\sigma$ errors for 
the charged leptons, evaluated at the GUT scale (for the running of these 
parameters, see e.g.~\cite{Antusch:2013jca}), assuming $\tan\beta=10$, $M_\mathrm{SUSY}=10$\,TeV, 
and $\bar\eta_b=0.09375$, as described in~\cite{Chen:2021zty,Ding:2021zbg}.
In the bottom block, the best-fit values and $1\sigma$ errors of neutrino-oscillation parameters are presented, 
as given by the global analysis NuFIT v5.1~\cite{Esteban:2020cvm}. These values include 
data on atmospheric neutrinos provided by the Super-Kamiokande collaboration. 
The table contains only data for normal ordering because a successful fit of our model with inverted 
ordering was not possible.
Note that the oscillation parameters are given at the low scale. It is common in the literature on modular flavor 
symmetries to assume that the running from low energies to the GUT scale of these parameters is negligible. 
This is justified by arguing that the effects of the running would be smaller than the experimental 
errors. We shall adopt this practice here.

The lepton sector of our model depends on a set $x$ of 7 parameters, i.e.\
\begin{equation}
\label{eq:paramsLepton}
x~=~ \left\{\re\,\vev T,\, \im\,\vev T,\, 
\vev{\tilde\varphi_{\mathrm{e},1}},\,
\vev{\tilde\varphi_{\mathrm{e},2}},\,
\vev{\tilde\varphi_{\nu,1}},\,
\vev{\tilde\varphi_{\nu,2}},\,
\Lambda_\nu \right\}\;,
\end{equation}
which include the VEVs of the two real components of the modulus $T$, and the VEVs 
of the four nontrivial (real) components of the flavon triplets $\varphi_\mathrm{e}$ and 
$\varphi_\nu$, and the neutrino mass scale $\Lambda_\nu$.
In addition, one might include the overall mass scale $\Lambda_\mathrm{e}$ of
charged leptons, but we omit it as we shall fit only the mass ratios of that sector.
For each choice of the values of the parameters~\eqref{eq:paramsLepton} one can numerically 
diagonalize the charged-lepton and neutrino mass matrices, eqs.~\eqref{eq:ElectronMassMatrix} 
and~\eqref{eq:NeutrinoMassMatrix}. From this process one can then extract
the physical masses as well as the mixing angles and \CP violation phase(s) that parametrize 
the lepton mixing matrix.\footnote{We use the PDG convention for 
the parametrization of the lepton mixing matrix~\cite{ParticleDataGroup:2020ssz}.} 

\begin{figure}[t]
	\begin{center}
		\subfloat[]{\label{fig:ErrorMismatchS23}
			\includegraphics[width=0.48\linewidth]{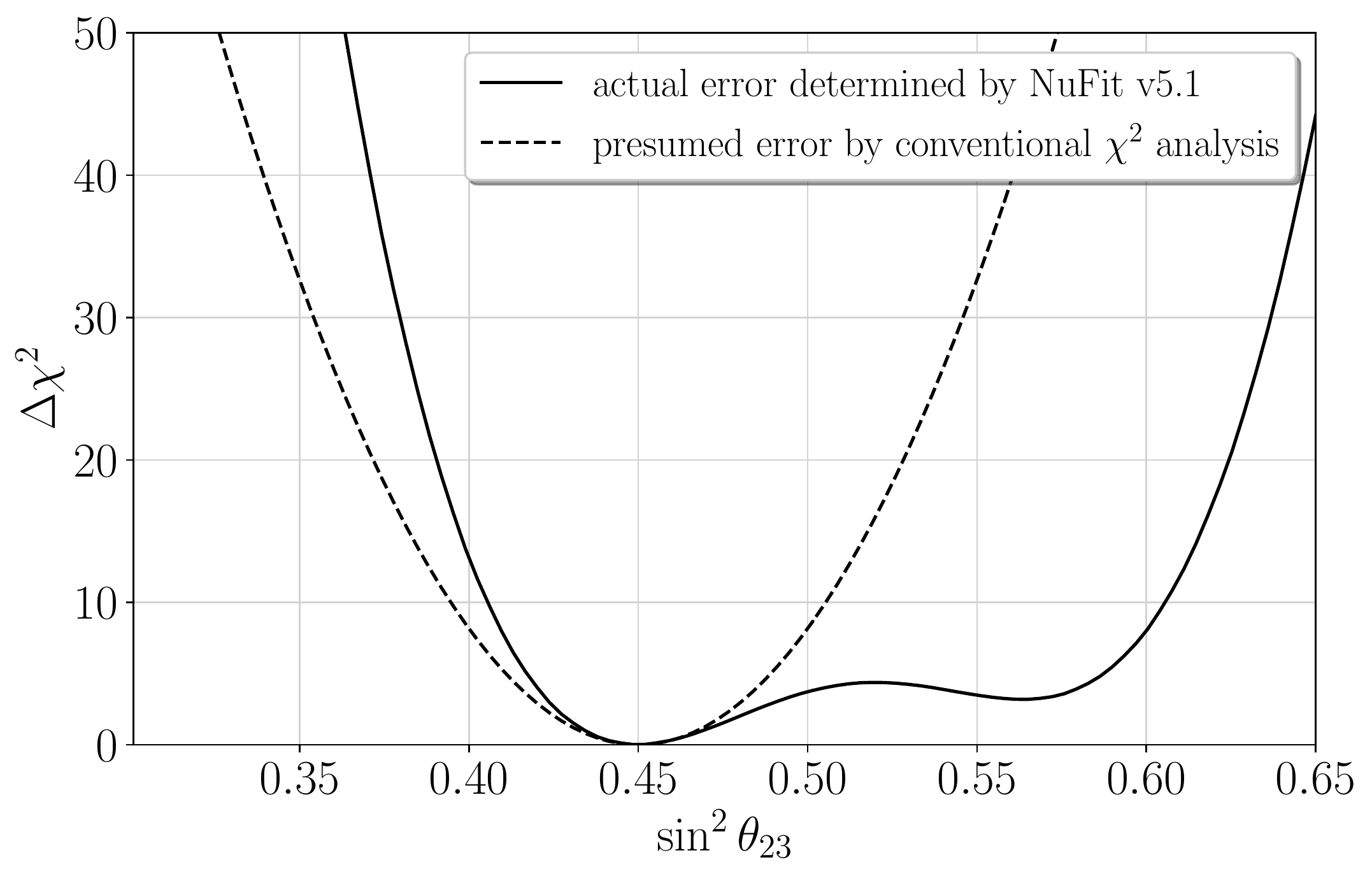}}
		\subfloat[]{\label{fig:ErrorMismatchDcp}
			\includegraphics[width=0.48\linewidth]{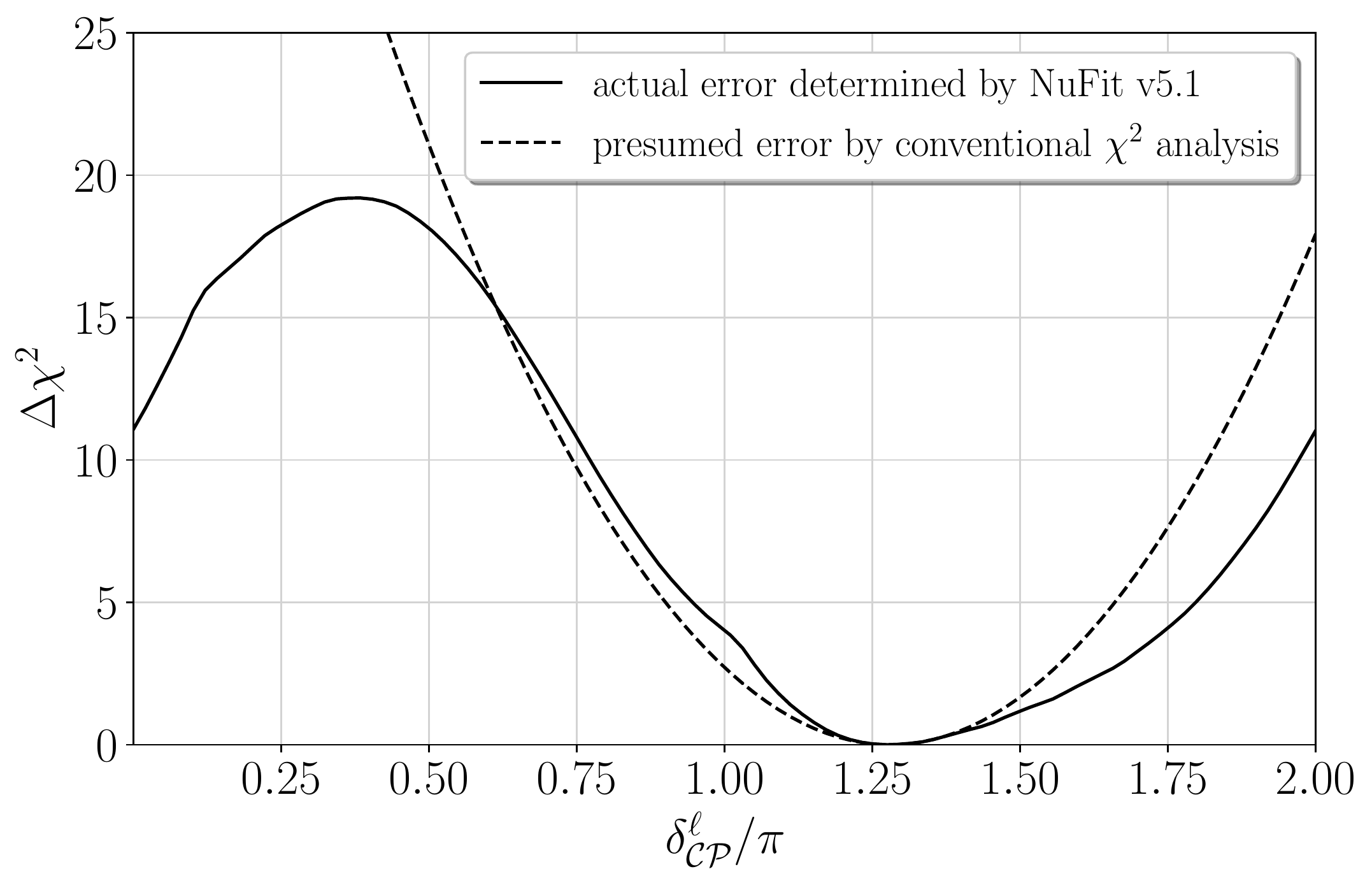}}
	\end{center}
	\vspace{-0.5cm}
	\caption{Comparison of the $\chi^2$ profile determined by the global analysis 
		NuFIT v5.1~\cite{Esteban:2020cvm} and the presumed profile computed with eq.~\eqref{eq:ConventionalChisquare} 
		in a conventional $\chi^2$ analysis for (a) $\sin^2\theta_{23}$ and (b) $\delta^\ell_{\CP}$. 
		\label{fig:ErrorMismatch}}
\end{figure}

As a quantitative measurement for the goodness of our fit, we perform a $\chi^2$ analysis.
We define a $\chi^2$ function
\begin{equation}
\chi^2(x) ~=~ \sum_i \Delta\chi^2_i(x)\;,
\end{equation}
where we sum over charged-lepton mass ratios and all observables listed in table~\ref{tab:ExpDataLeptons}.
For the charged-lepton mass ratios we use
\begin{equation}
\Delta \chi_i(x) ~=~ \dfrac{\mu_{i,\mathrm{exp}} - \mu_{i,\mathrm{model}}(x)}{\sigma_i}\;,
\label{eq:ConventionalChisquare}
\end{equation}
where $\mu_\mathrm{model}$ is the prediction of the model and $\mu_\mathrm{exp}$ and 
$\sigma$ are its corresponding experimental best-fit value and the size of its $1\sigma$ 
error, respectively. 
For the neutrino-oscillation parameters, we use the profiles of the one dimensional 
$\Delta\chi^2$ projections obtained by the global analysis NuFIT v5.1.%
\footnote{The data for the one dimensional $\Delta\chi^2$ projections 
is conveniently accessible on the NuFIT website~\cite{Esteban:2020cvm}.}
This makes a difference especially for $\sin^2\theta_{23}$ and $\delta^{\ell}_{\CP}$, 
as can be directly appreciated from figure~\ref{fig:ErrorMismatch}.
For instance, by using a conventional $\Delta\chi^2$ obtained from eq.~\eqref{eq:ConventionalChisquare}, 
one would underestimate the goodness of the fit by multiple sigma ranges for the second 
octant of $\theta_{23}$ and also for small values of $\delta^{\ell}_{\CP}$. 
For $\sin^2\theta_{23}<0.45$ the goodness of the fit would be overestimated.
We included $\delta^{\ell}_{\CP}$ when calculating $\chi^2$ because, even 
though no values could be excluded with $5\sigma$ by now, experiments do seem to favor 
some values of $\delta^{\ell}_{\CP}$ over others.
We numerically minimize the function $\chi^2(x)$ as described in appendix~\ref{app:FitDetails}.

\begin{figure}[t]
	\centering
	\includegraphics[width=0.96\linewidth]{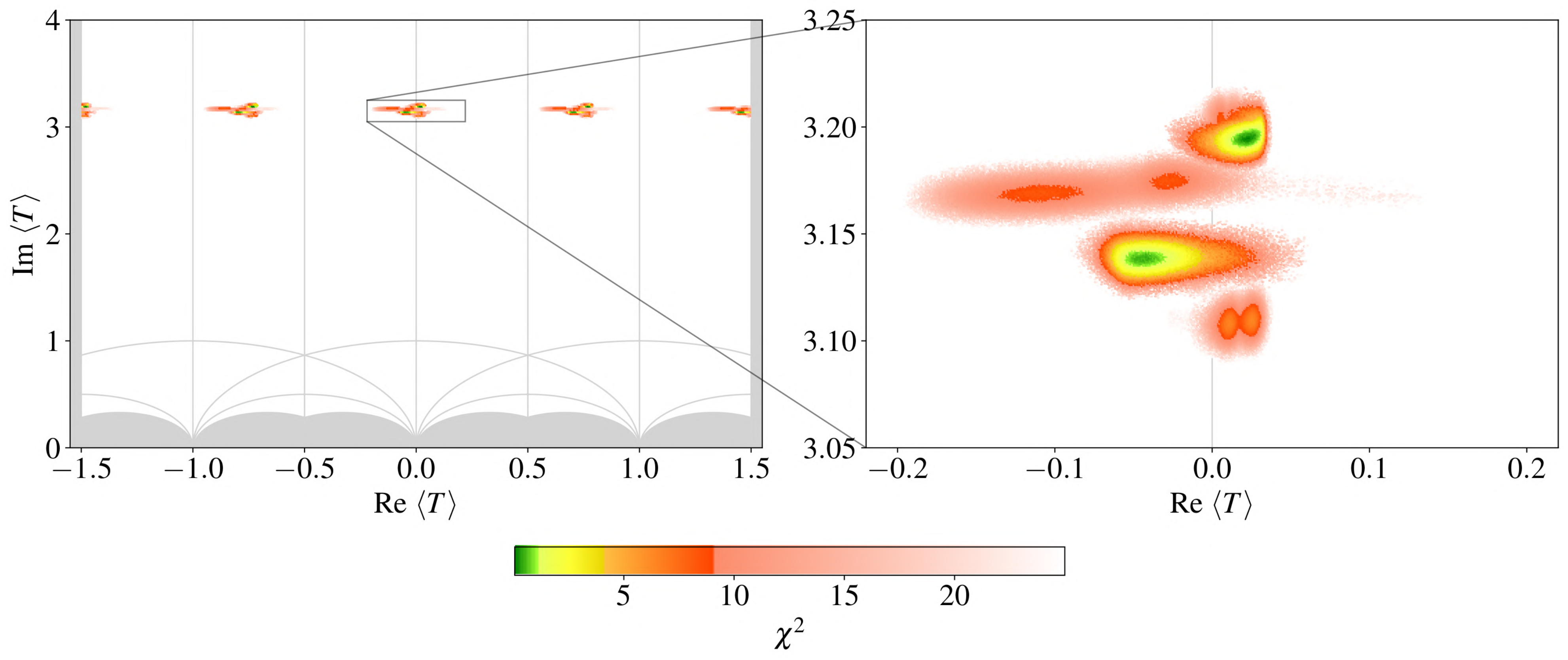}
	\caption{Regions in the fundamental domain of $\Gamma(3)$ that yield fits with $\chi^2\leq25$. 
		Note that a mapping into the fundamental domain of $\SL{2,\Z{}}$ with a modular transformation 
		$\gamma\in T'$ is not possible for this model, since we require the flavon VEVs to be real, 
		i.e.\ to respect the \CP-like symmetry. The analogous flavon VEVs after performing a $T'$ 
		transformation would in general be complex. The colors green, yellow, and orange may be interpreted 
		as the $1\sigma$, $2\sigma$, and $3\sigma$ confidence levels, while the opaque red fades out to white 
		until the $5\sigma$ barrier is reached. 
		Note that there are two disconnected $1\sigma$ regions on the right-hand side plot.
		In the right green region, the best point is $\vev T = 0.02279+3.195\,\I$, which yields $\chi^2=0.08$. 
		In the left green region, $\vev T = -0.04283 + 3.139\,\I$ yields $\chi^2=0.45$. Therefore, 
		the best-fit value of the model lies in the right green region. 
		\label{fig:modulispace}}
\end{figure}

This numerical scan yields a successful fit to current experimental data with
an overall $\chi^2 = 0.08$.
The regions in moduli space that yield good fits, with $\chi^2\leq25$,
are depicted in figure~\ref{fig:modulispace}.  As we see, there are multiple 
clusters that yield good fits. Interestingly, they have roughly the same shape 
but are shifted by $T\rightarrow T+\nicefrac{3}{4}$ while also 
$\vev{\tilde\varphi_{\mathrm{e},1}} \rightarrow - \vev{\tilde\varphi_{\mathrm{e},1}}$, $\vev{\tilde\varphi_{\nu,1}} \rightarrow - \vev{\tilde\varphi_{\nu,1}}$. 
Note that this transformation is {\it not} part of the eclectic flavor group. 
It therefore turns out to be an accidental approximate symmetry of the model. 
This symmetry originates from the properties under modulus shifts of the modular
forms of weight 1 that appear in our model. Namely, $T\rightarrow T+\nicefrac{3}{4}$ 
results in $\hat{Y}_1(T) \rightarrow -\I\,\hat{Y}_1(T)$, up to $\mathcal{O}(q)$ 
corrections, as shown in eq.~\eqref{eq:ModformApproximateSymmetry}.
Moreover, every cluster has two $1\sigma$ (green) regions. As we shall shortly see, this 
bimodality is inherited by most observables. 
For the two green regions of the cluster in the fundamental domain of $\SL{2,\Z{}}$, 
the best-fit values and $1\sigma$ intervals for the parameters $x$ of the model are 
listed in table~\ref{tab:LeptonFitParameters}.
Note that the best-fit values are very close to the predictions
from the analytical approximate analysis for the mass ratios given in 
eqs.~\eqref{eq:Varphi2Prediction} and~\eqref{eq:ImtauPrediction}.

\begin{table}[t]
	\vspace*{10pt}
	\centering
	\resizebox{\textwidth}{!}{
		\begin{tabular}{llclc}
			\toprule
			&\multicolumn{2}{c}{right green region} & \multicolumn{2}{c}{left green region}\\\cmidrule(r{4pt}l{4pt}){2-3}\cmidrule(l{4pt}r{4pt}){4-5}
			parameter~~                          & best-fit value~        & $1\sigma$ interval & best-fit value~        & $1\sigma$ interval \\\midrule
			$\re\, \vev T$                       & $\phantom{-}0.02279$    & $0.01345 \rightarrow 0.03087$ & $-0.04283$    & $-0.05416 \rightarrow -0.02926$\\
			$\im\, \vev T$                       & $\phantom{-}3.195$      & $3.191 \rightarrow 3.199$ & $\phantom{-}3.139$      & $3.135 \rightarrow 3.142$\\
			$\vev{\tilde\varphi_{\mathrm{e,1}}}$ & $-4.069 \cdot 10^{-5}$ & $-4.321 \cdot 10^{-5}\rightarrow -3.947 \cdot 10^{-5}$ & $\phantom{-}2.311 \cdot 10^{-5}$ & $2.196 \cdot 10^{-5}\rightarrow 2.414 \cdot 10^{-5}$\\
			$\vev{\tilde\varphi_{\mathrm{e,2}}}$ & $\phantom{-}0.05833$    & $0.05793 \rightarrow 0.05876$ & $\phantom{-}0.05826$    & $0.05792 \rightarrow 0.05863$\\
			$\vev{\tilde\varphi_{\mathrm{\nu,1}}}$ & $\phantom{-}0.001224$ & $0.001201 \rightarrow 0.001248$  & $-0.001274$ & $-0.001304 \rightarrow -0.001248$ \\ 
			$\vev{\tilde\varphi_{\mathrm{\nu,2}}}$ & $-0.9857$             & $-1.0128 \rightarrow -0.9408$ & $\phantom{-}0.9829$             & $0.9433 \rightarrow 1.0122$\\
			$\Lambda_\nu~[\mathrm{eV}]$          & $\phantom{-}0.05629$    & $0.05442 \rightarrow 0.05888$ & $\phantom{-}0.05591$    & $0.05408 \rightarrow 0.05850$ \\\midrule
			$\,\chi^2$                           & $\phantom{-}0.08$     && $\phantom{-}0.45$ & \\
			\bottomrule
		\end{tabular}
	}
	\caption{Best-fit values and their corresponding $1\sigma$ intervals for the two green regions 
		displayed in the plot on the right-hand side of figure~\ref{fig:modulispace}.
		\label{tab:LeptonFitParameters}}
	\vspace*{30pt}
\end{table}

\begin{table}
	\centering
	\resizebox{\textwidth}{!}{
		\begin{tabular}{lrrrrrr}
			\toprule
			& \multicolumn{3}{c}{model} &\multicolumn{3}{c}{experiment}\\\cmidrule(r{4pt}l{4pt}){2-4}\cmidrule(l{4pt}r{4pt}){5-7}
			observable & best fit & $1\sigma$ interval & $3\sigma$ interval & best fit & $1\sigma$ interval & $3\sigma$ interval\\\midrule
			$m_\mathrm{e}/m_\mu$ & $0.00473$ & $0.00470\rightarrow0.00477$ & $0.00462\rightarrow0.00485$ & $0.00474$ & $0.00470\rightarrow0.00478$ & $0.00462\rightarrow0.00486$\\
			$m_\mu/m_\tau$ & $0.0586$ & $0.0581\rightarrow0.0590$ & $0.0572\rightarrow0.0600$ & $0.0586$ & $0.0581\rightarrow0.0590$ & $0.0572\rightarrow0.0600$\\\midrule
			$\sin^2\theta_{12}$ & $0.303$ & $0.294\rightarrow0.315$ & $0.275\rightarrow0.335$ & $0.304$ & $0.292\rightarrow0.316$ & $0.269\rightarrow0.343$ \\
			$\sin^2\theta_{13}$ & $0.02254$ & $0.02189\rightarrow0.02304$ & $0.02065\rightarrow0.02424$ & $0.02246$ & $0.02184\rightarrow0.02308$  & $0.02060\rightarrow0.02435$\\
			$\sin^2\theta_{23}$ & $0.449$ & $0.436\rightarrow0.468$ & $0.414\rightarrow0.593$ & $0.450$ & $0.434 \rightarrow 0.469$ & $0.408 \rightarrow 0.603$ \\\midrule
			$\delta^\ell_\CP/\pi$ & $1.28$ & $1.15\rightarrow1.47$ & $0.81\rightarrow1.94$ & $1.28$ & $1.14 \rightarrow 1.48$ & $0.80 \rightarrow 1.94$ \\
			$\eta_1/\pi \mod 1$ & $0.029$ & $0.018\rightarrow0.048$ & $-0.031\rightarrow0.090$ & - & - & - \\
			$\eta_2/\pi \mod 1$ & $0.994$ & $0.992\rightarrow0.998$ & $0.935\rightarrow1.004$ & - & - & - \\
			$J_\CP$ & $-0.026$ & $-0.033\rightarrow-0.015$ & $-0.035\rightarrow0.019$ & $-0.026$ & $-0.033\rightarrow-0.016$ & $-0.033\rightarrow0.000$ \\
			$J_\CP^\mathrm{max}$ & $0.0335$ & $0.0330\rightarrow0.0341$ & $0.0318\rightarrow0.0352$ & $0.0336$ & $0.0329\rightarrow0.0341$ & $0.0317\rightarrow0.0353$ \\\midrule
			$\Delta m_{21}^2/10^{-5}~[\mathrm{eV}^2]$ & $7.39$ & $7.35\rightarrow7.49$ & $7.21\rightarrow7.65$ & $7.42$ & $ 7.22\rightarrow 7.63$ & $6.82 \rightarrow 8.04$ \\
			$\Delta m_{31}^2/10^{-3}~[\mathrm{eV}^2]$ & $2.508$ & $2.488\rightarrow2.534$ & $2.437\rightarrow2.587$ & $2.521$ & $ 2.483\rightarrow 2.537$ & $2.430 \rightarrow 2.593$ \\
			$m_1~[\mathrm{eV}]$ & $0.0042$ & $0.0039\rightarrow0.0049$ & $0.0034\rightarrow0.0131$ & $<0.037$ & - & - \\
			$m_2~[\mathrm{eV}]$ & $0.0095$ & $0.0095\rightarrow0.0099$ & $0.0092\rightarrow0.0157$ & - & - & - \\
			$m_3~[\mathrm{eV}]$ & $0.0504$ & $0.0501\rightarrow0.0505$ & $0.0496\rightarrow0.0519$ & - & - & - \\
			$\sum_i m_i~[\mathrm{eV}]$ & $0.0641$ & $0.0636\rightarrow0.0652$ & $0.0628\rightarrow0.0806$ & $<0.120$ & - & - \\
			$m_{\beta\beta}~[\mathrm{eV}]$ & $0.0055$ & $0.0045\rightarrow0.0064$ & $0.0040\rightarrow0.0145$ & $<0.036$ & - & - \\
			$m_{\beta}~[\mathrm{eV}]$ & $0.0099$ & $0.0097\rightarrow0.0102$ & $0.0094\rightarrow0.0159$ & $<0.8$ & - & - \\\midrule
			$\chi^2$ & $0.08$ & & & & & \\
			\bottomrule
		\end{tabular}
	}
	\caption{Comparison of the best-fit values in the lepton sector of our model against the 
		experimental data. In the columns 2--4 we present the best values from our fit with their $1\sigma$ and $3\sigma$-error
		intervals. We have added $\mod1$ for $\eta_{1,2}$ because there are two disconnected 
		$1\sigma$ regions shifted by $\pi$, cf.\ figure~\ref{fig:majoranaphases}. In the last three
		columns, we include the experimental best fit and $1\sigma$ ranges for the
		charged-lepton mass ratios at the GUT scale, assuming $\tan\beta=10$, $M_\mathrm{SUSY}=10$\,TeV, and $\bar\eta_b=0.09375$, 
		taken from~\cite{Antusch:2013jca}. In addition, we give the best-fit values and error intervals for 
		the neutrino-oscillation parameters as obtained by the global analysis NuFIT v5.1~\cite{Esteban:2020cvm} 
		with Super-Kamiokande data for normal ordering.
		\label{tab:FitLeptons}}
\end{table}

In table~\ref{tab:FitLeptons}, we summarize the best-fit values for the observables resulting from 
our numerical scan.
At the best-fit point, all observables (i.e.\ the charged lepton mass ratios, the neutrino mass-squared 
differences, and the four lepton mixing matrix parameters) are within the $1\sigma$ interval
of the current experimental data. In addition, even though we did not demand it in our fit, it turns out
that the results of the fit are in agreement with the experimental bounds for the lightest 
neutrino mass $m_1$, the sum of neutrino masses $\sum_i m_i$, the effective mass for 
neutrino-less double beta decay $m_{\beta\beta}$, and the neutrino mass observable in ${}^3\mathrm{H}$ 
beta decay $m_\beta$, cf.~\cite{GAMBITCosmologyWorkgroup:2020rmf}, \cite{Planck:2018vyg}, 
\cite{KamLAND-Zen:2022tow}, and \cite{KATRIN:2021uub}, respectively.

\begin{figure}
	\centering
	\includegraphics[width=0.55\linewidth]{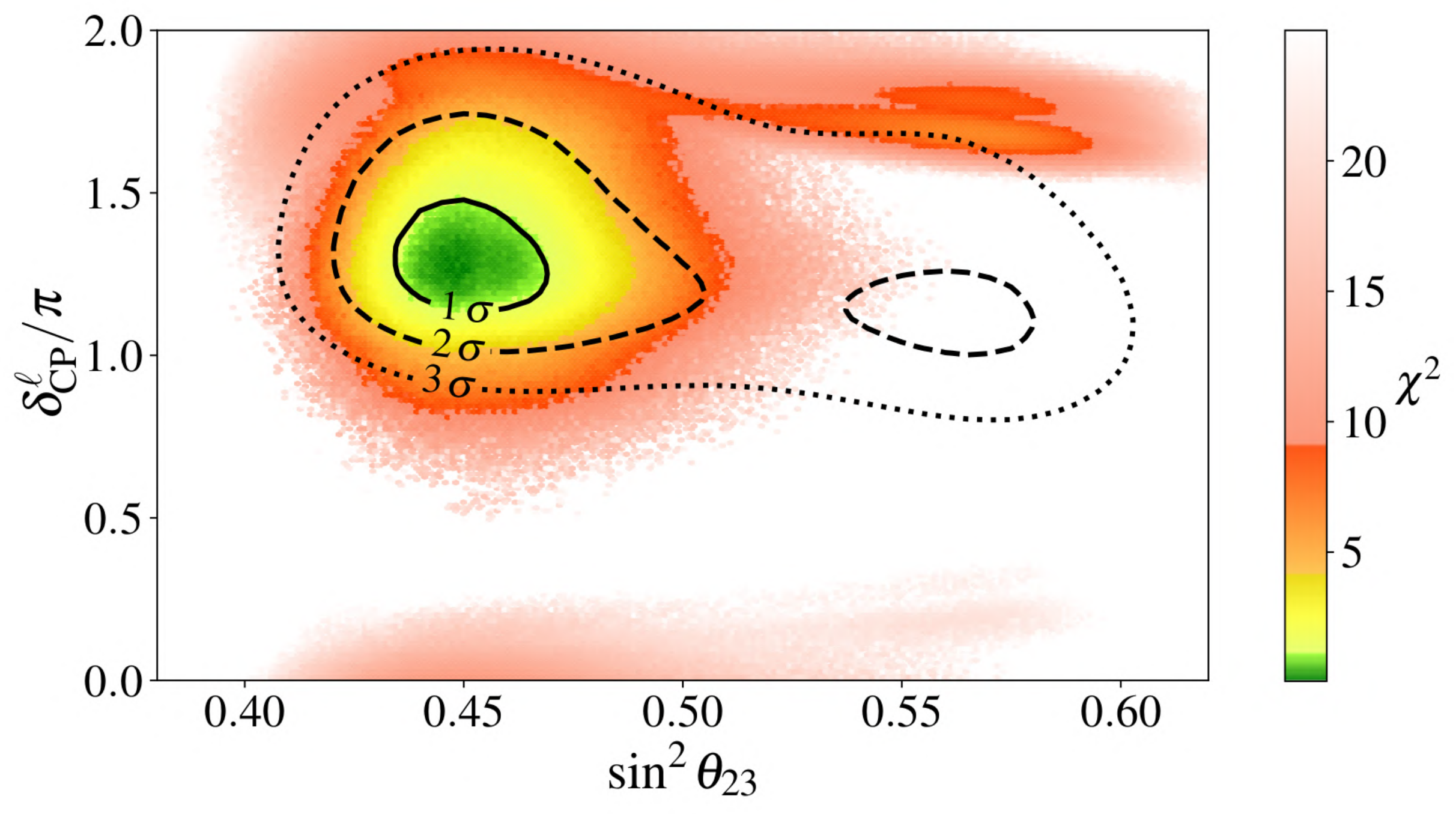}
	\caption{Fitted regions with $\chi^2\leq25$ in the space of $\sin^2\theta_{23}$ and 
		$\delta^\ell_{\CP}$ achieved in our model.
		The black lines delimit the experimental $1$, $2$, and $3\sigma$ regions as 
		determined by the global analysis NuFIT. 
		The bimodality appearing in moduli space, 
		cf.\ figure~\ref{fig:modulispace}, seems to be absent in the $\theta_{23}-\delta^\ell_\CP$ 
		plane, as the two green regions overlap and therefore appear as only one green 
		region here. \label{fig:s23dcp}}
\end{figure}

\begin{figure}
	\centering
	\includegraphics[width=0.6\linewidth]{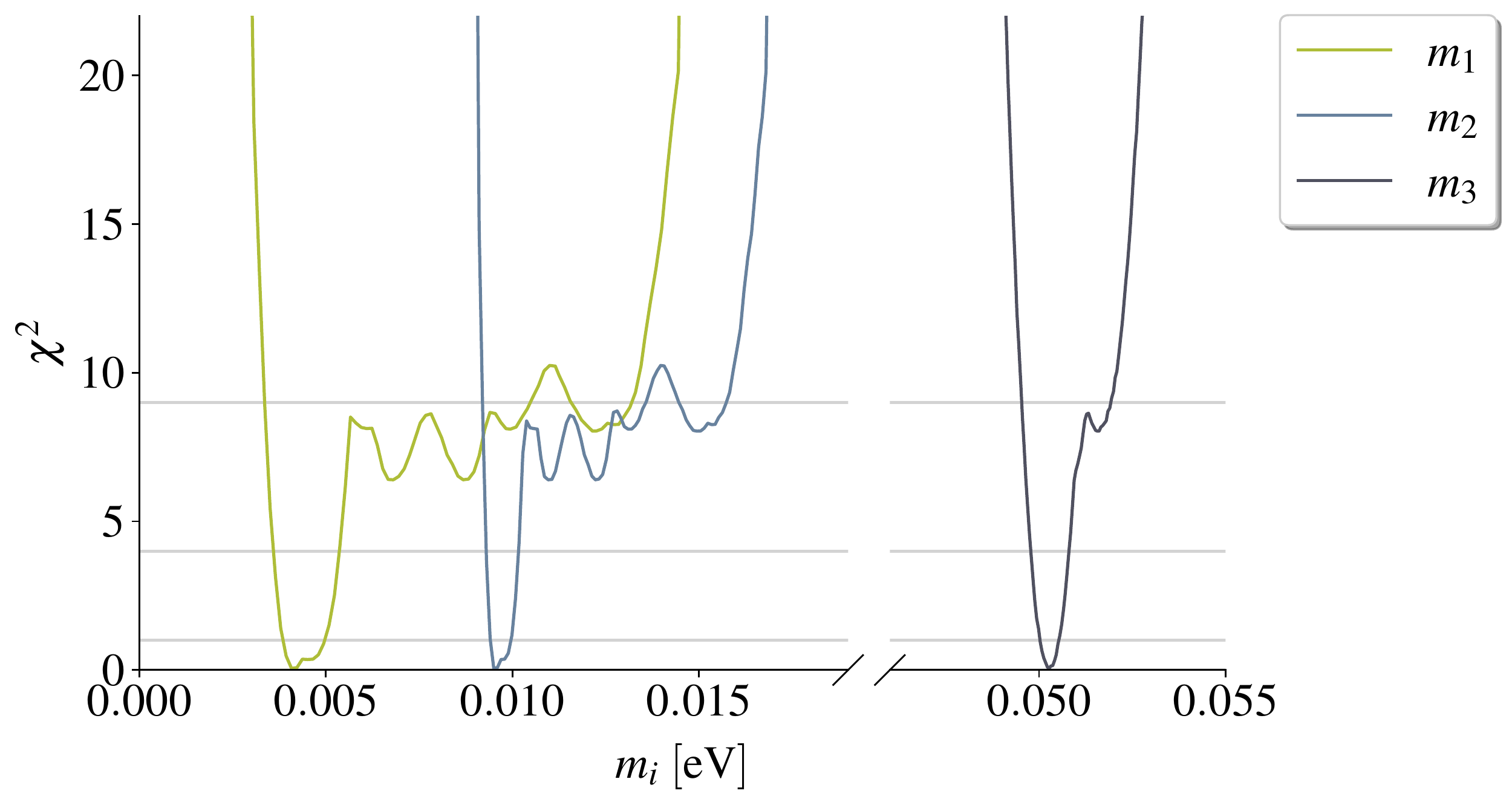}
	\caption{Projections of $\chi^2$ on the neutrino masses, 
		which are clearly normal ordered.\label{fig:neutrinomasses}}
\end{figure}

\begin{figure}[t]
	\centering
	\includegraphics[width=0.96\linewidth]{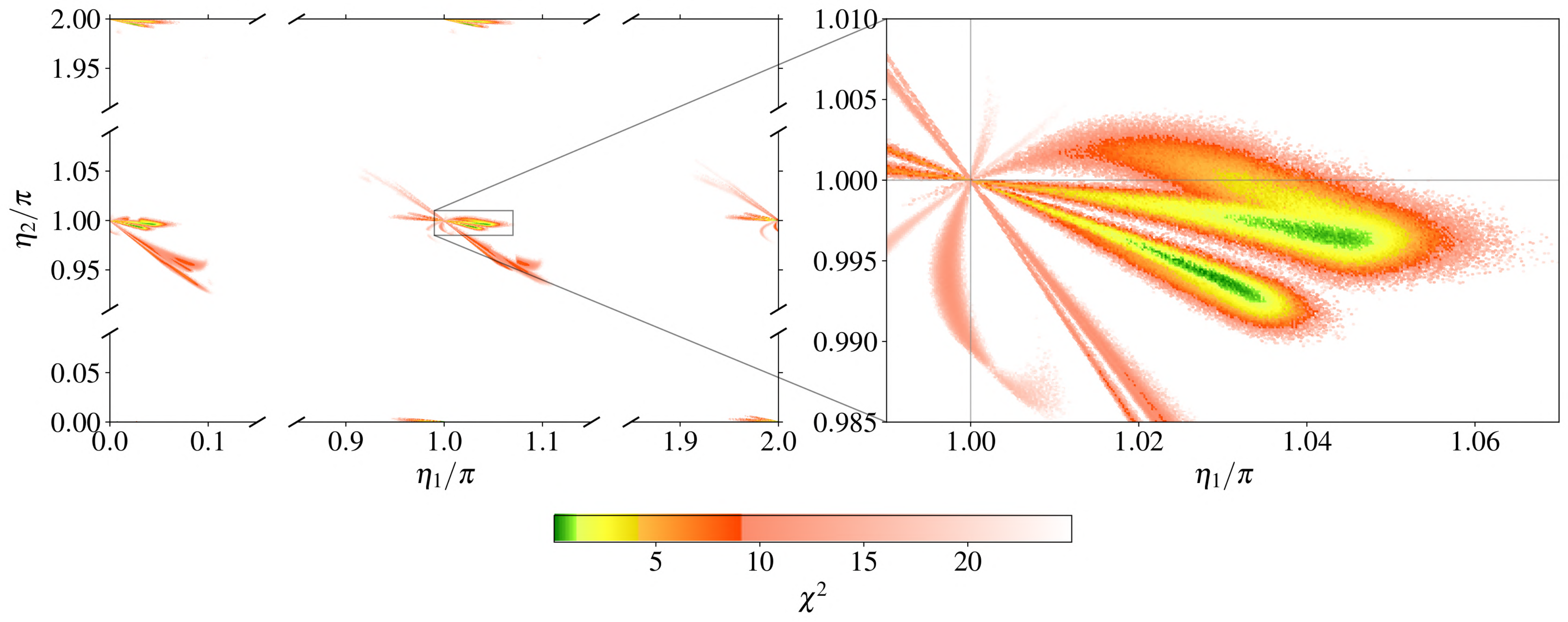}
	\caption{Majorana phases predicted by our model.
	Note that the Majorana phases are found to be in general near \CP-conserving values.
	The appearance of two $1\sigma$ (green) regions in this plot stems from
	the bimodality found in moduli space: each green region in the plot on 
	the right-hand side arises from a different $1\sigma$ region of the fundamental
	domain of \SL{2,\Z{}} in figure~\ref{fig:modulispace}.
	\label{fig:majoranaphases}}
\end{figure}

\begin{figure}[t]
	\centering
	\includegraphics[width=0.96\linewidth]{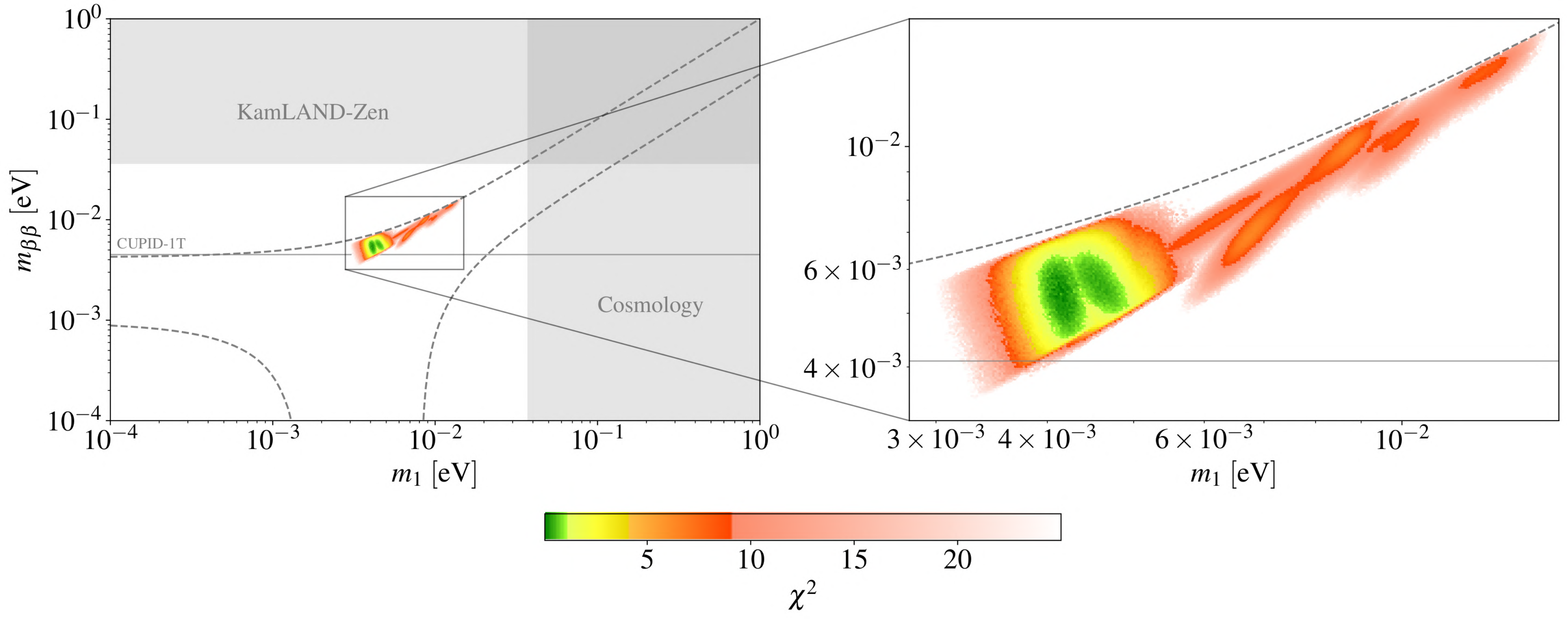}
	\caption{Effective neutrino mass for $0\nu\beta\beta$ as a function of 
	the lightest neutrino mass. The dashed lines delimit the experimentally 
	admissible region within $3\sigma$ for normal ordering. Gray-shaded areas 
	are excluded by KamLAND-Zen~\cite{KamLAND-Zen:2022tow} or cosmological 
	bounds~\cite{GAMBITCosmologyWorkgroup:2020rmf,Planck:2018vyg}. 
	The $1\sigma$ and $2\sigma$ (green and yellow) regions are within the 
	bounds that next-to-next generation experiments are aiming at. For example, 
	the stated preliminary exclusion sensitivity of the CUPID-1T experiment 
	goes down to $4.1~\mathrm{meV}$~\cite{CUPID:2022wpt}, which is indicated 
	in the plot by a thin gray line. As for Majorana phases, 
	cf.\ figure~\ref{fig:majoranaphases}, the appearance of two
	$1\sigma$ (green) regions in this plot is related to the bimodality in 
	moduli space, cf.\ figure~\ref{fig:modulispace}.
	\label{fig:m1mbb}}
\end{figure}

For observables whose values have not yet been determined by experiment, our model has the following 
predictions:
\begin{itemize}
	\item As shown in figure~\ref{fig:s23dcp}, in our model $\theta_{23}$ is preferably
	found in the first octant, i.e.\ $\theta_{23}<45^\circ$. Taking the atmospheric data 
	provided by Super-Kamiokande into account, this octant is currently also preferred by 
	experiment in the case of normal ordering. Unfortunately, for this octant, the model 
	does not provide a prediction for the \CP violating phase $\delta_{\mathrm{\CP}}^\ell$. 
	
	\item The model has a rather precise prediction for the neutrino masses, especially 
	for the heaviest neutrino mass, cf.\ figure~\ref{fig:neutrinomasses}. At $1\sigma$, 
	the neutrino masses are predicted to be $3.9~\mathrm{meV} < m_1 < 4.9~\mathrm{meV}$, 
	$9.5~\mathrm{meV} < m_2 < 9.9~\mathrm{meV}$, and $50.1~\mathrm{meV} < m_3 < 50.5~\mathrm{meV}$.
	
	\item Only Majorana phases that are close to \CP-conserving values are 
	compatible with the fit of our model. For more details, see figure~\ref{fig:majoranaphases}.
	
	\item The prediction for the effective neutrino mass $m_{\beta\beta}$ is, unfortunately, 
	not reachable by the next-generation experiments for neutrinoless double beta decay. 
	However, potential next-to-next generation experiments, e.g.\ CUPID-1T~\cite{CUPID:2022wpt}, 
	aim at covering the predicted region, see figure~\ref{fig:m1mbb}.
	
	\item We have performed a wide numerical scan and did not find any successful fit that
	accepts inverted ordered neutrino masses. Hence, we observe that our model clearly prefers
	normal ordering.
\end{itemize} 

\section{Simultaneous fit of quark and lepton sectors}
\label{sec:NumericalAnalysisBoth}

So far, we have discussed only the lepton sector. It has been fitted by
choosing appropriate VEVs for the modulus $T$ and the flavon triplets $\varphi_\mathrm{e}$
and $\varphi_\nu$. Let us now include in our analysis the masses and mixings of quarks.
Inspecting the superpotential~\eqref{eq:superpot}, we realize that up-type quark Yukawa 
couplings include an additional flavon triplet $\varphi_\mathrm{u}$ while down-type 
Yukawas share the flavon triplet $\varphi_\mathrm{e}$. Consequently, at leading order 
i) the structure of the mass matrices of up and down-type quarks are equal, and ii) the masses
of charged leptons and down-type quarks differ only by their overall scale. 
The latter contradicts experimental observations, but it can be amended
by taking into account contributions from the K\"ahler potential. As discussed 
in section~\ref{sec:Kaehler}, if flavons develop VEVs, there can be considerable
off-diagonal corrections to the K\"ahler metric already at next-to-leading order.

In principle, K\"ahler corrections can affect both leptons and quarks.
However, for simplicity, we assume that the parameters in the lepton sector 
yield negligible contributions to additional terms in the K\"ahler potential. That is, 
only the quark sector will be influenced by K\"ahler corrections. According to our
previous discussion in section~\ref{sec:Kaehler}, the next-to-leading order
corrections to the K\"ahler metric of quark fields $\Psi\in\{\bar{u},\bar{d},q\}$
take the form (see eq.~\eqref{eq:KahlerMetricParameterized})
\begin{equation}
K_{ij}^{f} \supset  
\lambda_{\varphi_\mathrm{eff}}^{f}\, \left(A_{ij}^{f}
+ \kappa_{\varphi_\mathrm{eff}}^{f}\, B_{ij}^{f}
\right) \;,
\label{eq:KahlerCorrectionsQuark}
\end{equation}
where $f\in\{\mathrm{u,d,q}\}$ labels the effective flavons and K\"ahler parameters associated 
with each quark field, explicitly defined in eqs.~\eqref{eq:KahlerMetricParameterized}--\eqref{eqs:effectiveParams}.
To simplify our notation, we have suppressed the arguments of the K\"ahler matrix elements, such that
\begin{equation}
A_{ij}^{f} := A_{ij}(\tilde{\varphi}_\mathrm{eff}^{(A),f})\qquad\text{and}\qquad
B_{ij}^{f} := B_{ij}(\tilde{\varphi}_\mathrm{eff}^{(B),f})\;.
\end{equation}
These matrix elements are quadratic in the VEVs of the components of the effective
flavon triplets. However, since these VEVs appear in the K\"ahler metric always accompanied by
the coefficients $\lambda_{\varphi_{\mathrm{eff}}}^{f}$, it is convenient to 
use instead the parameters
\begin{equation}
\label{eq:alpha-betaParameters}
\alpha_i^f := \sqrt{\lambda_{\varphi_{\mathrm{eff}}}^{f}} \, \vev{\tilde{\varphi}_{\mathrm{eff},i}^{(A),f}}
\qquad\text{and}\qquad
\beta_i^f := \sqrt{\lambda_{\varphi_{\mathrm{eff}}}^{f}} \, \vev{\tilde{\varphi}_{\mathrm{eff},i}^{(B),f}}\;,
\end{equation}
such that 
\begin{equation}
\lambda_{\varphi_{\mathrm{eff}}}^{f} A_{ij}^{f} = \alpha_i^{f}\,\alpha_j^{f}\;,
\end{equation}
and $\lambda_{\varphi_{\mathrm{eff}}}^{f} B_{ij}^{f}$ is quadratic in $\beta^{f}$ up to $\mathcal{O}(1)$ 
factors. Note that the parameters $\alpha_i^{f}$ and $\beta_i^{f}$ represent a good measure of the size 
of the K\"ahler corrections.

\begin{table}[!t!]
	\vspace*{10pt}
	\begin{minipage}{0.25\textwidth}%
		\centering
		\subfloat[(a)][]{
			\label{tab:SimultaneousFitParameters}
			\centering 
			\scalebox{0.8}{
				\begin{tabular}{cr|l}
					\toprule
					\multicolumn{2}{r|}{parameter} & best-fit value \\\midrule
					&$\im\, \vev T$ & $\phantom{-}3.195$ \\
					&$\re\,\vev T$ & $\phantom{-}0.02279$ \\
					&$\vev{\tilde{\varphi}_{\mathrm{u},1}}$ & $\phantom{-}2.0332\cdot10^{-4}$ \\
					&$\vev{\vartheta_{\mathrm{u},1}}$ & $\phantom{-}1.6481$ \\
					&$\vev{\tilde{\varphi}_{\mathrm{u},2}}$ & $\phantom{-}6.3011\cdot10^{-2}$ \\
					&$\vev{\vartheta_{\mathrm{u},2}}$ & $-1.5983$ \\
					&$\vev{\tilde{\varphi}_{\mathrm{e},1}}$ & $-4.069\cdot 10^{-5}$ \\
					&$\vev{\tilde{\varphi}_{\mathrm{e},2}}$ & $\phantom{-}5.833\cdot10^{-2}$ \\
					&$\vev{\tilde{\varphi}_{\nu,1}}$ & $\phantom{-}1.224\cdot10^{-3}$ \\
					&$\vev{\tilde{\varphi}_{\nu,2}}$ & $-0.9857$ \\
					\multirow{-11}{10pt}{\rotatebox{90}{superpotential}}&$\Lambda_\nu~[\mathrm{eV}]$   & $\phantom{-}0.05629$\\\midrule
					&$\alpha_1^{\mathrm{u}}$ & $-0.94917$ \\
					&$\alpha_2^{\mathrm{u}}$ & $\phantom{-}0.0016906$ \\
					&$\alpha_3^{\mathrm{u}}$ & $\phantom{-}0.31472$ \\
					&$\alpha_1^{\mathrm{d}}$ & $\phantom{-}0.95067$ \\
					&$\alpha_2^{\mathrm{d}}$ & $\phantom{-}0.0077533$ \\
					&$\alpha_3^{\mathrm{d}}$ & $\phantom{-}0.30283$ \\
					&$\alpha_1^{\mathrm{q}}$ & $-0.96952$ \\
					&$\alpha_2^{\mathrm{q}}$ & $-0.20501$ \\
					\multirow{-9}{10pt}{\rotatebox{90}{K\"ahler potential}}&$\alpha_3^{\mathrm{q}}$ & $\phantom{-}0.041643$ \\
					\bottomrule
			\end{tabular}}
		}
		\vspace*{44mm}
	\end{minipage}%
	\hspace*{28pt}
	\begin{minipage}{0.7\textwidth}%
		\centering
		\subfloat[(b)][]{
			\label{tab:SimultaneousFitObservables}
			\centering 
			\scalebox{0.8}{
				\begin{tabular}{cr|llc}
					\toprule
					\multicolumn{2}{r|}{observable} & model best fit & exp. best fit & exp. $1\sigma$ interval \\\midrule
					&$m_\mathrm{u}/m_\mathrm{c}$ & $\phantom{-}0.00193$ & $\phantom{-}0.00193$ & $0.00133 \rightarrow 0.00253$ \\
					&$m_\mathrm{c}/m_\mathrm{t}$ & $\phantom{-}0.00280$ & $\phantom{-}0.00282$ & $0.00270 \rightarrow 0.00294$\\
					&$m_\mathrm{d}/m_\mathrm{s}$ & $\phantom{-}0.0505$ & $\phantom{-}0.0505$ & $0.0443 \rightarrow 0.0567$\\
					&$m_\mathrm{s}/m_\mathrm{b}$ & $\phantom{-}0.0182$ & $\phantom{-}0.0182$ & $0.0172 \rightarrow 0.0192$ \\\cmidrule{2-5}
					&$\vartheta_{12}~[\mathrm{deg}]$ & $\,\;\!13.03$ & $\,\;\!13.03$ & $12.98 \rightarrow 13.07$\\
					&$\vartheta_{13}~[\mathrm{deg}]$ & $\phantom{-}0.200$ & $\phantom{-}0.200$ & $0.193 \rightarrow 0.207$ \\
					&$\vartheta_{23}~[\mathrm{deg}]$ & $\phantom{-}2.30$ & $\phantom{-}2.30$ & $2.26 \rightarrow 2.34$\\
					\multirow{-8}{10pt}{\rotatebox{90}{quark sector}}	&$\delta_\CP^\mathrm{q}~[\mathrm{deg}]$ & $\,\;\!69.2$ & $\,\;\!69.2$ & $66.1 \rightarrow 72.3$\\\midrule		
					&$m_\mathrm{e}/m_\mu$ & $\phantom{-}0.00473$ & $\phantom{-}0.00474$ & $0.00470\rightarrow0.00478$ \\
					&$m_\mu/m_\tau$ & $\phantom{-}0.0586$ & $\phantom{-}0.0586$ & $0.0581\rightarrow0.0590$ \\\cmidrule{2-5}
					&$\sin^2\theta_{12}$ & $\phantom{-}0.303$ & $\phantom{-}0.304$ & $0.292\rightarrow0.316$ \\
					&$\sin^2\theta_{13}$ & $\phantom{-}0.0225$ & $\phantom{-}0.0225$ & $0.0218\rightarrow0.0231$ \\
					&$\sin^2\theta_{23}$ & $\phantom{-}0.449$ & $\phantom{-}0.450$ & $0.434 \rightarrow 0.469$ \\\cmidrule{2-5}
					&$\delta_\CP^\ell/\pi$ & $\phantom{-}1.28$ & $\phantom{-}1.28$ & $1.14 \rightarrow 1.48$ \\
					&$\eta_1/\pi$ & $\phantom{-}0.029$ & ~~~\,~- & - \\
					&$\eta_2/\pi$ & $\phantom{-}0.994$ & ~~~\,~- & - \\
					&$J_\CP$ & $-0.026$ & $-0.026$ & $-0.033\rightarrow-0.016$ \\
					&$J_\CP^\mathrm{max}$ & $\phantom{-}0.0335$ & $\phantom{-}0.0336$ & $0.0329\rightarrow0.0341$ \\\cmidrule{2-5}
					&$\Delta m_{21}^2/10^{-5}~[\mathrm{eV}^2]$ & $\phantom{-}7.39$ & $\phantom{-}7.42$ & $ 7.22\rightarrow 7.63$ \\
					&$\Delta m_{31}^2/10^{-3}~[\mathrm{eV}^2]$ & $\phantom{-}2.521$ & $\phantom{-}2.510$ & $ 2.483\rightarrow 2.537$ \\
					&$m_1~[\mathrm{eV}]$ & $\phantom{-}0.0042$ & $\!<\!0.037$ & - \\
					&$m_2~[\mathrm{eV}]$ & $\phantom{-}0.0095$ & ~~~\,~- & -  \\
					&$m_3~[\mathrm{eV}]$ & $\phantom{-}0.0504$ & ~~~\,~- & - \\
					&$\sum_i m_i~[\mathrm{eV}]$ & $\phantom{-}0.0641$ & $\!<\!0.120$ & -  \\
					&$m_{\beta\beta}~[\mathrm{eV}]$ & $\phantom{-}0.0055$ & $\!<\!0.036$ & - \\
					\multirow{-18}{10pt}{\rotatebox{90}{lepton sector}}&$m_{\beta}~[\mathrm{eV}]$ & $\phantom{-}0.0099$ & $\!<\!0.8$ & - \\\midrule
					& $\chi^2$ & $\phantom{-}0.11$ & & \\
					\bottomrule	
				\end{tabular}
			}
		}
	\end{minipage}%
	\caption{Results of a simultaneous fit of the quark and lepton sectors with $\chi^2=0.11$.
		(a) Values of the model parameters at the best-fit point. The parameter values in the lepton sector 
		coincide with the modulus and flavon VEVs showed in table~\ref{tab:LeptonFitParameters}. In addition, 
		for the quark sector we provide the (complex) components of the flavon VEV $\vev{\tilde\varphi_\mathrm{u}}$ 
		appearing in the superpotential, along with the effective K\"ahler parameters $\alpha_i^f$, $f\in\{\mathrm{u,d,q}\}$ 
		and $i\in\{1,2,3\}$, defined in eq.~\eqref{eq:alpha-betaParameters}. 
		(b) Best-fit values of flavor observables obtained from our model. We compare them with the 
		corresponding experimental best-fit value; we include the experimental $1\sigma$ error.
		The quark-sector observables are successfully fitted while keeping untouched the lepton-sector 
		fit presented in table~\ref{tab:FitLeptons}.
		\label{tab:SimultaneousFit}}
\end{table}

The additional parameters of the quark sector include first the complex components of the normalized 
up-type flavon triplet
\begin{equation}
\label{eq:uflavon}
   \vev{\tilde\varphi_\mathrm{u}} ~=~ \Big(
   \vev{\tilde\varphi_{\mathrm{u},1}} \,\exp\;\!\!\left(\I\vev{\vartheta_{\mathrm{u},1}}\right), 
   \vev{\tilde\varphi_{\mathrm{u},2}} \,\exp\;\!\!\left(\I\vev{\vartheta_{\mathrm{u},2}}\right), 
   1\Big)\;.
\end{equation}
Furthermore, the K\"ahler corrections introduce 9 parameters $\alpha_i^f$, 9 $\beta_i^f$ and 3 
$\kappa_{\varphi_\mathrm{eff}}^{f}$. 
In order to simplify somewhat our fit, we impose the following constraints:
\begin{itemize}
  \item $\kappa_{\varphi_\mathrm{eff}}^{f}=1$ for all $f\in\{\mathrm{u,d,q}\}$,
  \item $\alpha_i^{f} = \beta_i^{f}$ for all $f$ and $i\in\{1,2,3\}$, and
  \item all $\alpha_i^{f}$ are real.
\end{itemize}
While these constraints may appear ad-hoc, we stress that the philosophy here is not to
scan the \textit{full} parameter space but to demonstrate, in the first place, that there is a region in the parameter space that indeed agrees with a realistic low energy phenomenology.
Taking the constraints into account, we arrive at a remaining set of 13 quark parameters that we include in our numerical scan, 
aiming at a global fit of both leptons and quarks. 
The numerical procedure to achieve the global fit is based on a $\chi^2$ minimization, analogous to the one used in
the lepton sector, which is discussed in detail in appendix~\ref{app:FitDetails}. 
As for charged leptons, the experimental data we consider for quarks are the mass ratios and mixing 
parameters at the GUT scale~\cite{Antusch:2013jca}, assuming a running with $\tan\beta=10$, $M_\mathrm{SUSY}=10$\,TeV, 
and $\bar\eta_b=0.09375$, as in refs.~\cite{Chen:2021zty,Ding:2021zbg}.
These experimental best-fit values together with their respective errors are presented in the 
last two columns of table~\ref{tab:SimultaneousFitObservables}.

The resulting best-fit values are displayed in table~\ref{tab:SimultaneousFit}. The modulus and 
flavon VEVs of the model have the values shown in table~\ref{tab:SimultaneousFitParameters}.
We point out that the magnitude of the K\"ahler corrections needed 
to arrive at a successful global fit all satisfy $\alpha^f_i<1$.
Also, the VEVs of the modulus \vev{T} and the lepton flavons $\vev{\tilde\varphi_{\mathrm{e},i}}$ and 
$\vev{\tilde\varphi_{\nu,i}}$ preserve the values obtained in the lepton fit, cf.\ table~\ref{tab:FitLeptons}.
In table~\ref{tab:SimultaneousFitObservables} we compare our best fit against the experimental values 
of quark and lepton observables. Our global fit of all fermion mass ratios, mixing angles and 
\CP phases exhibits $\chi^2=0.11$. 
Although we do not provide any prediction in the quark sector, it is remarkable that 
the eclectic scenario arising from a string compactification can fit the observed data so well.

Before concluding, let us mention some caveats of our model. First, the VEV parameters of 
the model included in eqs.~\eqref{eq:paramsLepton} and~\eqref{eq:uflavon} as well as the K\"ahler 
parameters of eqs.~\eqref{eq:alpha-betaParameters} have been considered here to be free. However, 
in a full string model the computation of the couplings and the dynamic stabilization of the VEVs
are in principle achievable. Unfortunately, these tasks have not been solved so far, remaining
as open questions for our model. Secondly, notice that the 
values of the K\"ahler parameters in our fit, displayed in table~\ref{tab:SimultaneousFitParameters},
are all controllable in the sense that they arise in a K\"ahler potential that is explicitly 
constrained by the eclectic flavor group and, moreover, their magnitudes turn out to be smaller than 
unity ensuring the perturbativity of our model. Yet, because of its complexity, the rigorous string 
computation of these parameters lies still beyond the scope of our study. Finally, our focus is the 
flavor puzzle only, assuming that all other phenomenological questions of particle physics and cosmology 
can be solved by some methods introduced in many earlier influential works. For example, we have assumed 
that all exotic matter states appearing in table~\ref{tab:exotics-model447} can acquire masses much 
larger than the physical scale of the flavor sector in supersymmetric 
vacua~\cite{Buchmuller:2005jr,Buchmuller:2006ik,Lebedev:2007hv,Lebedev:2008un}. One might then 
argue that only the physical right-handed neutrinos and Higgs doublets are left massless
as a result of the existence of some unbroken ($R$-)symmetries either beyond the flavor 
sector~\cite{Kappl:2008ie,Brummer:2010fr,Kappl:2010yu} or intimately linked with it~\cite{Chen:2013dpa}. 
As shown in those works, such symmetries could also be relevant for proton stability and the 
suppression of the $\mu$-term. In addition, relaxing our assumption on the decoupling
of the extra right-handed neutrinos in table~\ref{tab:exotics-model447} might be instrumental
to arrive at a better understanding of the relation between the Majorana and the 
observable neutrino mass scales~\cite{Buchmuller:2007zd}.
Our scheme also admits proposals to solve the discrepancy between the GUT and string scale 
in heterotic models~\cite{Witten:1996mz,Buchmuller:2005sh} since it can be embedded in anisotropic 
compactifications. Furthermore, heterotic orbifolds seem to be equipped with useful properties
to achieve supersymmetry breakdown~\cite{Lebedev:2006tr}. All these aspects should be studied 
elsewhere in detail to complete our construction and extend it to other relevant phenomenological 
questions, such as identifying the cause of inflation, the origin of dark matter and the baryon asymmetry 
of the Universe.

\section{Conclusions and Outlook}
\label{sec:conclusions}

We have studied the flavor phenomenology of the lepton and quark sectors emerging 
from a specific $\mathbbm T^6/\Z3\x\Z3$ heterotic orbifold model that gives rise 
to the eclectic flavor group $\Omega(2)$. This TD scenario combines the virtues of 
a modular $T'$ and a traditional $\Delta(54)$ flavor symmetry, while avoiding the 
arbitrariness in the choice of quantum numbers of matter fields inherent to BU constructions.
The (traditional, modular and gauge symmetry) representations of matter fields as 
well as their modular weights are entirely fixed by the compactification. 
In our example model, SM fermions and flavons form identical flavor triplets and 
exhibit equal (fractional) modular weights, cf.\ table~\ref{tab:MSSM-model447}.
Hence, the structure of the superpotential and K\"ahler potential are determined 
by the theory, guaranteeing, in particular, a canonical leading-order K\"ahler 
potential, as is most frequently assumed in the BU approach. 
However, in addition, our setup also allows us to control non-canonical, 
higher-order, Planck-suppressed corrections to the K\"ahler potential that arise 
after the traditional flavor symmetry is spontaneously broken by flavons.
We computed these corrections (to next-to-leading order), which turn out to 
be instrumental for a successful phenomenological fit since they contribute to 
the structure of mass matrices.
Both, the modulus and some of the flavons inherent to the construction must attain 
non-trivial VEVs in order to break the modular and traditional components of 
the eclectic symmetry, as required by phenomenology. Special values of these VEVs 
lead to discrete remnants of the flavor group that can appear as approximate discrete 
symmetries at low energies~\cite{Baur:2021bly}. 

In our string-derived example model, we have explicitly computed the leading-order
superpotential~\eqref{eq:superpot} and confirmed the canonical leading-order 
structure of the K\"ahler potential~\eqref{eq:KahlerFirstOrder}. These results reveal 
that our model accommodates naturally a type-I see-saw mechanism as explanation for 
the neutrino masses. We have shown that points in moduli space perturbatively close 
to the symmetry-enhanced point $\vev T=\I\infty$ enjoy various approximate symmetries as 
remnants of the eclectic group. Their successive spontaneous breaking through the 
misaligned VEVs of the modulus and flavon fields can account for technically natural 
(symmetry-protected) correct hierarchies. The tight, symmetry-based constraints allow 
us to derive approximate analytical expressions for the mass hierarchies, as explained 
in section~\ref{sec:StepwiseBreaking}.

In order to fully explore the phenomenology of the model,
we have performed a numerical analysis of the charged-lepton and neutrino sectors. 
We found that the 11 independent observables listed in table~\ref{tab:FitLeptons} 
can be well fitted by adjusting seven free parameters 
corresponding to the VEVs of the modulus and flavons as well as the neutrino mass scale. 
Their values at the best-fit point are presented in table~\ref{tab:LeptonFitParameters}
and show that our analytical treatment is fairly accurate. The octant of
$\theta_{23}$, the normal ordering of neutrino masses, the observable 
values of $m_{\beta\beta}$, as well as the neutrino Majorana phases 
are predictions of the fit. 
These results are illustrated in figures~\ref{fig:s23dcp}--\ref{fig:m1mbb}.

Next-to-leading-order K\"ahler corrections turn out to be crucial to arrive
at a model of flavor that includes the quark sector in a phenomenologically viable 
manner.
This is another consequence of the highly constrained nature of TD constructions,
as our example model contains only a single non-singlet flavon field that is 
responsible for the structure and hierarchies of down-quark and charged-lepton 
Yukawa couplings, as well as of the neutrino Majorana mass term. This results 
in a particular kind of bottom-tau unification that must be modified in order 
to arrive at a realistic phenomenology. We have shown that this can be achieved 
thanks to the presence of next-to-leading-order K\"ahler corrections, which allowed us to obtain
a successful numerical fit to quark phenomenology that does not
change our predictions for the lepton sector.

In summary, we have presented for the first time a UV-complete, full string
theory model that exhibits a flavor scheme that can accomodate 
the experimentally observed pattern of quark and lepton flavor phenomenology. 
Reducing the number of free parameters was possible by taking into account the restrictive 
constraints on the effective superpotential and K\"ahler potential arising from the entire, 
partly non-linearly realized, eclectic flavor symmetry. Achieving the ambitious goal of a 
complete fit to the low-energy flavor data was
possible only as a consequence of the existence of controllable K\"ahler corrections.

This represents the first decisive step towards connecting the BU and TD efforts in the quest 
for an ultimate theory of flavor, and demonstrates the potential of this TD approach.
It would be interesting to compare our results to the outcome of similar TD
constructions, such as the orbifold models of type B--D classified in ref.~\cite{Baur:2021bly},
orbifold constructions endowed with a $\mathbbm T^2/\Z2$ sector~\cite{Baur:2020jwc,Baur:2021mtl},
or other TD scenarios that can admit three fermion generations and metaplectic flavor
symmetries~\cite{Almumin:2021fbk}, and also exhibit eclectic features~\cite{Ohki:2020bpo}.
Moreover, quasi--eclectic models~\cite{Chen:2021prl} offer another interesting possibility 
to explore in order to further connect the BU and TD approaches.

Future efforts should aim at further reducing the number of free parameters, 
either by rigorous string computations of some of the low energy parameters, 
or by identifying other potentially realistic string setups that are even more 
constrained by symmetry. Further attention should also be paid to the 
field-theoretical minimization of the flavon potential as well as to 
the longstanding question of modulus stabilization.

\section*{Acknowledgments}
It is a pleasure to thank Mu-Chun Chen, Ferruccio Feruglio, Steve King, Hajime Otsuka, 
Jo\~ao Penedo, Serguey Petcov, Michael Ratz, and Arsenii Titov for insightful discussions.
We are grateful to Miguel Levy and Jim Talbert for identifying an important typo in appendix~\ref{app:spectrum}.
A.B., S.R-S., and A.T.\ are grateful to the Bethe Center for Theoretical Physics (BCTP), 
Bonn, for hospitality and support during the Bethe Forum ``Modular Flavor Symmetries.'' 
A.B.\ is supported by the Deutsche Forschungsgemeinschaft (SFB1258).
A.T.\ is grateful to the Mainz Institute for Theoretical Physics (MITP) of the Cluster 
of Excellence PRISMA+ (Project ID 39083149), for its hospitality and partial support 
during the completion of this work.

\pagebreak
\appendix

\section{K\"ahler potential at next-to-leading order}
\label{app:Kahler}

In order to arrive at the next-to-leading order K\"ahler potential,
eq.~\eqref{eq:KahlerMetricParameterized}, one must compute the tensor products
of the relevant representations (by using e.g.\ ref.~\cite{Ishimori:2010au}).
Here we discuss in detail the results of the computation.

The first tensor product in eq.~\eqref{eq:KahlerSecondOrder} is given by
\begin{equation}
T_{1,a} ~=~ \left[\varphi^* \otimes \varphi \otimes \Psi^* \otimes \Psi\right]_{\rep{1},a}  \;.
\end{equation}
This product has two invariant singlet contractions, i.e.\ $a\in\{1,2\}$. For $a=1$ it reads
\begin{equation}
T_{1,a=1} ~=~
\Psi^\dagger
\begin{pmatrix}
|\varphi_1|^2 & \varphi_1\,\varphi_2^* & \varphi_1\,\varphi_3^* \\
\varphi_2\,\varphi_1^* & |\varphi_2|^2 & \varphi_2\,\varphi_3^* \\
\varphi_3\,\varphi_1^* & \varphi_3\,\varphi_2^* & |\varphi_3|^2
\end{pmatrix}
\Psi \;.
\end{equation}
Here $\varphi_i$ corresponds to the $i$-th component of the flavor triplet $\varphi$, or equivalently
\begin{equation}
\label{eq:firstcontraction}
T_{1,a=1} ~=~
\Psi^*_i ~ A_{ij}(\varphi) ~ \Psi_j\;,
\end{equation}
where the components of the matrix $A$ are given by
\begin{equation} \label{eq:Aij}
A_{ij}(\varphi) ~:=~ \varphi_i\,\varphi_j^*\;,
\end{equation}
and summation over repeated indices is implied.
The second invariant singlet contraction, i.e.\ $T_{1,a=2} = \Psi^*_j~\varphi_i\,\varphi^*_i~\Psi_j$, 
is irrelevant because it is proportional to the identity matrix and hence its contribution to 
the K\"ahler metric can be absorbed by the symmetry-invariant constant $\chi$ of the leading-order 
K\"ahler potential~\eqref{eq:KmetricIdterm}. Thus, we shall not discuss it here.

The second tensor product in the next-to-leading order K\"ahler potential is given by
\begin{equation}
T_{2,a} ~=~ \left[\left(\hat{Y}^{(1)}(T)\right)^* \otimes \hat{Y}^{(1)}(T) \otimes \varphi^* \otimes \varphi \otimes \Psi^* \otimes \Psi\right]_{\rep{1},a}\;,
\qquad a\in\{1,2,3\}\;.
\end{equation}
This tensor product yields three linearly-independent invariant terms,
but only two of them cannot be absorbed in~\eqref{eq:KmetricIdterm}.
The first nontrivial term reads
\begin{equation}
T_{2,a=1} ~=~ \Psi^*_i ~ |\hat{Y}^{(1)}(T)|^2 \, A_{ij}(\varphi) ~ \Psi_j\;.
\end{equation}
Note that this term, apart from the overall factor of $|\hat{Y}^{(1)}(T)|^2$, structurally 
yields the same K\"ahler metric as the first tensor product \eqref{eq:firstcontraction}.
The second invariant singlet contraction reads
\begin{equation}
\label{eq:T2a=2}
T_{2,a=2} ~=~ \Psi^*_i ~  \left( B_{ij}(\varphi) +  |\hat{Y}_2|^2 |\varphi|^2 \delta_{ij} \right) \Psi_j\;,
\end{equation}
where
\begin{equation} \label{eq:Bij}
B_{ij}(\varphi) ~=~ \begin{cases}
\left( |\hat{Y}_1|^2  - 2\,|\hat{Y}_2|^2 \right) \,\varphi_i\,\varphi_j^*\;, & \text{for\ } i = j\\
-|\hat{Y}_1|^2 \,\varphi_i\,\varphi_j^* + \sqrt{2} \left( \hat{Y}_1\,\hat{Y}_2^* \,\varphi_i^*\,\varphi_k + \hat{Y}_2\,\hat{Y}_1^*\,\varphi_k^*\,\varphi_j \right) \;, & 
\text{for\ }k \neq i \neq j \neq k\;. 
\end{cases}
\end{equation} 
As before, the term proportional to $\delta_{ij}$ in~\eqref{eq:T2a=2} can be absorbed
in~\eqref{eq:KmetricIdterm} and will thus be ignored.

Using eqs.~\eqref{eq:firstcontraction} and~\eqref{eq:T2a=2}, we find that the next-to-leading
order contributions to the K\"ahler metric~\eqref{eq:KahlerSecondOrder} 
that are not proportional to $\delta_{ij}$, are given by
\begin{align}
\label{eq:KahlerMetric2}
K_{ij}^{(\mathrm{non-id})} ~\supset&~ \sum_\varphi \left[
\left(\left(-\I T+\I\bar T\right)^{\nicefrac{-4}{3}} \, \zeta_1^{(\varphi)} + \left(-\I T+\I\bar T\right)^{\nicefrac{-1}{3}} \, \zeta_1^{(Y\varphi)} |\hat{Y}^{(1)}(T)|^2 \right) \, A_{ij}(\varphi) \right. \\
&\qquad~+
\left.\left(-\I T+\I\bar T\right)^{\nicefrac{-1}{3}} \, \zeta_2^{(Y\varphi)} ~ B_{ij}(\varphi)\right]\;,\nonumber
\end{align}
where we sum over all flavon triplets $\varphi$ (with all possible modular weights)
that develop VEVs. We stress that the noncanonical contributions~\eqref{eq:KahlerMetric2} 
arise only as a result of the breaking of the traditional flavor symmetry by flavon VEVs, and that 
they are clearly Planck suppressed.

\section{Numerical procedure}
\label{app:FitDetails}

Let us describe here in detail the numerical procedure we follow to arrive at
the fit of the lepton sector.
The goal of the numerical procedure is to explore the parameter space of 
the model parameters $x$ defined in eq.~\eqref{eq:paramsLepton}
in order to find the regions that yield values of lepton masses and mixings that
are in agreement with experimental observations. In detail, we search for parameters
that yield $\chi^2\leq 25$ corresponding to a compatibility with $5\sigma$. 
Moreover, we also want to identify the point in parameter space that yields 
the best match to the experimental data. We therefore split the numerical 
analysis in two steps: i) First, we find all minima with $\chi^2\leq 25$; and
ii) then we explore the regions around these minima.

The first step is a typical optimization problem that can be conveniently 
approached by using the non-linear optimization interface \texttt{lmfit}~\cite{lmfit}. 
We start by picking a random start-point in the parameter space, whose boundaries we set to
\begin{subequations}
	\begin{align}
	&0 ~<~ \left|\vev{\tilde\varphi_{\mathrm{e},1}}\right|,~ \left|\vev{\tilde\varphi_{\mathrm{e},2}}\right| ~<~1\;, &&\quad
	0 ~<~ \left|\vev{\tilde\varphi_{\nu,1}}\right|,~ \left|\vev{\tilde\varphi_{\nu,2}}\right| ~<~2\;,  \\
	&0 ~<~ \left|\re\,\vev T\right| ~<~ 1.5\;, &&\quad 0 ~<~ \im\,\vev T ~<~ 5\;.
	\end{align}
\end{subequations}
As we expect the flavon VEVs to be hierarchically ordered, we sample them with 
a blend of a uniform and a logarithmic distribution.
Moreover, we use the analytical result 
$|\vev{\tilde{\varphi}_\mathrm{e,2}}| ~\approx~ \frac{m_\mu}{m_\tau}~=~ 0.0586$ 
obtained in section~\ref{sec:AnalyticalLeptons} and sample 
$|\vev{\tilde{\varphi}_\mathrm{e,2}}|$ only in the vicinity of this value.
To the chosen start-point, we then consecutively apply five randomly chosen
minimization algorithms included in the \texttt{lmfit} interface. 
For our setup, especially the algorithms `Constrained trust-region' and `L-BFGS-B' 
deliver good results. We repeat this procedure 
until roughly 1000 points with $\chi^2 \leq 25$ and no new 
minima are found by the algorithms.

Finally, we explore the neighborhood of each minimum using the Markov Chain Monte 
Carlo (MCMC) sampler \texttt{emcee}~\cite{emcee}, which is also supported by \texttt{lmfit}.
The MCMC sampler chooses random points with a probability function that it tries 
to couple to $\chi^2$. They are therefore well suited to provide information on the 
vicinity of the minima and hence the boundaries of the respective confidence levels.

Although similar methods have been thoroughly explained in other works, see 
e.g.~\cite{Novichkov:2018ovf}, we make our \texttt{python} code available upon request
to be applied both in BU and TD constructions. Please, send your inquiries preferably 
to \href{mailto:alexander.baur@tum.de}{\tt alexander.baur@tum.de}.

\begin{landscape}

\section[Complete spectrum of a model with Omega(2) eclectic flavor symmetry]{\boldmath Complete spectrum of a model with $\Omega(2)$ eclectic flavor symmetry\unboldmath}
\label{app:spectrum}

We provide all quantum numbers of the massless spectrum of our example $\mathbbm T^6/\Z3\x\Z3$ heterotic orbifold model,
including the representations under $G_{SM}=\SU3_c\x\SU2_L\x\U1_Y$, the eclectic flavor group 
$\Omega(2)=\Delta(54)\cup T'\cup\Z9^R$ (along with the associated modular weights $n$),
and the extra $\Z3{}^3$ flavor and `hidden' $\SU4\x\U1_\textrm{anom}\x\U1^8$ gauge factors.
\enlargethispage{\baselineskip}


{\centering{
\footnotesize
\setlength\LTleft{0pt}
\setlength\LTright{0pt}
\begin{longtable}{|l|l|ccc|c|ccc||c|ccccccccc|c|}
\cline{3-19}
\multicolumn{2}{c}{} & \multicolumn{7}{|c||}{Flavor charges} & \multicolumn{10}{c|}{`Hidden' gauge charges} & \multicolumn{1}{c}{}\\
\hline
sector & $G_{SM}^{\phantom{A^{A}}}$ & $\Delta(54)$ & $T'$ & $\Z9^R$ & $n$ & \Z3 & \Z3 & \Z3 & \SU4 & 
   $q_\text{anom}$ & $q_2$ & $q_3$ & $q_4$ & $q_5$ & $q_6$ & $q_7$ & $q_8$ & $q_9$ & labels \\
\hline\hline
\endhead
\hline
\endfoot
$U_1^{\phantom{A^{A^A}}}$ & $(\rep1,\rep2)_{\frac12}$  & $\rep1$ & $\rep1$ & $0$ & $0$ & $1$ & $1$ & $1$ & $\rep1$ &
   $    0$ & $    1$ & $    0$ & $    0$ & $    0$ & $    0$ & $  -59$ & $   32$ & $ -124$ & $H_u$ \\
                          & $(\rep1,\rep2)_{-\frac12}$ & $\rep1$ & $\rep1$ & $0$ & $0$ & $1$ & $1$ & $1$ & $\rep1$ &
   $    0$ & $    1$ & $    0$ & $    0$ & $    0$ & $    0$ & $   59$ & $  -32$ & $  124$ & $H_d$ \\
                          & $(\rep1,\rep1)_{0}$ & $\rep1$ & $\rep1$ & $0$ & $0$ & $1$ & $1$ & $1$ & $\rep1$ &
   $    0$ & $   -2$ & $    0$ & $    0$ & $    0$ & $    0$ & $    0$ & $    0$ & $    0$ & $\phi^0$ \\
                          & $(\rep3,\rep1)_{\frac23}$ & $\rep1$ & $\rep1$ & $0$ & $0$ & $1$ & $1$ & $1$ & $\rep1$ &
   $    0$ &  $    0$ & $    0$ & $    0$ & $    0$ & $  -73$ & $  -90$ & $  -24$ & $   93$ & $U_1$ \\
\hline
$U_2^{\phantom{A^{A}}}$   & $(\rep1,\rep1)_{0}$ & $\rep1$ & $\rep1$ & $0$ & $0$ & $1$ & $1$ & $1$ & $\rep1$ &
   $   -2$ & $   -1$ & $    0$ & $    0$ & $    0$ & $   73$ & $  149$ & $   -8$ & $ -125$ & $\phi^0_\mathrm{M}$ \\
                          & $(\rep1,\rep1)_{0}$ & $\rep1$ & $\rep1$ & $0$ & $0$ & $1$ & $1$ & $1$ & $\rep1$ &
   $   -2$ & $    0$ & $    2$ & $  -38$ & $   32$ & $   42$ & $  -84$ & $  264$ & $   30$ & $s_{1}$ \\                          
                          & $(\rep1,\rep1)_{0}$ & $\rep1$ & $\rep1$ & $0$ & $0$ & $1$ & $1$ & $1$ & $\rep1$ &
   $    0$ & $    0$ & $   -7$ & $   16$ & $  -73$ & $    0$ & $    0$ & $    0$ & $    0$ & $s_{2}$ \\                          
                          & $(\rep1,\rep1)_{0}$ & $\rep1$ & $\rep1$ & $0$ & $0$ & $1$ & $1$ & $1$ & $\rep1$ &
   $    4$ & $   -1$ & $    0$ & $    0$ & $    0$ & $   73$ & $   31$ & $   56$ & $   95$ & $s_{3}$ \\                          
                          & $(\rep1,\rep1)_{0}$ & $\rep1$ & $\rep1$ & $0$ & $0$ & $1$ & $1$ & $1$ & $\rep1$ &
   $    2$ & $    0$ & $    5$ & $   22$ & $   41$ & $  -42$ & $   84$ & $ -264$ & $  -30$ & $s_{5}$ \\
                          & $(\rep3,\rep1)_{\frac23}$ & $\rep1$ & $\rep1$ & $0$ & $0$ & $1$ & $1$ & $1$ & $\rep1$ &
   $   -2$ & $    1$ & $    0$ & $    0$ & $    0$ & $    0$ & $   59$ & $  -32$ & $  -32$ & $U_2$ \\
                          & $(\crep3,\rep1)_{-\frac23}$ & $\rep1$ & $\rep1$ & $0$ & $0$ & $1$ & $1$ & $1$ & $\rep1$ &
   $   -2$ & $    0$ & $    0$ & $    0$ & $    0$ & $  -73$ & $  -90$ & $  -24$ & $  -63$ & $\bar{U}_{1}$ \\
\hline
$U_3^{\phantom{A^{A}}}$   & $(\rep1,\rep1)_{0}$ & $\rep1'$ & $\rep1$ & $3$ & $-1$ & $1$ & $1$ & $1$ & $\rep1$ &
   $    2$ & $    0$ & $    0$ & $    0$ & $    0$ & $  146$ & $  180$ & $   48$ & $  -30$ & $s_7$ \\
                          & $(\crep3,\rep1)_{-\frac23}$ & $\rep1'$ & $\rep1$ & $3$ & $-1$ & $1$ & $1$ & $1$ & $\rep1$ &
   $    2$ & $    1$ & $    0$ & $    0$ & $    0$ & $    0$ & $  -59$ & $   32$ & $   32$ & $\bar{U}_2$ \\
                          & $(\rep1,\rep2)_{-\frac12}$ & $\rep1'$ & $\rep1$ & $3$ & $-1$ & $1$ & $1$ & $1$ & $\rep1$ &
   $   -4$ & $    0$ & $    0$ & $    0$ & $    0$ & $  -73$ & $   28$ & $  -88$ & $   29$ & $L_{1}$ \\
                          & $(\rep1,\rep2)_{\frac12}$  & $\rep1'$ & $\rep1$ & $3$ & $-1$ & $1$ & $1$ & $1$ & $\rep1$ &
   $    2$ & $    0$ & $    0$ & $    0$ & $    0$ & $  -73$ & $ -208$ & $   40$ & $    1$ & $\bar{L}_{1}$ \\
\hline
\hline
$T_{(0,1)}^{\phantom{A^{A}}}$ & $(\rep3,\rep2)_{\frac16}$ & $\rep3_2$ & $\rep2'\oplus\rep1$ & $1$ & $\nicefrac{-2}3$ & $1$ & $\w^2$ & $1$ & $\rep1$ &
   $\tfrac{2}{3}$ & $\tfrac{1}{3}$ & $-\tfrac{2}{3}$ & $-\tfrac{40}{3}$ & $   -2$ & $-\tfrac{17}{3}$ & $\tfrac{329}{3}$ & $\tfrac{40}{3}$ & $\tfrac{1}{3}$ & $(q_1,q_2,q_3)$ \\
                              & $(\crep3,\rep1)_{-\frac23}$ & $\rep3_2$ & $\rep2'\oplus\rep1$ & $1$ & $\nicefrac{-2}3$ & $1$ & $\w^2$ & $1$ & $\rep1$ &
   $\tfrac23$  & $\tfrac13$ & $-\tfrac23$ & $-\tfrac{40}{3}$ & $   -2$ & $-\tfrac{17}{3}$ & $\tfrac{329}{3}$ & $\tfrac{40}{3}$ & $\tfrac13$ & $(\bar{u}_1,\bar{u}_2,\bar{u}_3)$ \\
                              & $(\rep1,\rep{1})_{1}$ & $\rep3_2$ & $\rep2'\oplus\rep1$ & $1$ & $\nicefrac{-2}3$ & $1$ & $\w^2$ & $1$ & $\rep1$ &
   $\tfrac23$ & $\tfrac13$ & $-\tfrac23$ & $-\tfrac{40}{3}$ & $   -2$ & $-\tfrac{17}{3}$ & $\tfrac{329}{3}$ & $\tfrac{40}{3}$ & $\tfrac13$ & $(\bar{e}_1,\bar{e}_2,\bar{e}_3)$ \\ 
                              & $(\rep1,\rep{1})_{0}$ & $\rep3_2$ & $\rep2'\oplus\rep1$ & $1$ & $\nicefrac{-2}3$ & $1$ & $\w^2$ & $1$ & $\rep1$ &
   $\tfrac23$ & $\tfrac13$ & $-\tfrac23$ & $-\tfrac{40}{3}$ & $   -2$ & $-\tfrac{236}{3}$ & $\tfrac{59}{3}$ & $-\tfrac{32}{3}$ & $\tfrac{280}{3}$ & $(\bar\nu_1,\bar\nu_2,\bar\nu_3)$ \\
                              & $(\rep1,\rep{1})_{0}$ & $\rep3_2$ & $\rep2'\oplus\rep1$ & $1$ & $\nicefrac{-2}3$ & $1$ & $\w^2$ & $1$ & $\rep1$ &
   $-\tfrac43$ & $-\tfrac23$ & $-\tfrac23$ & $-\tfrac{40}{3}$ & $   -2$ & $-\tfrac{17}{3}$ & $\tfrac{506}{3}$ & $-\tfrac{56}{3}$ & $-\tfrac{95}{3}$ & $(s_{10},s_{13},s_{16})$ \\
                              & $(\rep1,\rep{1})_{0}$ & $\rep3_2$ & $\rep2'\oplus\rep1$ & $1$ & $\nicefrac{-2}3$ & $1$ & $\w^2$ & $1$ & $\rep1$ &
   $\tfrac23$ & $\tfrac13$ & $\tfrac43$ & $\tfrac{80}{3}$ & $   4$ & $\tfrac{253}{3}$ & $-\tfrac{565}{3}$ & $\tfrac{88}{3}$ & $-\tfrac{185}{3}$ 
   & $(\varphi_{\mathrm{e},1},\varphi_{\mathrm{e},2},\varphi_{\mathrm{e},3})$ \\
                              & $(\rep3,\rep1)_{-\frac13}$ & $\rep3_2$ & $\rep2'\oplus\rep1$ & $1$ & $\nicefrac{-2}3$ & $1$ & $\w^2$ & $1$ & $\rep1$ &
   $\tfrac{2}{3}$ & $-\tfrac{2}{3}$ & $-\tfrac{2}{3}$ & $-\tfrac{40}{3}$ & $   -2$ & $-\tfrac{236}{3}$ & $-\tfrac{118}{3}$ & $\tfrac{64}{3}$ & $-\tfrac{92}{3}$ & $(D_1,D_2,D_3)$ \\
                              & $(\rep1,\rep2)_{\frac12}$  & $\rep3_2$ & $\rep2'\oplus\rep1$ & $1$ & $\nicefrac{-2}3$ & $1$ & $\w^2$ & $1$ & $\rep1$ &
   $\tfrac23$ & $-\tfrac23$ & $-\tfrac23$ & $-\tfrac{40}{3}$ & $   -2$ & $-\tfrac{236}{3}$ & $-\tfrac{118}{3}$ & $\tfrac{64}{3}$ & $-\tfrac{92}{3}$ & $(\bar{L}_2,\bar{L}_3,\bar{L}_4)$ \\
                              & $(\rep1,\rep1)_{-\frac13}$ & $\rep3_2$ & $\rep2'\oplus\rep1$ & $1$ & $\nicefrac{-2}3$ & $1$ & $1$ & $1$ & $\rep1$ & 
   $    0$ & $\tfrac{1}{3}$ & $\tfrac{7}{3}$ & $\tfrac{62}{3}$ & $-\tfrac{11}{3}$ & $\tfrac{127}{3}$ & $   53$ & $ -320$ & $   31$ & $(V_1, V_2, V_3)$ \\
                              & $(\rep1,\rep1)_{\frac13}$ & $\rep3_2$ & $\rep2'\oplus\rep1$ & $1$ & $\nicefrac{-2}3$ & $1$ & $\w$ & $1$ & $\rep1$ & 
   $-\tfrac23$ & $-\tfrac23$ & $\tfrac{10}{3}$ & $\tfrac{44}{3}$ & $\tfrac{140}{3}$ & $-\tfrac{236}{3}$ & $\tfrac{118}{3}$ & $-\tfrac{64}{3}$ & $\tfrac{92}{3}$ & $(\bar{V}_1,\bar{V}_2,\bar{V}_3)$ \\
   
                              & $(\rep1,\rep2)_{-\frac16}$ & $\rep3_2$ & $\rep2'\oplus\rep1$ & $1$ & $\nicefrac{-2}3$ & $1$ & $\w$ & $1$ & $\rep1$ & 
   $\tfrac43$ & $\tfrac13$ & $-\tfrac{11}{3}$ & $\tfrac{92}{3}$ & $-\tfrac{79}{3}$ & $-\tfrac{17}{3}$ & $\tfrac{211}{3}$ & $\tfrac{104}{3}$ & $-\tfrac{91}{3}$ & $(W_1,W_2,W_3)$ \\   
\hline
\hline
 $T_{(0,2)}^{\phantom{A^{A}}}$ & $(\rep1,\rep1)_{0}$ & $\crep3_1$ & $\rep2''\oplus\rep1$ & $2$ & $\nicefrac{-1}3$ & $1$ & $\w$ & $1$ & $\rep1$ & 
   $\tfrac{4}{3}$ & $-\tfrac{1}{3}$ & $-\tfrac{4}{3}$ & $-\tfrac{80}{3}$ & $   -4$ & $-\tfrac{34}{3}$ & $\tfrac{481}{3}$ & $\tfrac{176}{3}$ & $-\tfrac{370}{3}$ & $(s_{17},s_{21},s_{25})$ \\ 
                               & $(\rep1,\rep1)_{0}$ & $\crep3_1$ & $\rep2''\oplus\rep1$ & $2$ & $\nicefrac{-1}3$ & $1$ & $\w$ & $1$ & $\rep1$ & 
   $\tfrac43$ & $-\tfrac43$ & $\tfrac23$ & $\tfrac{40}{3}$ & $    2$ & $\tfrac{17}{3}$ & $-\tfrac{506}{3}$ & $\tfrac{56}{3}$ & $\tfrac{95}{3}$ & $(s_{18},s_{22},s_{26})$ \\
                               & $(\rep1,\rep1)_{0}$ & $\crep3_1$ & $\rep2''\oplus\rep1$ & $2$ & $\nicefrac{-1}3$ & $1$ & $\w$ & $1$ & $\rep1$ & 
   $\tfrac43$ & $\tfrac23$ & $\tfrac23$ & $\tfrac{40}{3}$ & $    2$ & $\tfrac{17}{3}$ & $-\tfrac{506}{3}$ & $\tfrac{56}{3}$ & $\tfrac{95}{3}$ & $(s_{19},s_{23},s_{27})$ \\
                               & $(\rep1,\rep1)_{0}$ & $\crep3_2$ & $\rep2''\oplus\rep1$ & $5$ & $\nicefrac23$ & $1$ & $\w$ & $1$ & $\rep1$ & 
   $-\tfrac83$ & $-\tfrac13$ & $\tfrac23$ & $\tfrac{40}{3}$ & $    2$ & $\tfrac{17}{3}$ & $\tfrac{25}{3}$ & $-\tfrac{232}{3}$ & $\tfrac{275}{3}$ & $(s_{20},s_{24},s_{28})$ \\
                               & $(\rep3,\rep1)_{-\frac13}$ & $\crep3_1$ & $\rep2''\oplus\rep1$ & $2$ & $\nicefrac{-1}3$ & $1$ & $\w$ & $1$ & $\rep1$ & 
   $\tfrac43$ & $-\tfrac13$ & $\tfrac23$ & $\tfrac{40}{3}$ & $    2$ & $\tfrac{236}{3}$ & $-\tfrac{59}{3}$ & $\tfrac{32}{3}$ & $\tfrac{188}{3}$ & $(D_4,D_5,D_6)$ \\
                               & $(\rep1,\rep2)_{\frac12}$ & $\crep3_1$ & $\rep2''\oplus\rep1$ & $2$ & $\nicefrac{-1}3$ & $1$ & $\w$ & $1$ & $\rep1$ & 
   $\tfrac43$ & $-\tfrac13$ & $\tfrac23$ & $\tfrac{40}{3}$ & $    2$ & $\tfrac{236}{3}$ & $-\tfrac{59}{3}$ & $\tfrac{32}{3}$ & $\tfrac{188}{3}$ & $(\bar{L}_5,\bar{L}_6,\bar{L}_7)$ \\
                               & $(\rep1,\rep1)_{-\frac13}$ & $\crep3_1$ & $\rep2''\oplus\rep1$ & $2$ & $\nicefrac{-1}3$ & $1$ & $\w^2$ & $1$ & $\rep1$ & 
   $\tfrac23$ & $\tfrac23$ & $\tfrac{11}{3}$ & $-\tfrac{92}{3}$ & $\tfrac{79}{3}$ & $\tfrac{236}{3}$ & $-\tfrac{118}{3}$ & $\tfrac{64}{3}$ & $-\tfrac{92}{3}$ & $(V_4,V_5,V_6)$ \\
                               & $(\rep1,\rep1)_{\frac13}$ & $\crep3_1$ & $\rep2''\oplus\rep1$ & $2$ & $\nicefrac{-1}3$ & $1$ & $1$ & $1$ & $\rep1$ & 
   $    2$ & $-\tfrac13$ & $-\tfrac{13}{3}$ & $\tfrac{52}{3}$ & $-\tfrac{85}{3}$ & $-\tfrac{253}{3}$ & $   31$ & $   56$ & $  -61$ & $(\bar{V}_4,\bar{V}_5,\bar{V}_6)$\\
                               & $(\rep1,\rep2)_{\frac16}$ & $\crep3_1$ & $\rep2''\oplus\rep1$ & $2$ & $\nicefrac{-1}3$ & $1$ & $\w^2$ & $1$ & $\rep1$ & 
   $\tfrac23$ & $-\tfrac13$ & $\tfrac53$ & $\tfrac{22}{3}$ & $-\tfrac{17}{3}$ & $-\tfrac{109}{3}$ & $\tfrac{41}{3}$ & $-\tfrac{896}{3}$ & $\tfrac13$ & $(\overline{W}_1,\overline{W}_2,\overline{W}_3)$\\
\hline
\hline
 $T_{(1,0)}^{\phantom{A^{A}}}$ & $(\crep3,\rep1)_{\frac13}$ & $\rep3_2$ & $\rep2'\oplus\rep1$ & $1$ & $\nicefrac{-2}3$ & $\w$ & $1$ & $\w$ & $\rep1$ &  
   $   -2$ & $    0$ & $-\tfrac53$ & $\tfrac{56}{3}$ & $-\tfrac{67}{3}$ & $-\tfrac{56}{3}$ & $-\tfrac{242}{3}$ & $-\tfrac{160}{3}$ & $\tfrac{152}{3}$ & $(\bar{d}_1,\bar{d}_2,\bar{d}_3)$ \\
                               & $(\rep1,\rep2)_{-\frac12}$ & $\rep3_2$ & $\rep2'\oplus\rep1$ & $1$ & $\nicefrac{-2}3$ & $\w$ & $1$ & $1$ & $\rep1$ &  
   $-\tfrac83$ & $    0$ & $\tfrac43$ & $\tfrac23$ & $\tfrac{38}{3}$ & $-\tfrac{14}{3}$ & $-\tfrac{326}{3}$ & $\tfrac{104}{3}$ & $\tfrac{182}{3}$ & $(\ell_1,\ell_2,\ell_3)$ \\
                               & $(\rep1,\rep1)_{0}$ & $\rep3_2$ & $\rep2'\oplus\rep1$ & $1$ & $\nicefrac{-2}3$ & $\w$ & $1$ & $\w^2$ & $\rep1$ &  
   $-\tfrac43$ & $    1$ & $\tfrac73$ & $\tfrac{62}{3}$ & $\tfrac{47}{3}$ & $\tfrac{121}{3}$ & $\tfrac{289}{3}$ & $-\tfrac{448}{3}$ & $\tfrac{215}{3}$ & $(\varphi_{\mathrm u,1},\varphi_{\mathrm u,2},\varphi_{\mathrm u,3})$\\
                               & $(\rep1,\rep1)_0$ & $\rep3_2$ & $\rep2'\oplus\rep1$ & $1$ & $\nicefrac{-2}3$ & $\w$ & $1$ & $1$ & $\rep4$ &
   $\tfrac{11}{3}$ & $    0$ & $-\tfrac23$ & $-\tfrac13$ & $\tfrac{68}{3}$ & $-\tfrac23$ & $\tfrac43$ & $-\tfrac{166}{3}$ & $-\tfrac{10}{3}$ & $(s_{29},s_{37},s_{45})$ \\
                               & $(\rep1,\rep1)_{0}$ & $\rep3_2$ & $\rep2'\oplus\rep1$ & $1$ & $\nicefrac{-2}3$ & $\w$ & $1$ & $1$ & $\rep1$ &
   $    3$ & $    0$ & $-\tfrac{17}{3}$ & $-\tfrac{67}{3}$ & $-\tfrac{55}{3}$ & $\tfrac{124}{3}$ & $-\tfrac{248}{3}$ & $-\tfrac{448}{3}$ & $\tfrac{20}{3}$ & $(s_{30},s_{38},s_{46})$ \\
                               & $(\rep1,\rep1)_{0}$ & $\rep3_  1$ & $\rep2'\oplus\rep1$ & $-2$ & $\nicefrac{-5}3$ & $\w$ & $1$ & $1$ & $\rep1$ &
   $\tfrac43$ & $    0$ & $\tfrac43$ & $\tfrac23$ & $\tfrac{38}{3}$ & $\tfrac{205}{3}$ & $-\tfrac{410}{3}$ & $\tfrac{368}{3}$ & $\tfrac{95}{3}$ & $(s_{31},s_{39},s_{47})$ \\
                               & $(\rep1,\rep1)_{0}$ & $\rep3_2$ & $\rep2'\oplus\rep1$ & $1$ & $\nicefrac{-2}3$ & $\w$ & $1$ & $\w$ & $\crep4$ &
   $\tfrac{11}{3}$  & $    0$ & $\tfrac{10}{3}$ & $\tfrac53$ & $\tfrac83$ & $-\tfrac{26}{3}$ & $\tfrac{52}{3}$ & $-\tfrac{10}{3}$ & $-\tfrac{10}{3}$ & $(s_{32},s_{40},s_{48})$ \\
                               & $(\rep1,\rep1)_{0}$ & $\rep3_2$ & $\rep2'\oplus\rep1$ & $1$ & $\nicefrac{-2}3$ & $\w$ & $1$ & $\w$ & $\rep1$ &
   $\tfrac73$ & $    0$ & $\tfrac{10}{3}$ & $\tfrac53$ & $\tfrac83$ & $-\tfrac{26}{3}$ & $\tfrac{52}{3}$ & $\tfrac{1064}{3}$ & $\tfrac{50}{3}$ & $(s_{33},s_{41},s_{49})$ \\
                               & $(\rep1,\rep1)_{0}$ & $\rep3_2$ & $\rep2'\oplus\rep1$ & $1$ & $\nicefrac{-2}3$ & $\w$ & $1$ & $\w^2$ & $\rep1$ &
   $    3$ & $    0$ & $\tfrac73$ & $-\tfrac{55}{3}$ & $-\tfrac{175}{3}$ & $\tfrac{76}{3}$ & $-\tfrac{152}{3}$ & $-\tfrac{136}{3}$ & $\tfrac{20}{3}$ & $(s_{34},s_{42},s_{50})$ \\
                               & $(\rep1,\rep1)_0$ & $\rep3_2$ & $\rep2'\oplus\rep1$ & $1$ & $\nicefrac{-2}3$ & $\w$ & $1$ & $\w^2$ & $\rep1$ &
   $\tfrac73$ & $    0$ & $-\tfrac{14}{3}$ & $-\tfrac73$ & $\tfrac{128}{3}$ & $\tfrac{22}{3}$ & $-\tfrac{44}{3}$ & $\tfrac{752}{3}$ & $\tfrac{50}{3}$ & $(s_{36},s_{44},s_{52})$ \\
\hline
\hline
 $T_{(1,2)}^{\phantom{A^{A}}}$ & $(\rep1,\rep1)_0$ & $\rep1$ & $\rep1$ & $0$ & $0$ & $\w$ & $\w$ & $1$ & $\rep1$ &
   $\tfrac83$ & $-\tfrac13$ & $    0$ & $  -26$ & $\tfrac{26}{3}$ & $   57$ & $\tfrac{71}{3}$ & $\tfrac{544}{3}$ & $-\tfrac{275}{3}$ & $s_{53}$ \\
                                & $(\rep1,\rep1)_0$ & $\rep1$ & $\rep1$ & $0$ & $0$ & $\w$ & $\w$ & $1$ & $\rep1$ &
   $-\tfrac{10}{3}$ & $\tfrac23$ & $    0$ & $  -26$ & $\tfrac{26}{3}$ & $  -89$ & $\tfrac{62}{3}$ & $\tfrac{112}{3}$ & $-\tfrac{5}{3}$ & $s_{54}$ \\
                                & $(\rep1,\rep1)_0$ & $\rep1$ & $\rep1$ & $0$ & $0$ & $\w$ & $\w$ & $1$ & $\rep1$ &
   $\tfrac83$ & $-\tfrac13$ & $   -2$ & $   12$ & $-\tfrac{70}{3}$ & $  -58$ & $\tfrac{407}{3}$ & $-\tfrac{512}{3}$ & $\tfrac{190}{3}$ & $s_{55}$ \\
                                & $(\rep1,\rep1)_0$ & $\rep1$ & $\rep1$ & $0$ & $0$ & $\w$ & $\w$ & $1$ & $\rep1$ &
   $\tfrac23$ & $-\tfrac13$ & $    0$ & $  -26$ & $\tfrac{26}{3}$ & $  -16$ & $\tfrac{155}{3}$ & $\tfrac{280}{3}$ & $\tfrac{280}{3}$ & $s_{56}$ \\
                                & $(\rep1,\rep1)_0$ & $\rep1$ & $\rep1$ & $0$ & $0$ & $\w$ & $\w$ & $\w$ & $\rep1$ &
   $\tfrac{10}{3}$ & $-\tfrac13$ & $   -3$ & $   -8$ & $-\tfrac{79}{3}$ & $   43$ & $\tfrac{155}{3}$ & $\tfrac{280}{3}$ & $-\tfrac{305}{3}$ & $s_{57}$ \\
                                & $(\rep1,\rep1)_0$ & $\rep1$ & $\rep1$ & $0$ & $0$ & $\w$ & $\w$ & $\w$ & $\rep1$ &
   $-\tfrac83$ & $\tfrac23$ & $   -3$ & $   -8$ & $-\tfrac{79}{3}$ & $ -103$ & $\tfrac{146}{3}$ & $-\tfrac{152}{3}$ & $-\tfrac{35}{3}$ & $s_{58}$ \\
                                & $(\rep1,\rep1)_0$ & $\rep1$ & $\rep1$ & $0$ & $0$ & $\w$ & $\w$ & $\w$ & $\rep1$ &
   $\tfrac43$ & $-\tfrac13$ & $    4$ & $  -24$ & $\tfrac{140}{3}$ & $  -30$ & $\tfrac{239}{3}$ & $\tfrac{16}{3}$ & $\tfrac{250}{3}$ & $s_{59}$ \\
                                & $(\rep1,\rep1)_0$ & $\rep1$ & $\rep1$ & $0$ & $0$ & $\w$ & $\w$ & $\w$ & $\rep1$ &
   $\tfrac43$ & $-\tfrac13$ & $   -3$ & $   -8$ & $-\tfrac{79}{3}$ & $  -30$ & $\tfrac{239}{3}$ & $\tfrac{16}{3}$ & $\tfrac{250}{3}$ & $s_{60}$ \\
                                & $(\rep1,\rep1)_0$ & $\rep1$ & $\rep1$ & $0$ & $0$ & $\w$ & $\w$ & $\w^2$ & $\rep1$ &
   $    0$ & $-\tfrac13$ & $   -4$ & $  -28$ & $-\tfrac{88}{3}$ & $   -2$ & $\tfrac{71}{3}$ & $\tfrac{544}{3}$ & $\tfrac{310}{3}$ & $s_{61}$ \\
                                & $(\rep1,\rep1)_0$ & $\rep1$ & $\rep1$ & $0$ & $0$ & $\w$ & $\w$ & $\w^2$ & $\rep1$ &
   $   -2$ & $\tfrac23$ & $    1$ & $   -6$ & $\tfrac{35}{3}$ & $ -117$ & $\tfrac{230}{3}$ & $-\tfrac{416}{3}$ & $-\tfrac{65}{3}$ & $s_{62}$ \\
                                & $(\rep1,\rep1)_0$ & $\rep1$ & $\rep1$ & $0$ & $0$ & $\w$ & $\w$ & $\w^2$ & $\rep1$ &
   $    4$ & $-\tfrac13$ & $    1$ & $   -6$ & $\tfrac{35}{3}$ & $   29$ & $\tfrac{239}{3}$ & $\tfrac{16}{3}$ & $-\tfrac{335}{3}$ & $s_{63}$ \\
                                & $(\rep1,\rep1)_0$ & $\rep1$ & $\rep1$ & $0$ & $0$ & $\w$ & $\w$ & $\w^2$ & $\rep1$ &
   $    2$ & $-\tfrac13$ & $    1$ & $   -6$ & $\tfrac{35}{3}$ & $  -44$ & $\tfrac{323}{3}$ & $-\tfrac{248}{3}$ & $\tfrac{220}{3}$ & $s_{64}$ \\
                                & $(\rep1,\rep1)_{-\frac13}$ & $\rep1$ & $\rep1$ & $0$ & $0$ & $\w$ & $\w^2$ & $1$ & $\rep1$ &
   $-\tfrac{5}{3}$ & $-\tfrac13$ & $    1$ & $    7$ & $  -41$ & $ -104$ & $   31$ & $   56$ & $   56$ & $V_{7}$ \\
                                & $(\rep1,\rep1)_{-\frac13}$ & $\rep1$ & $\rep1$ & $0$ & $0$ & $\w$ & $\w^2$ & $\w$ & $\rep1$ &
   $-\tfrac{10}{3}$ & $-\tfrac13$ & $    2$ & $  -12$ & $    4$ & $  -13$ & $ -151$ & $  -88$ & $   81$ & $V_{8}$ \\
                                & $(\rep1,\rep1)_{-\frac13}$ & $\rep1$ & $\rep1$ & $0$ & $0$ & $\w$ & $\w^2$ & $\w$ & $\rep1$ &
   $    1$ & $-\tfrac{1}{3}$ & $    5$ & $    9$ & $   -3$ & $   28$ & $  239$ & $   16$ & $   16$ & $V_{9}$ \\
                                & $(\rep1,\rep1)_{-\frac13}$ & $\rep1$ & $\rep1$ & $0$ & $0$ & $\w$ & $\w^2$ & $\w$ & $\rep1$ &
   $    1$ & $-\tfrac13$ & $    5$ & $    9$ & $   -3$ & $  -45$ & $   31$ & $   56$ & $ -139$ & $V_{10}$ \\
                                & $(\rep1,\rep1)_{-\frac13}$ & $\rep1$ & $\rep1$ & $0$ & $0$ & $\w$ & $\w^2$ & $\w$ & $\rep1$ &
   $-\tfrac53$ & $-\tfrac13$ & $   -7$ & $    3$ & $   -1$ & $  -88$ & $   -1$ & $  -48$ & $   56$ & $V_{11}$ \\
                                & $(\rep1,\rep1)_{-\frac13}$ & $\rep1$ & $\rep1$ & $0$ & $0$ & $\w$ & $\w^2$ & $\w^2$ & $\rep1$ &
   $-\tfrac23$ & $-\tfrac13$ & $   -1$ & $    6$ & $  -31$ & $  119$ & $   57$ & $ -128$ & $   41$ & $V_{12}$ \\
                                & $(\rep1,\rep1)_{-\frac13}$ & $\rep1$ & $\rep1$ & $0$ & $0$ & $\w$ & $\w^2$ & $\w^2$ & $\rep1$ &
   $-\tfrac23$ & $-\tfrac13$ & $   -1$ & $    6$ & $  -31$ & $   46$ & $ -151$ & $  -88$ & $ -114$ & $V_{13}$ \\
                                & $(\rep1,\rep1)_{-\frac13}$ & $\rep1$ & $\rep1$ & $0$ & $0$ & $\w$ & $\w^2$ & $\w^2$ & $\rep1$ &
   $    1$ & $-\tfrac13$ & $   -3$ & $    5$ & $   37$ & $   44$ & $  207$ & $  -88$ & $   16$ & $V_{14}$ \\
                                & $(\rep1,\rep1)_{-\frac13}$ & $\rep1$ & $\rep1$ & $0$ & $0$ & $\w$ & $\w^2$ & $\w^2$ & $\rep1$ &
   $    1$ & $-\tfrac13$ & $   -3$ & $    5$ & $   37$ & $  -29$ & $   -1$ & $  -48$ & $ -139$ & $V_{15}$ \\
                                & $(\rep1,\rep1)_{\frac13}$ & $\rep1$ & $\rep1$ & $0$ & $0$ & $\w$ & $1$ & $1$ & $\rep1$ &
   $-\tfrac73$ & $\tfrac23$ & $    6$ & $    3$ & $-\tfrac{32}{3}$ & $   59$ & $-\tfrac{236}{3}$ & $\tfrac{128}{3}$ & $\tfrac{167}{3}$ & $\bar{V}_{7}$ \\
                                & $(\rep1,\rep1)_{\frac13}$ & $\rep1$ & $\rep1$ & $0$ & $0$ & $\w$ & $1$ & $1$ & $\rep1$ &
   $\tfrac{10}{3}$ & $-\tfrac13$ & $   -3$ & $   18$ & $-\tfrac{47}{3}$ & $  -16$ & $-\tfrac{317}{3}$ & $\tfrac{536}{3}$ & $-\tfrac{88}{3}$ & $\bar{V}_{8}$ \\
                                & $(\rep1,\rep1)_{\frac13}$ & $\rep1$ & $\rep1$ & $0$ & $0$ & $\w$ & $1$ & $1$ & $\rep1$ &
   $    1$ & $\tfrac23$ & $   -3$ & $  -21$ & $\tfrac{79}{3}$ & $  -43$ & $\tfrac{376}{3}$ & $-\tfrac{568}{3}$ & $\tfrac{17}{3}$ & $\bar{V}_{9}$ \\
                                & $(\rep1,\rep1)_{\frac13}$ & $\rep1$ & $\rep1$ & $0$ & $0$ & $\w$ & $1$ & $\w$ & $\rep1$ &
   $-\tfrac73$ & $\tfrac23$ & $   -2$ & $   -1$ & $\tfrac{88}{3}$ & $   75$ & $-\tfrac{332}{3}$ & $-\tfrac{184}{3}$ & $\tfrac{167}{3}$ & $\bar{V}_{10}$ \\
                                & $(\rep1,\rep1)_{\frac13}$ & $\rep1$ & $\rep1$ & $0$ & $0$ & $\w$ & $1$ & $\w$ & $\crep4$ &
   $-\tfrac23$ & $\tfrac23$ & $   -4$ & $   -2$ & $-\tfrac{56}{3}$ & $   12$ & $\tfrac{46}{3}$ & $-\tfrac{298}{3}$ & $\tfrac{92}{3}$ & $\bar{V}_{11}$ \\
                                & $(\rep1,\rep1)_{\frac13}$ & $\rep1$ & $\rep1$ & $0$ & $0$ & $\w$ & $1$ & $\w$ & $\rep1$ &
   $   -2$ & $\tfrac23$ & $   -4$ & $   -2$ & $-\tfrac{56}{3}$ & $   12$ & $\tfrac{46}{3}$ & $\tfrac{776}{3}$ & $\tfrac{152}{3}$ & $\bar{V}_{12}$ \\
                                & $(\rep1,\rep1)_{\frac13}$ & $\rep1$ & $\rep1$ & $0$ & $0$ & $\w$ & $1$ & $\w$ & $\rep1$ &
   $    0$ & $\tfrac23$ & $    1$ & $   20$ & $\tfrac{67}{3}$ & $  -30$ & $\tfrac{298}{3}$ & $-\tfrac{16}{3}$ & $\tfrac{62}{3}$ & $\bar{V}_{13}$ \\
                                & $(\rep1,\rep1)_{\frac13}$ & $\rep1$ & $\rep1$ & $0$ & $0$ & $\w$ & $1$ & $\w^2$ & $\rep4$ &
   $-\tfrac23$ & $\tfrac23$ & $    0$ & $    0$ & $-\tfrac{116}{3}$ & $    4$ & $\tfrac{94}{3}$ & $-\tfrac{142}{3}$ & $\tfrac{92}{3}$ & $\bar{V}_{14}$ \\
                                & $(\rep1,\rep1)_{\frac13}$ & $\rep1$ & $\rep1$ & $0$ & $0$ & $\w$ & $1$ & $\w^2$ & $\rep1$ &
   $-\tfrac43$ & $\tfrac23$ & $    2$ & $  -38$ & $-\tfrac{20}{3}$ & $   46$ & $-\tfrac{158}{3}$ & $-\tfrac{424}{3}$ & $\tfrac{122}{3}$ & $\bar{V}_{15}$ \\
                                & $(\rep1,\rep1)_{\frac13}$ & $\rep1$ & $\rep1$ & $0$ & $0$ & $\w$ & $1$ & $\w^2$ & $\rep1$ &
   $-\tfrac43$ & $-\tfrac43$ & $    0$ & $    0$ & $\tfrac{58}{3}$ & $   -2$ & $\tfrac{130}{3}$ & $\tfrac{512}{3}$ & $\tfrac{122}{3}$ & $\bar{V}_{16}$ \\
                                & $(\rep1,\rep1)_{\frac13}$ & $\rep1$ & $\rep1$ & $0$ & $0$ & $\w$ & $1$ & $\w^2$ & $\rep1$ &
   $    1$ & $\tfrac23$ & $    5$ & $  -17$ & $-\tfrac{41}{3}$ & $  -59$ & $\tfrac{472}{3}$ & $-\tfrac{256}{3}$ & $\tfrac{17}{3}$ & $\bar{V}_{17}$ \\
                                & $(\rep1,\rep1)_{\frac13}$ & $\rep1$ & $\rep1$ & $0$ & $0$ & $\w$ & $1$ & $\w^2$ & $\rep1$ &
   $-\tfrac43$ & $\tfrac23$ & $    0$ & $    0$ & $\tfrac{58}{3}$ & $   -2$ & $\tfrac{130}{3}$ & $\tfrac{512}{3}$ & $\tfrac{122}{3}$ & $\bar{V}_{18}$ \\
                                & $(\rep1,\rep2)_{-\frac16}$ & $\rep1$ & $\rep1$ & $0$ & $0$ & $\w$ & $1$ & $\w$ & $\rep1$ &
   $2$ & $-\tfrac13$ & $    1$ & $   20$ & $\tfrac{67}{3}$ & $   43$ & $\tfrac{391}{3}$ & $\tfrac{152}{3}$ & $-\tfrac{121}{3}$ & $W_{4}$ \\
                                & $(\rep1,\rep2)_{\frac16}$ & $\rep1$ & $\rep1$ & $0$ & $0$ & $\w$ & $\w^2$ & $1$ & $\rep1$ &
   $    2$ & $-\tfrac13$ & $    3$ & $    8$ & $    7$ & $   32$ & $ -123$ & $ -176$ & $   32$ & $\overline{W}_{4}$ \\   
                                & $(\rep1,\rep2)_{\frac16}$ & $\rep1$ & $\rep1$ & $0$ & $0$ & $\w$ & $\w^2$ & $1$ & $\rep1$ &
   $\tfrac53$ & $-\tfrac13$ & $   -4$ & $  -15$ & $   34$ & $   -1$ & $  -57$ & $  128$ & $   37$ & $\overline{W}_{5}$ \\
                                & $(\rep1,\rep2)_{\frac16}$ & $\rep1$ & $\rep1$ & $0$ & $0$ & $\w$ & $\w^2$ & $\w^2$ & $\rep1$ &
   $\tfrac53$ & $-\tfrac13$ & $  4$ & $  -11$ & $   -6$ & $  -17$ & $  -25$ & $  232$ & $   37$ & $\overline{W}_{6}$ \\
                                & $(\rep1,\rep1)_{-\frac23}$ & $\rep1$ & $\rep1$ & $0$ & $0$ & $\w$ & $1$ & $1$ & $\rep1$ &
   $-\tfrac83$ & $-\tfrac13$ & $   -3$ & $   18$ & $-\tfrac{47}{3}$ & $  -16$ & $\tfrac{391}{3}$ & $\tfrac{152}{3}$ & $-\tfrac43$ & $X_{1}$ \\
                                & $(\rep1,\rep1)_{-\frac23}$ & $\rep1$ & $\rep1$ & $0$ & $0$ & $\w$ & $1$ & $\w$ & $\rep1$ &
   $    2$ & $\tfrac23$ & $    1$ & $   20$ & $\tfrac{67}{3}$ & $  -30$ & $-\tfrac{56}{3}$ & $\tfrac{176}{3}$ & $-\tfrac{214}{3}$ & $X_{2}$ \\
                                & $(\rep1,\rep1)_{\frac23}$ & $\rep1$ & $\rep1$ & $0$ & $0$ & $\w$ & $\w^2$ & $1$ & $\rep1$ &
   $-\tfrac53$ & $-\tfrac13$ & $    1$ & $    7$ & $  -41$ & $  -31$ & $  121$ & $   80$ & $  -37$ & $\bar{X}_{1}$ \\
                                & $(\rep1,\rep1)_{\frac23}$ & $\rep1$ & $\rep1$ & $0$ & $0$ & $\w$ & $\w^2$ & $\w$ & $\rep1$ &
   $-\tfrac{10}{3}$ & $-\tfrac13$ & $    2$ & $  -12$ & $    4$ & $   60$ & $  -61$ & $  -64$ & $  -12$ & $\bar{X}_{2}$ \\
                                & $(\rep1,\rep1)_{\frac23}$ & $\rep1$ & $\rep1$ & $0$ & $0$ & $\w$ & $\w^2$ & $\w$ & $\rep1$ &
   $-\tfrac53$ & $-\tfrac13$ & $   -7$ & $    3$ & $   -1$ & $  -15$ & $   89$ & $  -24$ & $  -37$ & $\bar{X}_{3}$ \\
                                & $(\crep3,\rep1)_{-\frac13}$ & $\rep1$ & $\rep1$ & $0$ & $0$ & $\w$ & $1$ & $\w$ & $\rep1$ &
   $    2$ & $-\tfrac13$ & $    1$ & $   20$ & $\tfrac{67}{3}$ & $  -30$ & $\tfrac{121}{3}$ & $\tfrac{80}{3}$ & $\tfrac{158}{3}$ & $Y$ \\
                                & $(\crep3,\rep1)_0$ & $\rep1$ & $\rep1$ & $0$ & $0$ & $\w$ & $\w^2$ & $\w$ & $\rep1$ &
   $    1$ & $\tfrac23$ & $    5$ & $    9$ & $   -3$ & $  -45$ & $   90$ & $   24$ & $  -15$ & $Z_{1}$ \\
                                & $(\crep3,\rep1)_0$ & $\rep1$ & $\rep1$ & $0$ & $0$ & $\w$ & $\w^2$ & $\w^2$ & $\rep1$ &
   $-\tfrac23$ & $\tfrac23$ & $   -1$ & $    6$ & $  -31$ & $   46$ & $  -92$ & $ -120$ & $   10$ & $Z_{2}$ \\
                                & $(\crep3,\rep1)_0$ & $\rep1$ & $\rep1$ & $0$ & $0$ & $\w$ & $\w^2$ & $\w^2$ & $\rep1$ &
   $    1$ & $\tfrac23$ & $   -3$ & $    5$ & $   37$ & $  -29$ & $   58$ & $  -80$ & $  -15$ & $Z_{3}$ \\
                                & $(\rep3,\rep1)_0$ & $\rep1$ & $\rep1$ & $0$ & $0$ & $\w$ & $1$ & $\w^2$ & $\rep1$ &
   $-\tfrac43$ & $-\tfrac13$ & $    0$ & $    0$ & $\tfrac{58}{3}$ & $   -2$ & $-\tfrac{47}{3}$ & $\tfrac{608}{3}$ & $-\tfrac{250}{3}$ & $\bar{Z}_{1}$ \\
\hline
\hline
  $T_{(2,0)}^{\phantom{A^{A}}}$ & $(\rep1,\rep1)_0$ & $\crep3_1$ & $\rep2''\oplus\rep1$ & $2$ & $\nicefrac{-1}3$ & $\w^2$ & $1$ & $1$ & $\rep1$ & 
   $-\tfrac{10}{3}$ & $    0$ & $\tfrac{2}{3}$ & $-\tfrac{116}{3}$ & $\tfrac{58}{3}$ & $-\tfrac{79}{3}$ & $\tfrac{158}{3}$ & $\tfrac{424}{3}$ & $-\tfrac{5}{3}$ & $(s_{65},s_{69},s_{73})$ \\
                                & $(\rep1,\rep1)_0$ & $\crep3_1$ & $\rep2''\oplus\rep1$ & $2$ & $\nicefrac{-1}3$ & $\w^2$ & $1$ & $1$ & $\rep1$ & 
   $\tfrac83$ & $   -1$ & $-\tfrac43$ & $-\tfrac23$ & $-\tfrac{38}{3}$ & $\tfrac{14}{3}$ & $\tfrac{503}{3}$ & $-\tfrac{200}{3}$ & $\tfrac{190}{3}$ & $(s_{66},s_{70},s_{74})$ \\
                                & $(\rep1,\rep1)_0$ & $\crep3_1$ & $\rep2''\oplus\rep1$ & $2$ & $\nicefrac{-1}3$ & $\w^2$ & $1$ & $1$ & $\rep1$ & 
   $-\tfrac43$ & $    0$ & $-\tfrac43$ & $-\tfrac23$ & $-\tfrac{38}{3}$ & $-\tfrac{205}{3}$ & $\tfrac{410}{3}$ & $-\tfrac{368}{3}$ & $-\tfrac{95}{3}$ & $(s_{67},s_{71},s_{75})$ \\
                                & $(\rep1,\rep1)_0$ & $\crep3_1$ & $\rep2''\oplus\rep1$ & $2$ & $\nicefrac{-1}3$ & $\w^2$ & $1$ & $\w$ & $\rep1$ & 
   $\tfrac43$ & $    1$ & $-\tfrac73$ & $-\tfrac{62}{3}$ & $-\tfrac{47}{3}$ & $-\tfrac{121}{3}$ & $-\tfrac{289}{3}$ & $\tfrac{448}{3}$ & $-\tfrac{215}{3}$ & $(s_{68},s_{72},s_{76})$ \\
\hline
\hline
  $T_{(2,1)}^{\phantom{A^{A}}}$ & $(\rep1,\rep1)_0$ & $\rep1$ & $\rep1$ & $0$ & $0$ & $\w^2$ & $\w^2$ & $\w$ & $\rep1$ &
   $    0$ & $-\tfrac23$ & $   -1$ & $    6$ & $-\tfrac{35}{3}$ & $  -29$ & $-\tfrac{770}{3}$ & $\tfrac{272}{3}$ & $\tfrac{155}{3}$ & $\phi^0_\mathrm{u}$ \\
                                & $(\rep1,\rep1)_0$ & $\rep1$ & $\rep1$ & $0$ & $0$ & $\w^2$ & $\w^2$ & $\w^2$ & $\rep1$ & 
   $\tfrac23$ & $\tfrac13$ & $    1$ & $  -32$ & $\tfrac{61}{3}$ & $  -60$ & $\tfrac{301}{3}$ & $\tfrac{128}{3}$ & $-\tfrac{340}{3}$ & $\phi^0_\mathrm{d}$ \\
                                & $(\rep1,\rep1)_0$ & $\rep1$ & $\rep1$ & $0$ & $0$ & $\w^2$ & $\w^2$ & $1$ & $\rep1$ &
   $\tfrac43$ & $\tfrac13$ & $   -2$ & $  -14$ & $-\tfrac{44}{3}$ & $  -74$ & $\tfrac{385}{3}$ & $-\tfrac{136}{3}$ & $-\tfrac{370}{3}$ & $\phi^0_\mathrm{e}$ \\
                                & $(\rep1,\rep1)_0$ & $\rep1$ & $\rep1$ & $0$ & $0$ & $\w^2$ & $\w^2$ & $1$ & $\rep1$ &
   $\tfrac43$ & $-\tfrac23$ & $    0$ & $   26$ & $-\tfrac{26}{3}$ & $  -57$ & $-\tfrac{602}{3}$ & $-\tfrac{256}{3}$ & $\tfrac{95}{3}$ & $s_{78}$ \\ 
                                & $(\rep1,\rep1)_0$ & $\rep1$ & $\rep1$ & $0$ & $0$ & $\w^2$ & $\w^2$ & $\w$ & $\rep1$ &
   $    0$ & $\tfrac13$ & $   -3$ & $  -34$ & $-\tfrac{53}{3}$ & $  -46$ & $\tfrac{217}{3}$ & $\tfrac{392}{3}$ & $-\tfrac{310}{3}$ & $s_{79}$ \\
                                & $(\rep1,\rep1)_0$ & $\rep1$ & $\rep1$ & $0$ & $0$ & $\w^2$ & $\w^2$ & $\w^2$ & $\rep1$ &
   $\tfrac23$ & $-\tfrac23$ & $    3$ & $    8$ & $\tfrac{79}{3}$ & $  -43$ & $-\tfrac{686}{3}$ & $\tfrac{8}{3}$ & $\tfrac{125}{3}$ & $s_{82}$ \\
                                & $(\rep3,\rep1)_{-\frac13}$ & $\rep1$ & $\rep1$ & $0$ & $0$ & $\w^2$ & $\w^2$ & $1$ & $\rep1$ &
   $\tfrac43$ & $\tfrac13$ & $    0$ & $   26$ & $-\tfrac{26}{3}$ & $   16$ & $-\tfrac{155}{3}$ & $-\tfrac{280}{3}$ & $\tfrac{188}{3}$ & $D_{7}$ \\
                                & $(\rep3,\rep1)_{-\frac13}$ & $\rep1$ & $\rep1$ & $0$ & $0$ & $\w^2$ & $\w^2$ & $\w$ & $\rep1$ &
   $    0$ & $\tfrac13$ & $   -1$ & $    6$ & $-\tfrac{35}{3}$ & $   44$ & $-\tfrac{323}{3}$ & $\tfrac{248}{3}$ & $\tfrac{248}{3}$ & $D_{8}$ \\
                                & $(\rep3,\rep1)_{-\frac13}$ & $\rep1$ & $\rep1$ & $0$ & $0$ & $\w^2$ & $\w^2$ & $\w^2$ & $\rep1$ &
   $\tfrac23$ & $\tfrac13$ & $    3$ & $    8$ & $\tfrac{79}{3}$ & $   30$ & $-\tfrac{239}{3}$ & $-\tfrac{16}{3}$ & $\tfrac{218}{3}$ & $D_{9}$ \\
                                & $(\rep1,\rep2)_{\frac12}$ & $\rep1$ & $\rep1$ & $0$ & $0$ & $\w^2$ & $\w^2$ & $1$ & $\rep1$ &
   $\tfrac43$ & $\tfrac13$ & $    0$ & $   26$ & $-\tfrac{26}{3}$ & $   16$ & $-\tfrac{155}{3}$ & $-\tfrac{280}{3}$ & $\tfrac{188}{3}$ & $\bar{L}_{8}$ \\
                                & $(\rep1,\rep2)_{\frac12}$ & $\rep1$ & $\rep1$ & $0$ & $0$ & $\w^2$ & $\w^2$ & $\w$ & $\rep1$ &
   $    0$ & $\tfrac13$ & $   -1$ & $    6$ & $-\tfrac{35}{3}$ & $   44$ & $-\tfrac{323}{3}$ & $\tfrac{248}{3}$ & $\tfrac{248}{3}$ & $\bar{L}_{9}$ \\
                                & $(\rep1,\rep2)_{\frac12}$ & $\rep1$ & $\rep1$ & $0$ & $0$ & $\w^2$ & $\w^2$ & $\w^2$ & $\rep1$ &
   $\tfrac23$ & $\tfrac13$ & $    3$ & $    8$ & $\tfrac{79}{3}$ & $   30$ & $-\tfrac{239}{3}$ & $-\tfrac{16}{3}$ & $\tfrac{218}{3}$ & $\bar{L}_{10}$ \\
                                & $(\rep1,\rep1)_{-\frac13}$ & $\rep1$ & $\rep1$ & $0$ & $0$ & $\w^2$ & $1$ & $1$ & $\rep1$ &
   $\tfrac13$ & $-\tfrac23$ & $   -4$ & $   37$ & $\tfrac{50}{3}$ & $   31$ & $-\tfrac{304}{3}$ & $-\tfrac{272}{3}$ & $-\tfrac{77}{3}$ & $V_{16}$ \\
                                & $(\rep1,\rep1)_{-\frac13}$ & $\rep1$ & $\rep1$ & $0$ & $0$ & $\w^2$ & $1$ & $1$ & $\crep4$ &
   $    0$ & $-\tfrac23$ & $   -4$ & $   -2$ & $\tfrac23$ & $   10$ & $-\tfrac{178}{3}$ & $\tfrac{406}{3}$ & $-\tfrac{62}{3}$ & $V_{17}$ \\
                                & $(\rep1,\rep1)_{-\frac13}$ & $\rep1$ & $\rep1$ & $0$ & $0$ & $\w^2$ & $1$ & $1$ & $\rep1$ &
   $\tfrac23$ & $-\tfrac23$ & $    1$ & $   20$ & $\tfrac{125}{3}$ & $  -32$ & $\tfrac{74}{3}$ & $\tfrac{688}{3}$ & $-\tfrac{92}{3}$ & $V_{18}$ \\
                                & $(\rep1,\rep1)_{-\frac13}$ & $\rep1$ & $\rep1$ & $0$ & $0$ & $\w^2$ & $1$ & $1$ & $\rep1$ &
   $\tfrac23$ & $\tfrac43$ & $    3$ & $  -18$ & $\tfrac{47}{3}$ & $   16$ & $-\tfrac{214}{3}$ & $-\tfrac{248}{3}$ & $-\tfrac{92}{3}$ & $V_{19}$ \\
                                & $(\rep1,\rep1)_{-\frac13}$ & $\rep1$ & $\rep1$ & $0$ & $0$ & $\w^2$ & $1$ & $1$ & $\rep1$ &
   $\tfrac23$ & $-\tfrac23$ & $    3$ & $  -18$ & $\tfrac{47}{3}$ & $   16$ & $-\tfrac{214}{3}$ & $-\tfrac{248}{3}$ & $-\tfrac{92}{3}$ & $V_{20}$ \\
                                & $(\rep1,\rep1)_{-\frac13}$ & $\rep1$ & $\rep1$ & $0$ & $0$ & $\w^2$ & $1$ & $\w$ & $\rep4$ &
   $    0$ & $-\tfrac23$ & $    0$ & $    0$ & $-\tfrac{58}{3}$ & $    2$ & $-\tfrac{130}{3}$ & $\tfrac{562}{3}$ & $-\tfrac{62}{3}$ & $V_{21}$ \\
                                & $(\rep1,\rep1)_{-\frac13}$ & $\rep1$ & $\rep1$ & $0$ & $0$ & $\w^2$ & $1$ & $\w$ & $\rep1$ &
   $-\tfrac23$ & $-\tfrac23$ & $    2$ & $  -38$ & $\tfrac{38}{3}$ & $   44$ & $-\tfrac{382}{3}$ & $\tfrac{280}{3}$ & $-\tfrac{32}{3}$ & $V_{22}$ \\
                                & $(\rep1,\rep1)_{-\frac13}$ & $\rep1$ & $\rep1$ & $0$ & $0$ & $\w^2$ & $1$ & $\w$ & $\rep1$ &
   $\tfrac53$ & $-\tfrac23$ & $    5$ & $  -17$ & $\tfrac{17}{3}$ & $  -61$ & $\tfrac{248}{3}$ & $\tfrac{448}{3}$ & $-\tfrac{137}{3}$ & $V_{23}$ \\
                                & $(\rep1,\rep1)_{-\frac13}$ & $\rep1$ & $\rep1$ & $0$ & $0$ & $\w^2$ & $1$ & $\w$ & $\rep1$ &
   $\tfrac43$ & $-\tfrac23$ & $    0$ & $    0$ & $-\tfrac{58}{3}$ & $    2$ & $-\tfrac{130}{3}$ & $-\tfrac{512}{3}$ & $-\tfrac{122}{3}$ & $V_{24}$ \\
                                & $(\rep1,\rep1)_{-\frac13}$ & $\rep1$ & $\rep1$ & $0$ & $0$ & $\w^2$ & $1$ & $\w^2$ & $\rep1$ &
   $\tfrac13$ & $-\tfrac23$ & $    4$ & $   41$ & $-\tfrac{70}{3}$ & $   15$ & $-\tfrac{208}{3}$ & $\tfrac{40}{3}$ & $-\tfrac{77}{3}$ & $V_{25}$ \\
                                & $(\rep1,\rep1)_{-\frac13}$ & $\rep1$ & $\rep1$ & $0$ & $0$ & $\w^2$ & $1$ & $\w^2$ & $\rep1$ &
   $   -4$ & $\tfrac13$ & $   -1$ & $  -20$ & $-\tfrac{67}{3}$ & $   30$ & $\tfrac{233}{3}$ & $-\tfrac{272}{3}$ & $\tfrac{118}{3}$ & $V_{26}$ \\
                                & $(\rep1,\rep1)_{-\frac13}$ & $\rep1$ & $\rep1$ & $0$ & $0$ & $\w^2$ & $1$ & $\w^2$ & $\rep1$ &
   $\tfrac53$ & $-\tfrac23$ & $   -3$ & $  -21$ & $\tfrac{137}{3}$ & $  -45$ & $\tfrac{152}{3}$ & $\tfrac{136}{3}$ & $-\tfrac{137}{3}$ & $V_{27}$ \\
                                & $(\rep1,\rep1)_{\frac13}$ & $\rep1$ & $\rep1$ & $0$ & $0$ & $\w^2$ & $\w$ & $1$ & $\rep1$ &
   $\tfrac73$ & $\tfrac13$ & $    4$ & $   15$ & $  -34$ & $   74$ & $   29$ & $  -40$ & $  -66$ & $\bar{V}_{19}$ \\
                                & $(\rep1,\rep1)_{\frac13}$ & $\rep1$ & $\rep1$ & $0$ & $0$ & $\w^2$ & $\w$ & $1$ & $\rep1$ &
   $-\tfrac{1}{3}$ & $\tfrac{1}{3}$ & $   -1$ & $   -7$ & $   41$ & $   31$ & $   -3$ & $ -144$ & $  129$ & $\bar{V}_{20}$ \\
                                & $(\rep1,\rep1)_{\frac13}$ & $\rep1$ & $\rep1$ & $0$ & $0$ & $\w^2$ & $\w$ & $1$ & $\rep1$ &
   $-\tfrac{1}{3}$ & $\tfrac{1}{3}$ & $   -1$ & $   -7$ & $   41$ & $  -42$ & $ -211$ & $ -104$ & $  -26$ & $\bar{V}_{21}$ \\
                                & $(\rep1,\rep1)_{\frac13}$ & $\rep1$ & $\rep1$ & $0$ & $0$ & $\w^2$ & $\w$ & $1$ & $\rep1$ &
   $    2$ & $\tfrac{1}{3}$ & $   -3$ & $   -8$ & $   -7$ & $   41$ & $   95$ & $  264$ & $  -61$ & $\bar{V}_{22}$ \\
                                & $(\rep1,\rep1)_{\frac13}$ & $\rep1$ & $\rep1$ & $0$ & $0$ & $\w^2$ & $\w$ & $\w$ & $\rep1$ &
   $\tfrac{7}{3}$ & $\tfrac{1}{3}$ & $   -4$ & $   11$ & $    6$ & $   90$ & $   -3$ & $ -144$ & $  -66$ & $\bar{V}_{23}$ \\
                                & $(\rep1,\rep1)_{\frac13}$ & $\rep1$ & $\rep1$ & $0$ & $0$ & $\w^2$ & $\w$ & $\w^2$ & $\rep1$ &
   $-\tfrac{1}{3}$ & $\tfrac{1}{3}$ & $    7$ & $   -3$ & $    1$ & $   15$ & $   29$ & $  -40$ & $  129$ & $\bar{V}_{24}$ \\
                                & $(\rep1,\rep1)_{\frac13}$ & $\rep1$ & $\rep1$ & $0$ & $0$ & $\w^2$ & $\w$ & $\w^2$ & $\rep1$ &
   $-\tfrac{1}{3}$ & $\tfrac{1}{3}$ & $    7$ & $   -3$ & $    1$ & $  -58$ & $ -179$ & $    0$ & $  -26$ & $\bar{V}_{25}$ \\
                                & $(\rep1,\rep1)_{\frac13}$ & $\rep1$ & $\rep1$ & $0$ & $0$ & $\w^2$ & $\w$ & $\w^2$ & $\rep1$ &
   $\tfrac{4}{3}$ & $\tfrac{1}{3}$ & $   -2$ & $   12$ & $   -4$ & $  -60$ & $  179$ & $    0$ & $  104$ & $\bar{V}_{26}$ \\
                                & $(\rep1,\rep1)_{\frac13}$ & $\rep1$ & $\rep1$ & $0$ & $0$ & $\w^2$ & $\w$ & $\w^2$ & $\rep1$ &
   $\tfrac{4}{3}$ & $\tfrac{1}{3}$ & $   -2$ & $   12$ & $   -4$ & $ -133$ & $  -29$ & $   40$ & $  -51$ & $\bar{V}_{27}$ \\
                                & $(\rep1,\rep2)_{-\frac16}$ & $\rep1$ & $\rep1$ & $0$ & $0$ & $\w^2$ & $\w$ & $\w$ & $\rep1$ &
   $   -3$ & $\tfrac13$ & $    3$ & $   -5$ & $  -37$ & $   29$ & $    1$ & $   48$ & $  -17$ & $W_{6}$ \\ 
                                & $(\rep1,\rep2)_{-\frac16}$ & $\rep1$ & $\rep1$ & $0$ & $0$ & $\w^2$ & $\w$ & $\w$ & $\rep1$ &
   $-\tfrac43$ & $\tfrac13$ & $    1$ & $   -6$ & $   31$ & $  -46$ & $  151$ & $   88$ & $  -42$ & $W_{7}$ \\
                                & $(\rep1,\rep2)_{-\frac16}$ & $\rep1$ & $\rep1$ & $0$ & $0$ & $\w^2$ & $\w$ & $\w^2$ & $\rep1$ &
   $   -3$ & $\tfrac13$ & $   -5$ & $   -9$ & $    3$ & $   45$ & $  -31$ & $  -56$ & $  -17$ & $W_{8}$ \\
                                & $(\rep1,\rep2)_{\frac16}$ & $\rep1$ & $\rep1$ & $0$ & $0$ & $\w^2$ & $1$ & $1$ & $\rep1$ &
   $\tfrac23$ & $\tfrac13$ & $    3$ & $  -18$ & $\tfrac{47}{3}$ & $   89$ & $\tfrac{233}{3}$ & $-\tfrac{272}{3}$ & $\tfrac{1}{3}$ & $\overline{W}_{7}$ \\
                                & $(\rep1,\rep2)_{\frac16}$ & $\rep1$ & $\rep1$ & $0$ & $0$ & $\w^2$ & $1$ & $\w$ & $\rep1$ &
   $-\tfrac23$ & $\tfrac13$ & $    0$ & $    0$ & $-\tfrac{58}{3}$ & $  -71$ & $-\tfrac{223}{3}$ & $-\tfrac{680}{3}$ & $\tfrac{61}{3}$ & $\overline{W}_{8}$ \\
                                & $(\rep1,\rep1)_{-\frac23}$ & $\rep1$ & $\rep1$ & $0$ & $0$ & $\w^2$ & $\w$ & $1$ & $\rep1$ &
   $\tfrac73$ & $\tfrac13$ & $    4$ & $   15$ & $  -34$ & $    1$ & $  -61$ & $  -64$ & $   27$ & $X_{3}$ \\
                                & $(\rep1,\rep1)_{-\frac23}$ & $\rep1$ & $\rep1$ & $0$ & $0$ & $\w^2$ & $\w$ & $1$ & $\rep1$ &
   $    2$ & $\tfrac13$ & $   -3$ & $   -8$ & $   -7$ & $  -32$ & $    5$ & $  240$ & $   32$ & $X_{4}$ \\
                                & $(\rep1,\rep1)_{-\frac23}$ & $\rep1$ & $\rep1$ & $0$ & $0$ & $\w^2$ & $\w$ & $\w$ & $\rep1$ &
   $\tfrac73$ & $\tfrac13$ & $   -4$ & $   11$ & $    6$ & $   17$ & $  -93$ & $ -168$ & $   27$ & $X_{5}$ \\
                                & $(\rep1,\rep1)_{\frac23}$ & $\rep1$ & $\rep1$ & $0$ & $0$ & $\w^2$ & $1$ & $\w$ & $\rep1$ &
   $-\tfrac23$ & $-\tfrac23$ & $    0$ & $    0$ & $-\tfrac{58}{3}$ & $    2$ & $\tfrac{224}{3}$ & $-\tfrac{704}{3}$ & $\tfrac{154}{3}$ & $\bar{X}_{4}$ \\
                                & $(\rep1,\rep1)_{\frac23}$ & $\rep1$ & $\rep1$ & $0$ & $0$ & $\w^2$ & $1$ & $\w^2$ & $\rep1$ & 
   $    2$ & $\tfrac13$ & $   -1$ & $  -20$ & $-\tfrac{67}{3}$ & $   30$ & $-\tfrac{475}{3}$ & $\tfrac{112}{3}$ & $\tfrac{34}{3}$ & $\bar{X}_{5}$ \\
                                & $(\rep3,\rep1)_{\frac13}$ & $\rep1$ & $\rep1$ & $0$ & $0$ & $\w^2$ & $1$ & $\w$ & $\rep1$ &
   $-\tfrac23$ & $\tfrac13$ & $    0$ & $    0$ & $-\tfrac{58}{3}$ & $    2$ & $\tfrac{47}{3}$ & $-\tfrac{608}{3}$ & $-\tfrac{218}{3}$ & $\bar{Y}$ \\   
                                & $(\crep3,\rep1)_0$ & $\rep1$ & $\rep1$ & $0$ & $0$ & $\w^2$ & $1$ & $1$ & $\rep1$ &
   $\tfrac{2}{3}$ & $\tfrac{1}{3}$ & $    3$ & $  -18$ & $\tfrac{47}{3}$ & $   16$ & $-\tfrac{37}{3}$ & $-\tfrac{344}{3}$ & $\tfrac{280}{3}$ & $Z_{4}$ \\
                                & $(\rep3,\rep1)_0$ & $\rep1$ & $\rep1$ & $0$ & $0$ & $\w^2$ & $\w$ & $1$ & $\rep1$ &
   $-\tfrac{1}{3}$ & $-\tfrac{2}{3}$ & $   -1$ & $   -7$ & $   41$ & $   31$ & $  -62$ & $ -112$ & $    5$ & $\bar{Z}_{2}$ \\
                                & $(\rep3,\rep1)_0$ & $\rep1$ & $\rep1$ & $0$ & $0$ & $\w^2$ & $\w$ & $\w^2$ & $\rep1$ &
   $-\tfrac{1}{3}$ & $-\tfrac{2}{3}$ & $    7$ & $   -3$ & $    1$ & $   15$ & $  -30$ & $   -8$ & $    5$ & $\bar{Z}_{3}$ \\
                                & $(\rep3,\rep1)_0$ & $\rep1$ & $\rep1$ & $0$ & $0$ & $\w^2$ & $\w$ & $\w^2$ & $\rep1$ &
   $\tfrac{4}{3}$ & $-\tfrac{2}{3}$ & $   -2$ & $   12$ & $   -4$ & $  -60$ & $  120$ & $   32$ & $  -20$ & $\bar{Z}_{4}$ \\
\hline
\hline  
  $T_{(2,2)}^{\phantom{A^{A}}}$ & $(\rep1,\rep1)_0$ & $\rep3_2$ & $\rep2'\oplus\rep1$ & $1$ & $\nicefrac{-2}3$ & $\w^2$ & $\w$ & $1$ & $\rep1$ &  
   $    2$ & $\tfrac{2}{3}$ & $-\tfrac{2}{3}$ & $\tfrac{38}{3}$ & $-\tfrac{32}{3}$ & $\tfrac{250}{3}$ & $  148$ & $  -56$ & $  -30$ & $(\varphi_{\nu,1},\varphi_{\nu,2},\varphi_{\nu,3})$ \\
                                & $(\rep1,\rep1)_0$ & $\rep3_1$ & $\rep2'\oplus\rep1$ & $1$ & $\nicefrac{-5}3$ & $\w^2$ & $\w$ & $1$ & $\rep1$ &  
   $    0$ & $\tfrac23$ & $-\tfrac23$ & $\tfrac{38}{3}$ & $-\tfrac{32}{3}$ & $-\tfrac{188}{3}$ & $  -32$ & $ -104$ & $    0$ & $(s_{84},s_{90},s_{96})$ \\
                                & $(\rep1,\rep1)_0$ & $\rep3_2$ & $\rep2'\oplus\rep1$ & $1$ & $\nicefrac{-2}3$ & $\w^2$ & $\w$ & $\w$ & $\rep1$ &  
   $\tfrac{2}{3}$ & $\tfrac{2}{3}$ & $-\tfrac{5}{3}$ & $-\tfrac{22}{3}$ & $-\tfrac{41}{3}$ & $\tfrac{334}{3}$ & $   92$ & $  120$ & $  -10$ & $(s_{85},s_{91},s_{97})$ \\
                                & $(\rep1,\rep1)_0$ & $\rep3_1$ & $\rep2'\oplus\rep1$ & $1$ & $\nicefrac{-5}3$ & $\w^2$ & $\w$ & $\w$ & $\rep1$ &  
   $-\tfrac{4}{3}$ & $\tfrac{2}{3}$ & $-\tfrac{5}{3}$ & $-\tfrac{22}{3}$ & $-\tfrac{41}{3}$ & $-\tfrac{104}{3}$ & $  -88$ & $   72$ & $   20$ & $(s_{86},s_{92},s_{98})$ \\
                                & $(\rep1,\rep1)_0$ & $\rep3_2$ & $\rep2'\oplus\rep1$ & $1$ & $\nicefrac{-2}3$ & $\w^2$ & $\w$ & $\w^2$ & $\rep1$ &  
   $\tfrac{4}{3}$ & $\tfrac{2}{3}$ & $\tfrac{7}{3}$ & $-\tfrac{16}{3}$ & $\tfrac{73}{3}$ & $\tfrac{292}{3}$ & $  120$ & $   32$ & $  -20$ & $(s_{87},s_{93},s_{99})$ \\
                                & $(\rep1,\rep1)_0$ & $\rep3_1$ & $\rep2'\oplus\rep1$ & $1$ & $\nicefrac{-5}3$ & $\w^2$ & $\w$ & $\w^2$ & $\rep1$ &
   $-\tfrac23$ & $\tfrac{2}{3}$ & $\tfrac{7}{3}$ & $-\tfrac{16}{3}$ & $\tfrac{73}{3}$ & $-\tfrac{146}{3}$ & $  -60$ & $  -16$ & $   10$ & $(s_{88},s_{94},s_{100})$ \\
                                & $(\crep3,\rep1)_{\frac13}$ & $\rep3_2$ & $\rep2'\oplus\rep1$ & $1$ & $\nicefrac{-2}3$ & $\w^2$ & $\w$ & $1$ & $\rep1$ &  
   $    2$ & $-\tfrac{1}{3}$ & $-\tfrac{2}{3}$ & $\tfrac{38}{3}$ & $-\tfrac{32}{3}$ & $\tfrac{31}{3}$ & $   -1$ & $  -48$ & $  -61$ & $(\bar{D}_1,\bar{D}_4,\bar{D}_7)$ \\
                                & $(\crep3,\rep1)_{\frac13}$ & $\rep3_2$ & $\rep2'\oplus\rep1$ & $1$ & $\nicefrac{-2}3$ & $\w^2$ & $\w$ & $\w$ & $\rep1$ &  
   $\tfrac{2}{3}$ & $-\tfrac{1}{3}$ & $-\tfrac{5}{3}$ & $-\tfrac{22}{3}$ & $-\tfrac{41}{3}$ & $\tfrac{115}{3}$ & $  -57$ & $  128$ & $  -41$ & $(\bar{D}_2,\bar{D}_5,\bar{D}_8)$ \\
                                & $(\crep3,\rep1)_{\frac13}$ & $\rep3_2$ & $\rep2'\oplus\rep1$ & $1$ & $\nicefrac{-2}3$ & $\w^2$ & $\w$ & $\w^2$ & $\rep1$ &  
   $\tfrac{4}{3}$ & $-\tfrac{1}{3}$ & $\tfrac{7}{3}$ & $-\tfrac{16}{3}$ & $\tfrac{73}{3}$ & $\tfrac{73}{3}$ & $  -29$ & $   40$ & $  -51$ & $(\bar{D}_3,\bar{D}_6,\bar{D}_9)$ \\
                                & $(\rep1,\rep2)_{-\frac12}$ & $\rep3_2$ & $\rep2'\oplus\rep1$ & $1$ & $\nicefrac{-2}3$ & $\w^2$ & $\w$ & $1$ & $\rep1$ &  
   $    2$ & $-\tfrac13$ & $-\tfrac23$ & $\tfrac{38}{3}$ & $-\tfrac{32}{3}$ & $\tfrac{31}{3}$ & $   -1$ & $  -48$ & $  -61$ & $(L_2,L_5,L_8)$ \\
                                & $(\rep1,\rep2)_{-\frac12}$ & $\rep3_2$ & $\rep2'\oplus\rep1$ & $1$ & $\nicefrac{-2}3$ & $\w^2$ & $\w$ & $\w$ & $\rep1$ &  
   $\tfrac23$ & $-\tfrac13$ & $-\tfrac53$ & $-\tfrac{22}{3}$ & $-\tfrac{41}{3}$ & $\tfrac{115}{3}$ & $  -57$ & $  128$ & $  -41$ & $(L_3,L_6,L_9)$ \\
                                & $(\rep1,\rep2)_{-\frac12}$ & $\rep3_2$ & $\rep2'\oplus\rep1$ & $1$ & $\nicefrac{-2}3$ & $\w^2$ & $\w$ & $\w^2$ & $\rep1$ &  
   $\tfrac43$ & $-\tfrac13$ & $\tfrac73$ & $-\tfrac{16}{3}$ & $\tfrac{73}{3}$ & $\tfrac{73}{3}$ & $  -29$ & $   40$ & $  -51$ & $(L_4,L_7,L_{10})$ \\
   
   &$(\rep1,\rep1)_{-\frac13}$ & $\rep3_2$ & $\rep2'\oplus\rep1$ & $1$ & $\nicefrac{-2}3$ & $\w^2$ & $\w^2$ & $1$ & $\rep1$ &
   $-\tfrac73$ & $\tfrac23$ & $\tfrac73$ & $\tfrac{23}{3}$ & $-\tfrac{85}{3}$ & $\tfrac{112}{3}$ & $\tfrac{130}{3}$ & $\tfrac{512}{3}$ & $-\tfrac{112}{3}$ & $(V_{28},V_{34},V_{40})$ \\
   
   & $(\rep1,\rep1)_{-\frac13}$ & $\rep3_2$ & $\rep2'\oplus\rep1$ & $1$ & $\nicefrac{-2}3$ & $\w^2$ & $\w^2$ & $1$ & $\rep1$ &
   $    1$ & $-\tfrac13$ & $-\tfrac{14}{3}$ & $\tfrac{71}{3}$ & $\tfrac{44}{3}$ & $-\tfrac{143}{3}$ & $-\tfrac{245}{3}$ & $-\tfrac{304}{3}$ & $\tfrac{203}{3}$ & $(V_{29},V_{35},V_{41})$ \\
   
   & $(\rep1,\rep1)_{-\frac13}$ & $\rep3_2$ & $\rep2'\oplus\rep1$ & $1$ & $\nicefrac{-2}3$ & $\w^2$ & $\w^2$ & $1$ & $\rep1$ &
   $\tfrac43$ & $-\tfrac13$ & $\tfrac73$ & $-\tfrac{94}{3}$ & $\tfrac{41}{3}$ & $-\tfrac{188}{3}$ & $-\tfrac{155}{3}$ & $-\tfrac{280}{3}$ & $\tfrac{188}{3}$ & $(V_{30},V_{36},V_{42})$ \\
   
   & $(\rep1,\rep1)_{-\frac13}$ & $\rep3_2$ & $\rep2'\oplus\rep1$ & $1$ & $\nicefrac{-2}3$ & $\w^2$ & $\w^2$ & $\w$ & $\rep1$ &
   $-\tfrac73$ & $\tfrac23$ & $-\tfrac{17}{3}$ & $\tfrac{11}{3}$ & $\tfrac{35}{3}$ & $\tfrac{160}{3}$ & $\tfrac{34}{3}$ & $\tfrac{200}{3}$ & $-\tfrac{112}{3}$ & $(V_{31},V_{37},V_{43})$ \\
   
   & $(\rep1,\rep1)_{-\frac13}$ & $\rep3_2$ & $\rep2'\oplus\rep1$ & $1$ & $\nicefrac{-2}3$ & $\w^2$ & $\w^2$ & $\w$ & $\rep1$ &
   $    0$ & $\tfrac23$ & $-\tfrac23$ & $-\tfrac{40}{3}$ & $-\tfrac{64}{3}$ & $-\tfrac{11}{3}$ & $\tfrac{376}{3}$ & $-\tfrac{568}{3}$ & $-\tfrac{217}{3}$ & $(V_{32},V_{38},V_{44})$ \\
   
   & $(\rep1,\rep1)_{-\frac13}$ & $\rep3_2$ & $\rep2'\oplus\rep1$ & $1$ & $\nicefrac{-2}3$ & $\w^2$ & $\w^2$ & $\w^2$ & $\rep1$ &
   $    1$ & $-\tfrac13$ & $\tfrac{10}{3}$ & $\tfrac{83}{3}$ & $-\tfrac{76}{3}$ & $-\tfrac{191}{3}$ & $-\tfrac{149}{3}$ & $\tfrac{8}{3}$ & $\tfrac{203}{3}$ & $(V_{33},V_{39},V_{45})$ \\
   
   & $(\rep1,\rep1)_{\frac13}$ & $\rep3_2$ & $\rep2'\oplus\rep1$ & $1$ & $\nicefrac{-2}3$ & $\w^2$ & $1$ & $1$ & $\rep1$ &
   $-\tfrac{10}{3}$ & $-\tfrac13$ & $-\tfrac53$ & $\tfrac{56}{3}$ & $   -3$ & $-\tfrac{62}{3}$ & $-\tfrac{289}{3}$ & $\tfrac{448}{3}$ & $-\tfrac{98}{3}$ & $(\bar{V}_{28},\bar{V}_{34},\bar{V}_{40})$ \\ 
   
   & $(\rep1,\rep1)_{\frac13}$ & $\rep3_2$ & $\rep2'\oplus\rep1$ & $1$ & $\nicefrac{-2}3$ & $\w^2$ & $1$ & $1$ & $\rep1$ &
   $    3$ & $\tfrac23$ & $\tfrac{10}{3}$ & $\tfrac53$ & $  -36$ & $-\tfrac{14}{3}$ & $\tfrac{146}{3}$ & $-\tfrac{152}{3}$ & $\tfrac{82}{3}$ & $(\bar{V}_{29},\bar{V}_{35},\bar{V}_{41})$ \\
   
   & $(\rep1,\rep1)_{\frac13}$ & $\rep3_2$ & $\rep2'\oplus\rep1$ & $1$ & $\nicefrac{-2}3$ & $\w^2$ & $1$ & $\w$ & $\rep1$ &
   $-\tfrac{7}{3}$ & $-\tfrac{1}{3}$ & $\tfrac{7}{3}$ & $-\tfrac{55}{3}$ & $  -39$ & $-\tfrac{149}{3}$ & $-\tfrac{115}{3}$ & $\tfrac{208}{3}$ & $-\tfrac{143}{3}$ & $(\bar{V}_{30},\bar{V}_{36},\bar{V}_{42})$ \\
   & $(\rep1,\rep1)_{\frac13}$ & $\rep3_2$ & $\rep2'\oplus\rep1$ & $1$ & $\nicefrac{-2}3$ & $\w^2$ & $1$ & $\w$ & $\rep1$ &
   $    3$ & $\tfrac23$ & $-\tfrac{14}{3}$ & $-\tfrac73$ & $    4$ & $\tfrac{34}{3}$ & $\tfrac{50}{3}$ & $-\tfrac{464}{3}$ & $\tfrac{82}{3}$ & $(\bar{V}_{31},\bar{V}_{37},\bar{V}_{43})$ \\
   & $(\rep1,\rep1)_{\frac13}$ & $\rep3_2$ & $\rep2'\oplus\rep1$ & $1$ & $\nicefrac{-2}3$ & $\w^2$ & $1$ & $\w^2$ & $\rep1$ &
   $    2$ & $\tfrac23$ & $-\tfrac23$ & $\tfrac{116}{3}$ & $    0$ & $\tfrac{73}{3}$ & $-\tfrac{28}{3}$ & $\tfrac{88}{3}$ & $\tfrac{127}{3}$ & $(\bar{V}_{32},\bar{V}_{38},\bar{V}_{44})$ \\
   & $(\rep1,\rep1)_{\frac13}$ & $\rep3_2$ & $\rep2'\oplus\rep1$ & $1$ & $\nicefrac{-2}3$ & $\w^2$ & $1$ & $\w^2$ & $\rep1$ &
   $-\tfrac73$ & $-\tfrac13$ & $-\tfrac{17}{3}$ & $-\tfrac{67}{3}$ & $    1$ & $-\tfrac{101}{3}$ & $-\tfrac{211}{3}$ & $-\tfrac{104}{3}$ & $-\tfrac{143}{3}$ & $(\bar{V}_{33},\bar{V}_{39},\bar{V}_{45})$ \\
   & $(\rep1,\rep1)_{-\frac23}$ & $\rep3_2$ & $\rep2'\oplus\rep1$ & $1$ & $\nicefrac{-2}3$ & $\w^2$ & $1$ & $1$ & $\rep1$ &
   $\tfrac13$ & $-\tfrac13$ & $-\tfrac53$ & $-\tfrac{61}{3}$ & $   39$ & $\tfrac{76}{3}$ & $\tfrac{143}{3}$ & $-\tfrac{296}{3}$ & $\tfrac{16}{3}$ & $(X_6,X_9,X_{12})$ \\
   & $(\rep1,\rep1)_{-\frac23}$ & $\rep3_2$ & $\rep2'\oplus\rep1$ & $1$ & $\nicefrac{-2}3$ & $\w^2$ & $1$ & $\w$ & $\rep1$ &
   $-\tfrac23$ & $-\tfrac13$ & $\tfrac{7}{3}$ & $\tfrac{62}{3}$ & $   35$ & $\tfrac{115}{3}$ & $\tfrac{65}{3}$ & $\tfrac{256}{3}$ & $\tfrac{61}{3}$ & $(X_7,X_{10},X_{13})$ \\
   & $(\rep1,\rep1)_{-\frac23}$ & $\rep3_2$ & $\rep2'\oplus\rep1$ & $1$ & $\nicefrac{-2}3$ & $\w^2$ & $1$ & $\w^2$ & $\rep1$ &
   $\tfrac13$ & $-\tfrac13$ & $\tfrac{19}{3}$ & $-\tfrac{49}{3}$ & $   -1$ & $\tfrac{28}{3}$ & $\tfrac{239}{3}$ & $\tfrac{16}{3}$ & $\tfrac{16}{3}$ & $(X_8,X_{11},X_{14})$ \\
   & $(\rep1,\rep1)_{\frac23}$ & $\rep3_2$ & $\rep2'\oplus\rep1$ & $1$ & $\nicefrac{-2}3$ & $\w^2$ & $\w^2$ & $1$ & $\rep1$ &
   $    1$ & $-\tfrac13$ & $-\tfrac{14}{3}$ & $\tfrac{71}{3}$ & $\tfrac{44}{3}$ & $\tfrac{76}{3}$ & $\tfrac{25}{3}$ & $-\tfrac{232}{3}$ & $-\tfrac{76}{3}$ & $(\bar{X}_{6},\bar{X}_{9},\bar{X}_{12})$ \\
   & $(\rep1,\rep1)_{\frac23}$ & $\rep3_2$ & $\rep2'\oplus\rep1$ & $1$ & $\nicefrac{-2}3$ & $\w^2$ & $\w^2$ & $1$ & $\rep1$ &
   $\tfrac43$ & $-\tfrac13$ & $\tfrac73$ & $-\tfrac{94}{3}$ & $\tfrac{41}{3}$ & $\tfrac{31}{3}$ & $\tfrac{115}{3}$ & $-\tfrac{208}{3}$ & $-\tfrac{91}{3}$ & $(\bar{X}_{7},\bar{X}_{10},\bar{X}_{13})$ \\
   & $(\rep1,\rep1)_{\frac23}$ & $\rep3_2$ & $\rep2'\oplus\rep1$ & $1$ & $\nicefrac{-2}3$ & $\w^2$ & $\w^2$ & $\w^2$ & $\rep1$ &
   $    1$ & $-\tfrac13$ & $\tfrac{10}{3}$ & $\tfrac{83}{3}$ & $-\tfrac{76}{3}$ & $\tfrac{28}{3}$ & $\tfrac{121}{3}$ & $\tfrac{80}{3}$ & $-\tfrac{76}{3}$ & $(\bar{X}_{8},\bar{X}_{11},\bar{X}_{14})$ \\
\end{longtable}
}
}
\end{landscape}

{\small
\providecommand{\bysame}{\leavevmode\hbox to3em{\hrulefill}\thinspace}

}
\end{document}